\documentclass[preprintnumbers,twocolumn,groupedaddress,nofootinbib,aps]{revtex4}
\usepackage{float}
\usepackage{amsmath,amssymb}
\usepackage{graphicx}
\usepackage{color}
\usepackage{hyperref}
\usepackage{url}
\usepackage{slashed}
\usepackage{subfigure}
\usepackage{slashed,bbm}
\usepackage{graphics,psfrag,epsfig}
\usepackage{dsfont}
\usepackage{enumerate}
\usepackage{textcomp}
\usepackage{bm}
\usepackage{lmodern}
\usepackage{lmodern}
\usepackage[latin9]{inputenc}
\usepackage{bm}
\usepackage{amsmath}
\usepackage{amssymb}
\usepackage{graphicx}
\usepackage{esint}
\usepackage{verbatim}

\newcommand{\RNum}[1]{\uppercase\expandafter{\romannumeral #1\relax}}

\newcommand{\abs}[1]{ \left|  #1 \right| }

\newcommand{\beq}{\begin{equation}}
\newcommand{\eeq}{\end{equation}}
\newcommand{\bea}{\begin{eqnarray}}
\newcommand{\eea}{\end{eqnarray}}
\newcommand{\be}{\begin{equation}}
\newcommand{\ee}{\end{equation}}

\newcommand{\Ut}[1]{\tilde{U}}
\newcommand{\Utt}[1]{\tilde{\tilde{U}}}

 \definecolor{BLACK}{gray}{0}
 \definecolor{WHITE}{gray}{1}
 \definecolor{RED}{rgb}{1,0,0}
 \definecolor{GREEN}{rgb}{0,1,0}
 \definecolor{BLUE}{rgb}{0,0,1}
 \definecolor{CYAN}{cmyk}{1,0,0,0}
 \definecolor{MAGENTA}{cmyk}{0,1,0,0}
 \definecolor{YELLOW}{cmyk}{0,0,1,0}

\makeatother

\begin{document}

\title{Competing instabilities, orbital ordering and splitting of band degeneracies from a parquet renormalization group analysis of a 4-pocket model for iron-based superconductors: application to FeSe}

\author{Rui-Qi Xing$^{1}$,  Laura Classen$^{2}$, Maxim Khodas$^{3}$, and Andrey V. Chubukov$^{1}$
}
\affiliation{$^{1}$ School of Physics and Astronomy, University of Minnesota, Minneapolis, MN 55455, USA,\\
$^{2}$ Institut f\"ur Theoretische Physik, Universit\"at Heidelberg, 69120 Heidelberg, Germany,\\
$^{3}$  Racah Institute of Physics, The Hebrew University, Jerusalem 91904, Israel \\}

\begin{abstract}
We report the results of a parquet renormalization group (RG) study of competing instabilities in the full 2D four pocket, three orbital low-energy model for iron-based superconductors.
   We derive and analyze the RG flow of  the couplings, which describe all symmetry-allowed interactions between low-energy fermions. Despite that the number of the couplings is large, we argue that there are only two stable fixed trajectories of the RG flow
     and one weakly unstable fixed trajectory with a single unstable direction. Each fixed trajectory has a finite basin of attraction in the space of initial system parameters. On the stable trajectories, either interactions involving only $d_{xz}$ and $d_{yz}$  or only $d_{xy}$  orbital components on electron pockets dominate, while
on the weakly unstable  trajectory interactions involving $d_{xz}$ ($d_{yz}$) and $d_{xy}$ orbital states on electron pockets remain comparable.
The behavior along the two stable fixed trajectories has been analyzed earlier [A.V. Chubukov, M. Khodas, and R.M. Fernandes, arXiv:1602.05503]. Here we focus on the system behavior along the weakly unstable trajectory
     and apply the results to FeSe.
     We find, based on the analysis of susceptibilities along this trajectory,
    that the leading instability upon lowering the temperature is towards a {\it three-component} d-wave orbital nematic order.
    Two components are the differences between fermionic densities  on $d_{xz}$ and $d_{yz}$ orbitals on hole pockets and on electron pockets, and the third one is  the difference between the densities of $d_{xy}$ orbitals on the two electron pockets.  We argue that this order is consistent with the splitting of band degeneracies, observed in recent photoemission data on FeSe by A. Fedorov et al [arXiv:1606.03022].
 \end{abstract}

\maketitle

\section{Introduction}

The interplay and competition between different types of electronic order is at the focus of the research on iron based superconductors (FeSCs)~\cite{Johnston2010,kotliar,Dai2012,Chubukov_rev,review,dagotto}.
In most FeSCs  superconductivity (SC) emerges out of a stripe spin-density-wave (SDW) state upon either hole or electron doping, application of pressure, or by isovalent substitution of one pnictogen atom by the other (e.g., As by P). The SDW phase is often preceded by the nematic phase, in  which the system breaks  $C_4$ rotational symmetry down to $C_2$ but keeps spin-rotational symmetry intact.

 The nematic phase has been extensively studied both experimentally and theoretically~\cite{orb_order1,ku,kruger,leni,phillips,singh,brydon,rafael,stanev,moreo,gastiasoro,cfs_15,chu_r,Dumitrescu2015,Gallais2016,Wang2015,Thormsolle2016}.
The manifestations of spontaneous $C_4$ symmetry breaking include the anisotropy of resistivity~\cite{Fisher1,Tanatar2010,Fisher2}, spin susceptibility~\cite{spin_chi,spin_chi1,spin_chi2}, and optical conductivity~\cite{degeorgio1,degeorgio2},
 orthorhombic lattice distortion~\cite{lattice_dist,lattice_dist1}, and unequal occupation of Fe $d_{xz}$ and $d_{yz}$  orbitals~\cite{orbital_arpes,orbital_arpes1}.
   The majority of researchers believe that nematicity is driven by electronic degrees of freedom rather than by the lattice. There is no agreement, however, on the mechanism of the nematic order. It can be a composite Ising-nematic magnetic order~\cite{sdw_stripe}, preceding stripe SDW order, or a quantum-disordered spin state, which  breaks $C_4$ symmetry~\cite{lee}
   or a
     spontaneous orbital order~\cite{orb_order1,ku,kruger,phillips,singh,stanev,Dumitrescu2015}.  The Ising-nematic scenario likely applies to Fe-pnictides, in which the nematic phase is located in a close proximity to a stripe SDW phase.  However,  the application of this scenario  to  Fe-chalcogenide  FeSe is questionable because in FeSe at ambient pressure the
nematic transition  occurs at $T_s = 85$K, but there is no SDW transition down to $T=0$. The Ising-nematic scenario, particularly when combined with the idea of a weak dispersion of spin excitations along one direction in momentum space~\cite{mazin_1}, can still be the explanation because $T_s$ and $T_{SDW}$ do not have to be close to each other. However,
 NMR~\cite{Baek2014,Bohmer2015} and neutron scattering~\cite{neutrons_FeSe} experiments have found that the magnetic correlation length  does not show
 any notable  enhancement around $T_s$, which would be generally expected in the Ising-nematic scenario.  Substantial SDW fluctuations have been detected
 only at lower temperatures~\cite{kothapalli2016}, or upon applying pressure~\cite{pressure_2016}, when the system eventually develops an SDW order.

The fact that in FeSe at ambient pressure nematic order emerges without magnetism fuelled speculations that in this system nematicity may
 be due to a spontaneous orbital ordering.
   The most natural $C_4$ symmetry-breaking orbital order is associated with unequal occupation of $d_{xz}$ and $d_{yz}$ orbitals.
     In FeSe, these two orbitals are the building blocks for the low-energy states near both hole and electron pockets.
  The electronic structure of FeSe consists of two $\Gamma-$centered hole pockets and two electron pockets centered at $X = (\pi,0)$ and $Y = (0,\pi)$ in the 1Fe Brillouin zone (BZ)  (see Fig.~\ref{fig:FS_Sketch}).
   The $\Gamma-$centered hole pockets are made out of  $d_{xz}$ and $d_{yz}$ orbitals. The electron pockets are made out of these two orbitals and the $d_{xy}$ orbital.
More precisely, the pocket near X is made out of $d_{yz}$ and $d_{xy}$  orbitals, and the one near Y is made out of $d_{xz}$ and $d_{xy}$  orbitals.   Accordingly, one can introduce
 three $C_4$ breaking orbital order parameters. Two involve $d_{xz}$ and $d_{yz}$ orbitals: $\Gamma_{1,h} = \sum_k  d^\dagger_{xz} (k) d_{xz} (k) - d^\dagger_{yz} (k) d_{yz} (k)$  and $\Gamma_{1,e} = \sum_k d^\dagger_{xz} (k+Y) d_{xz} (k+Y) - d^\dagger_{yz} (k+X) d_{yz} (k+X)$,
 and the third one, $\Gamma_{2,e} = \sum_k d^\dagger_{xy} (k+Y) d_{xy} (k+Y) - d^\dagger_{xy} (k+X) d_{xy} (k+X)$,
  describes unequal occupation of the $d_{xy}$ orbital near $X$ and $Y$ electron pockets and induces an X/Y anisotropy of the hopping integral for the $d_{xy}$ orbital~\cite{fern_vaf}.
    Here and below the summation over $k$ is restricted to small $k$.

      All three order parameters, $\Gamma_{1,h}$, $\Gamma_{1,e}$, and $\Gamma_{2,e}$ belong to the same $B_{1g}$ representation of the point group $D_{4h}$~\cite{Cvetkovic2013}
       and break the same $C_4$ symmetry.
   The order parameter $\Gamma_{1,h}$ gives rise to elliptical elongation of the two hole pockets and  splits the two hole dispersions at the $\Gamma$ point.
    The order parameters $\Gamma_{1,e}$ and $\Gamma_{2,e}$ change the shape of  electron pockets and split the dispersions of $d_{xz}/d_{yz}$
       and $d_{xy}$ orbitals between $X$ and $Y$  pockets.

Recent ARPES experiments~\cite{borisenko,coldea,brouer,zhang2015} analyzed relative signs and magnitudes of the three order parameters $\Gamma_{1,h}, ~\Gamma_{1,e}$, and $\Gamma_{2,e}$, and the results of these experiments place constraints on theoretical considerations.  The ARPES data is taken in the
 2-Iron Brillouin Zone (2FeBZ), which is the physical BZ, because Se atoms in FeSe are located above and below the Fe plane in a ches-type order.
        In the 2FeBZ, both electron pockets are located at the $M$ point ($k_x=k_y = \pi$).
         Above $T_s$,  $d_{xz}$  and $d_{yz}$ dispersions are degenerate at $M$,
       even in the presence of spin-orbit coupling~\cite{Cvetkovic2013}. A non-zero $\Gamma_{1,e}$ splits the two dispersions by $\pm \Gamma_{1,e}$.  Similarly, the two
        $d_{xy}$ dispersions from $X$ and $Y$ pockets are degenerate at the M point above $T_s$, but split in the nematic phase by $\pm \Gamma_{2,e}$.
        The authors of Refs. [\onlinecite{borisenko}] reported that they detected the splitting of both, $d_{xz}/d_{yz}$ and $d^X_{xy}/d^Y_{xy}$, bands at the M-point. Both splittings are found to be around $15$meV,  what implies that the magnitudes of $\Gamma_{1,e}$ and $\Gamma_{2,e}$ are nearly equal
 ($|\Gamma_{1,e}| \sim |\Gamma_{2,e}| \sim 7.5$ meV).  These authors  also reported that they  detected a 20 meV spin-orbit induced splitting of $d_{xz}$ and $d_{yz}$ bands at the $\Gamma$-point above $T_s$, and  that this splitting increases to 25 meV in the nematic phase. The full splitting at $\Gamma$ is $\pm \sqrt{\Gamma^2_{so} + \Gamma^2_{1,h}}$
  (Ref. \cite{fern_vaf}).  Using this formula,  one extracts from the data
   $|\Gamma_{so}| =10$ meV and $|\Gamma_{1,h}| =7.5$ meV. The outcome is that all three  order parameters, $\Gamma_{1,h}$, $\Gamma_{1,e}$, and $\Gamma_{2,e}$ have about the same magnitude of $7.5$ meV.  Other ARPES groups~\cite{coldea,brouer,zhang2015} interpreted their data somewhat differently, and some reported larger $\Gamma_{1,e}$, and $\Gamma_{2,e}$.
     In a separate development,  the authors of  Ref.~\cite{suzuki} argued, based on their  ARPES results, that $\Gamma_{1,h}$ and $\Gamma_{1,e}$ have opposite signs.

   In this paper we analyze whether the near-equivalence of the magnitudes of $\Gamma_{1,h}$, $\Gamma_{1,e}$, and $\Gamma_{2,e}$  and the sign difference between
    $\Gamma_{1,h}$ and $\Gamma_{1,e}$
     can be understood theoretically.
       In our theory, we  obtain the ratios of the order parameters near $T_s$, when the magnitudes of all
     condensates are small.  We do find the near-equivalence of $\Gamma_{1,e}$ and $\Gamma_{2,e}$ and the sign change between $\Gamma_{1,h}$, and $\Gamma_{1,e}$. The ratio of $\Gamma_{1,h}$ and $\Gamma_{1,e}$ comes out larger in our analysis than in the ARPES data, but we caution that our calculations do not include spin-orbit coupling, which by itself splits $d_{xz}$ and $d_{yz}$ orbitals at the $\Gamma$ point.

  Our analysis is build on recent parquet renormalization group (RG) studies of orbital order in FeSCs. In Ref.~\cite{CKF2016},  Chubukov, Khodas, and Fernandes (CKF)
  analyzed the interplay between SDW, SC, and orbital order in two approximate 4-pocket models for FeSe.  In both models the hole pockets were treated without an approximation, but the  electron pockets were  assumed to be made entirely out of $d_{xz}/d_{yz}$ orbitals (model I), or entirely out of $d_{xy}$ orbitals (model II).  This was done to reduce the number of running RG couplings to 14, down from  30 in the full model (see below).  For both models, CKF found
 that the leading instability upon lowering the temperature is towards an orbital order, the subleading one is towards $s^{+-}$  superconductivity,
  and SDW order does not develop, despite that the SDW susceptibility is the largest at the beginning of the RG flow.
   This hierarchy of instabilities holds if the pockets are small enough
   and RG has a "space" to run, i.e., there is enough energy scales to integrate out.

      CKF  did  find that
       the sign of $\Gamma_{1,h}$ is opposite to that of $\Gamma_{1,e}$, in agreement with the ARPES analysis in ~\cite{suzuki}.  However, they could not explain the observed near-equivalence between  $\Gamma_{1,e}$ and $\Gamma_{2,e}$ at the M point
        because, by construction,
  in the two
   approximate models studied by CKF, either $\Gamma_{2,e} =0$ (model I) or $\Gamma_{1,e} =0$ (model II).

  In this paper we extend the analysis of CKF to the full 4-pocket, 3-orbital model of FeSCs.  The goal is two-fold: (1) verify whether the hierarchy of instabilities remains the same as in the approximate models studied by CKF, and (2) see whether the relations between $\Gamma_{1,e}, \Gamma_{1,h}$, and $\Gamma_{2,e}$  reproduce the ones extracted from the ARPES measurements.  The 4-pocket, 3-orbital low-energy model has been introduced by Cvetkovic and Vafek in Ref.~\cite{Cvetkovic2013}. These authors have shown that the number of different symmetry-allowed interactions between low-energy fermions is equal to 30. The initial values of all 30 couplings are expressed via local Hubbard and Hund interactions $U$, $U'$, $J$, $J'$.
   But the couplings evolve differently as one progressively integrates out fermions with higher energies, i.e., in the process of the RG flow the system self-generates longer-range interactions.
   We derive and analyze, both analytically and numerically, the set of
  30 coupled parquet RG equations, which describe  the flow of the couplings. We show that the flow  is towards universal fixed trajectories, along which the ratios between the couplings tend to fixed values.
    We then derive another set of RG equations for the  susceptibilities in different channels (SDW, SC, orbital) and solve them using the running couplings
  as inputs~\cite{fRG,rice,honerkamp,fRG_thomale2,fRG_lee}. We identify the channel in which the system first develops an instability as the one where the susceptibility diverges at the highest $T$, and, if critical $T$ are the same in several channels, as the one where the divergent susceptibility has the largest exponent.

We show that two of the universal fixed trajectories are stable, and that they are separated by several unstable fixed trajectories. [The system approaches a stable fixed trajectory from all directions within its basin of attraction,
 it  approaches an unstable fixed trajectory
from some directions
 and
  moves away from it along other directions].  We argue that on a
   stable fixed trajectory the system behavior
    effectively reduces to that of one of the two models considered by CKF. Specifically, on one stable trajectory, interactions involving $d_{xy}$ components of the electron pockets vanish compared to the interactions involving $d_{xz}$ (or $d_{yz}$) components (same as in model I of CKF), while along the other stable fixed trajectory interactions involving $d_{xz}$ (or $d_{yz}$) orbital components vanish compared to the interactions involving $d_{xy}$ components (model II of CKF).  Like we said,  each of these two models yields the same hierarchy of orderings (orbital order, then SC, but no SDW, if the pockets are small
 enough).
    However, neither model I, nor model II,  reproduces the observed near-equivalence of $\Gamma_{1,e}$ and $\Gamma_{2,e}$.

We next analyze the unstable fixed trajectories. In general, these  trajectories are irrelevant for the RG analysis, because the RG flow moves the system
  away from these trajectories towards the stable ones.  In our case,
    however,
     the system behavior is more nuanced. Namely, we show that there are several truly unstable fixed
    trajectories and one "weakly unstable'' fixed trajectory with just one
    direction, along which the system eventually moves away from it (i.e., the stability analysis yields
     one positive exponent).   This weakly unstable
     fixed trajectory is located in between
 the two stable fixed trajectories.
    We argue that under RG  the system first flows away from truly unstable trajectories  towards the weakly unstable trajectory,  and then
     flows towards one of the two stable fixed trajectories.
     However, the positive exponent, which characterizes
     how fast
        the system moves away from this trajectory,
        is quite small. This implies that the weakly unstable fixed trajectory behaves as a stable one  nearly up to the very end of the RG flow, when the
   hierarchy of susceptibilities is already established.
      We analyze
        the system behavior on this weakly unstable fixed trajectory and obtain
         the same sequence of orderings as the two stable trajectories, i.e., the leading instability is towards $C_4$-breaking orbital order, the subleading is towards $s^{+-}$ SC, and SDW order does not develop.
  In distinction to the stable fixed trajectories, however, now the interactions involving $d_{xz} (d_{yz})$ and $d_{xy}$ orbital components on electron pockets are of the same order. Therefore $\Gamma_{1,h}$,  $\Gamma_{1,e}$, and $\Gamma_{2,e}$ all become non-zero once the orbital order sets in.
   We solve the set of coupled equations for $\Gamma_{1,h}, \Gamma_{1,e}$, and $\Gamma_{2,e}$
     on the weakly unstable fixed trajectory and find that the magnitudes of $\Gamma_{1,e}$, and $\Gamma_{2,e}$  are nearly equal, and the  signs of  $\Gamma_{1,h}$ and $\Gamma_{1,e}$ are opposite.
    This is fully consistent with the ARPES data~\cite{borisenko,coldea,suzuki}.
      We view the agreement as an indication that the parquet RG analysis of the full 4-pocket, 3-orbital model  is capable to reproduce not only the sequence of phase transitions in FeSe upon lowering of temperature, but also the ARPES results for the magnitudes and signs of the nematic  orbital order parameters.
        At the same time, our analysis yields a larger ratio of $\Gamma_{1,h}/\Gamma_{1,e}$ than in the data. This may
         reveal
         a limited validity of
         the RG analysis.
            But the discrepancy  may also be due to the fact
            that, according to ARPES~\cite{borisenko},
             the largest splitting of $d_{xz}$ and $d_{yz}$ orbitals on hole pockets
             comes from spin-orbit coupling, which we did not include into the analysis. We conjecture that the feedback from spin-orbit-induced band splitting reduces the value of the orbital order parameter on hole pockets compared to our result, which, we reiterate, is obtained neglecting spin-orbit interaction.

 The paper is organized as follows.
 In Sec.~\ref{sec:model} we discuss the generic 4-band, 3-orbital model for FeSCs.
 We first present the kinetic energy and then introduce the 30 different $C_4$-symmetric interactions between low-energy fermions.  We argue that the structure of the
 interaction Hamiltonian remains invariant under RG, but the values of the couplings flow.
 In  Sec.~\ref{sec:pRG} we derive and solve the set of 30 coupled differential RG  equations for the flow of the couplings.
 In Sec.~\ref{sec:traj} we analyze the fixed trajectories resulting from the solution of the RG equations.
 Because the system behavior along the two stable fixed trajectories is the same as in the two approximate models studied by CKF, we do not
 re-derive the results here and instead
 focus on the system behavior along the weakly unstable fixed trajectory.
 In Sec.~\ref{sec:inst} we discuss the RG flow of the susceptibilities in different channels and analyze the   hierarchy of the instabilities on the weakly unstable fixed trajectory.
  In Sec.~\ref{sec:implication_exp} we discuss the interplay between the three orbital order parameters $\Gamma_{1,h}, \Gamma_{1,e}$, and $\Gamma_{2,e}$  and compare our results with the  ARPES data.
 We present our conclusions  in Sec.~\ref{sec:concl}.  Technical details of the RG analysis are presented in the Appendix.

In the complimentary work~\cite{cxkc}, we applied parquet RG to 5-band, 3-orbital  model with an additional $d_{xy}$ pocket at $(\pi,\pi)$ in the 1Fe BZ (at $\Gamma$ in the 2Fe BZ).
We argued that in some range of input parameters the fifth pocket does not affect the  low-energy behavior, and remains the same as in the 4-pocket, 3-orbital  model.

 \section{The Model}
 \label{sec:model}

We consider a four-band model with two hole pockets at the center of the 1FeBZ and two electron pockets at the zone edges,   and keep the actual orbital content of
 low-energy excitations near each pocket (See Fig.~\ref{fig:FS_Sketch}). Each of the two hole pockets have orbital character alternating between $d_{xz}$ and $d_{yz}$, with negligible contribution from  $d_{xy}$ and other orbitals. Of the two symmetry-related electron pockets, one is constructed from $d_{xz}$ and $d_{xy}$ orbitals, the other from $d_{yz}$ and $d_{xy}$
  orbitals, again with negligible contribution from other orbitals. Below we first consider the effective model for the low-energy band structure and then construct the interactions,
  consistent with the tetragonal crystal symmetry above the nematic transition.

 \subsection{Effective model of the band structure}
 \label{sec:bands}

\begin{figure}[b]
	\centering{}\includegraphics[width=0.95\columnwidth]{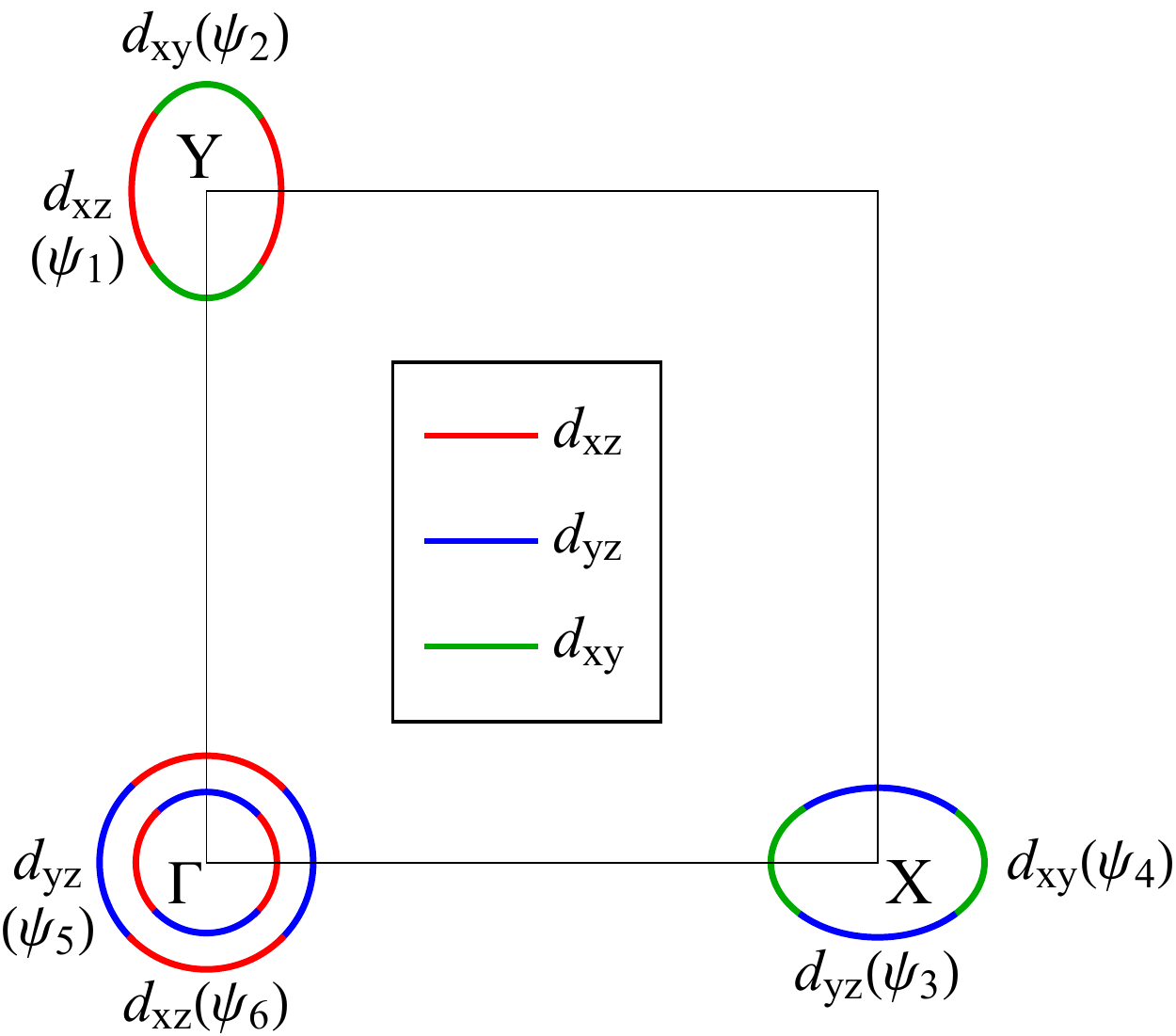} \protect\caption{The Fermi surfaces in 1Fe BZ with the orbital content of the interactions.
  The six $\psi$ fields are introduced in the text.}
		\label{fig:FS_Sketch}
\end{figure}

We follow the approach by Cvetkovic and Vafek,~\cite{Cvetkovic2013} who used
 Luttinger's method of invariants (also known as $k\cdot p$ theory) and symmetry constraints to construct the effective low energy model of the band structure.
  We  neglect spin-orbit coupling, assuming that is does not affect the RG flow at energies above $E_F$,  and perform calculations in the unfolded 1FeBZ.
  Because we are interested in the low-energy theory, fermions near different pockets  are treated as different species.
  Namely, we introduce a
 6-component
  spinor  $\psi=(\psi_1,\ldots,\psi_6)$ (See Fig.~\ref{fig:FS_Sketch}).
The components
$\psi_{1}({\bm{k}})$ and $\psi_{2}({\bm{k}})$ are the Bloch states of pure $d_{xz}$ and $d_{xy}$ orbital character, respectively, with momentum near $Y$ (spin indices are omitted for clarity).
The components $\psi_{3}({\bm{k}})$ and $\psi_{4}({\bm{k}})$ are the Bloch states of pure $d_{yz}$ and $d_{xy}$ orbital character near $X$, and $\psi_{5}({\bm{k}})$ and $\psi_{6}({\bm{k}})$ are the Bloch states of pure $d_{yz}$ and $d_{xz}$ orbital character near
 $\Gamma$.
We list  pocket and orbital "affiliations'' of $\psi_i$ in Table \ref{tab:affiliations} (see  also Fig.\ref{fig:FS_Sketch}).
\begin{table}[H]
	\begin{center}
		\begin{tabular}{ | c | c | c | }
			\hline
			$\psi_i$ & Pocket & Orbital \\ \hline
			$\psi_1$ & Y & $d_{xz}$ \\ \hline
			$\psi_2$ & Y & $d_{xy}$ \\ \hline
			$\psi_3$ & X & $d_{yz}$ \\ \hline
			$\psi_4$ & X & $d_{xy}$ \\ \hline
			$\psi_5$ & $\Gamma$ & $d_{yz}$ \\ \hline
			$\psi_6$ & $\Gamma$ & $d_{xz}$ \\
			\hline
		\end{tabular}
	\end{center}
	\caption{\label{tab:affiliations}Affiliation of $\psi_i$ with a pocket and an orbital.}
\end{table}
The non-interacting part of the effective Hamiltonian is expressed as
\begin{equation}\label{H_0}
H{}_{0}=\sum_{\bm{k},\sigma}\psi_{\sigma}^{\dagger}(\bm{k})\begin{pmatrix}h_{Y}(\bm{k})\\
& h_{X}(\bm{k})\\
&  & h_{\Gamma}(\bm{k})
\end{pmatrix}\psi_{\sigma}(\bm{k})\, ,
\end{equation}
In Eq.~\eqref{H_0} and in what follows all momenta $\bm{k}$ are counted relative to the respective high symmetry points $\Gamma$, $X$, or $Y$.
 The $d_{xz}$ dispersion at $Y$ and the $d_{yz}$ dispersion at $X$  are doubly degenerate, and the $d_{xy}$ dispersions at $X$ and $Y$ are also degenerate 
(see left panel in Fig.~\ref{fig:M_splitting}).
 These two degeneracies can be traced to the fact that chalcogen atoms in Fe-chalcogenides (Se in FeSe) or pnictogen atoms in  Fe-pnictides  are located above or below the Fe plane in chess-like order.
 In group theory language,  this "up-down" location of chalcogen/pnictogen atoms implies that  the symmetry group P4/nmm
  contains a glide plain symmetry element. The corresponding symmetry operation is a mirror reflection about the iron plane, followed by a translation by one lattice side along $X$ or $Y$ directions in the 1FeBZ or, equivalently, along the half of the unit cell diagonal in the actual 2FeBZ (see e.g., Ref.~\cite{Cvetkovic2013}).
 The symmetry group P4/nmm containing this glide plain symmetry is a nonsymmorphic group and therefore all physical irreducible representations are two-dimensional at M point (X and Y point in 1FeBZ map to M point in 2FeBZ; see left panel of Fig.~\ref{fig:M_pocket}), implying any states are doubly degenerate at M point. This is reflected in the expressions for $ h_{X}$ and $h_{Y}$ in  Eq. (\ref{H_0}).
 We  have
 \begin{equation}
	h_{Y,X}(\bm{k})\!=\!\!\begin{pmatrix}\!\epsilon_{1}\!+\!\frac{k^{2}}{2m_{1}}\!\!\pm\!a_{1}(k_{y}^{2}\!-\!k_{x}^{2}) & \!-\!iv_{\pm}(k)\\
		iv_{\pm}(k) & \!\!\epsilon_{2}\!+\!\frac{k^{2}}{2m_{2}} \!\pm \!a_{2}(k_{y}^{2}\!-\!k_{x}^{2})\!
	\end{pmatrix}
\label{ch_1_1}
\end{equation}
where the upper sign is for the $Y$ pocket and the lower one is for the $X$ pocket,
  $v{}_{+}(k)=v(2k_{x})+ O(k^3)$,
$v{}_{-}(k)=v(-2k_{y})+ O(k^3)$,
and  $v, \epsilon_{1,2}$, $a_{1,2}$, and $m_{1,2}$
 are parameters, which can be determined by fitting the band structure to ARPES data.
 The two pockets are interchangeable under a $90^{\circ}$ rotation, $h_{X}(k_x,k_y) = h_{Y}^{*}(-k_y,k_x)$.

The hole pockets are described by the effective Hamiltonian
 \begin{equation}
\small
\arraycolsep=-2pt
	h_{\Gamma}(\bm{k})\!=\!\begin{pmatrix}\epsilon_{3}-\frac{k^{2}}{2m_{3}}+b(k_{y}^{2}-k_{x}^{2}) & 2ck_{x}k_{y}\\
		2ck_{x}k_{y} & \epsilon_{3}-\frac{k^{2}}{2m_{3}}-b(k_{y}^{2}-k_{x}^{2})
\label{ch_1}
	\end{pmatrix}.
\end{equation}
The parameters $\epsilon_{3}$, $m_3$, $b$ and $c$ are again
 determined by fitting the band structure
  to ARPES data.  Note that these parameters generally differ from the ones obtained by  taking tight-binding LDA dispersion and expanding it near $\Gamma$, $X$, and $Y$ points,
   as Eqs.  (\ref{ch_1_1}) and (\ref{ch_1}) include regular self-energy corrections coming from
   high-energy fermions.

The band dispersions are obtained by diagonalizing the effective Hamiltonian Eq.~\eqref{H_0}. The result is
\begin{align}
H{}_{0}&=\sum_{k,\sigma}[\epsilon_{c}(k)c_{k\sigma}^{\dag}c_{k\sigma}+\epsilon_{d}(k)d_{k\sigma}^{\dag}d_{k\sigma}
\notag \\
+&\epsilon_{f_{1}}(k)f_{1,k\sigma}^{\dag}f_{1,k\sigma}+\epsilon_{f_{2}}(k)f_{2,k\sigma}^{\dag}f_{2,k\sigma}] \notag\\
 +&\epsilon_{g_{1}}(k)g_{1,k\sigma}^{\dag}g_{1,k\sigma}+\epsilon_{g_{2}}(k)g_{2,k\sigma}^{\dag}g_{2,k\sigma}]
\end{align}

The dispersions are
\begingroup
\allowdisplaybreaks
\begin{align}
&\epsilon_{c}(k) = \epsilon_3 - \frac{k^2}{2m_3} + \sqrt{b^2 \left(k^2_x-k^2_y \right)^2 + 4 c^2 k^2_x k^2_y}  \notag \\
&\epsilon_{d}(k) =  \epsilon_3 - \frac{k^2}{2m_3} - \sqrt{b^2 \left(k^2_x-k^2_y \right)^2 + 4 c^2 k^2_x k^2_y}  \notag \\
&\epsilon_{f_1} (k) = \frac{\epsilon_{1,Y} + \epsilon_{2,Y}}{2} + \sqrt{\left(\frac{\epsilon_{1,Y}-\epsilon_{2,Y}}{2}\right)^2 + 4 v^2 k^2_x} \notag \\
&\epsilon_{g_1} (k) = \frac{\epsilon_{1,Y} + \epsilon_{2,Y}}{2} - \sqrt{\left(\frac{\epsilon_{1,Y}-\epsilon_{2,Y}}{2}\right)^2 + 4 v^2 k^2_x} \notag \\
&\epsilon_{f_2} (k) = \frac{\epsilon_{1,X} + \epsilon_{2,X}}{2} + \sqrt{\left(\frac{\epsilon_{1,X}-\epsilon_{2,X}}{2}\right)^2 + 4 v^2 k^2_y} \notag \\
&\epsilon_{g_2} (k) = \frac{\epsilon_{1,X} + \epsilon_{2,X}}{2} - \sqrt{\left(\frac{\epsilon_{1,X}-\epsilon_{2,X}}{2}\right)^2 + 4 v^2 k^2_y} \notag \\
&
\label{nnn_2}
\end{align}
\endgroup
where
\bea
&&\epsilon_{1,Y(X)} = \epsilon_{1} + \frac{k^{2}}{2m_{1}} \pm a_{1}(k_{y}^{2} - k_{x}^{2}) \notag \\
&&\epsilon_{2,Y(X)} = \epsilon_{2} + \frac{k^{2}}{2m_{2}} \pm a_{2}(k_{y}^{2} - k_{x}^{2}) \notag \\
\eea
(upper sign for $Y$, lower for $X$).
 The transformation from orbital basis to band basis is a generalized rotation,
\begin{align}
	\begin{pmatrix}
		\psi_{1}(\bm{k})       \\
		\psi_{2}(\bm{k})
	\end{pmatrix}
	&=
	\begin{pmatrix}
		\cos\phi_{e,\bm{k}} & -i \sin\phi_{e,\bm{k}} \\
		-i\sin\phi_{e,\bm{k}} & \cos\phi_{e,\bm{k}}
	\end{pmatrix}
	\begin{pmatrix}
		f_{1,\bm{k}}       \\
		g_{1,\bm{k}}
	\end{pmatrix} \\
	\begin{pmatrix}
		\psi_{3}(\bm{k})        \\
		\psi_{4}(\bm{k})
	\end{pmatrix}
	&=
	\begin{pmatrix}
		\cos\phi^{'}_{e,\bm{k}} & -i \sin\phi^{'}_{e,\bm{k}}\\
		-i\sin\phi^{'}_{e,\bm{k}} & \cos\phi^{'}_{e,\bm{k}}
	\end{pmatrix}
	\begin{pmatrix}
		f_{2,\bm{k}}        \\
		g_{2,\bm{k}}
	\end{pmatrix} \\
	\begin{pmatrix}
		\psi_{5}(\bm{k})        \\
		\psi_{6}(\bm{k})
	\end{pmatrix}
	&=
	\begin{pmatrix}
		\cos\phi_{h,\bm{k}} & \sin\phi_{h,\bm{k}}\\
		-\sin\phi_{h,\bm{k}} & \cos\phi_{h,\bm{k}}
	\end{pmatrix}
	\begin{pmatrix}
		c_{\bm{k}}       \\
		d_{\bm{k}}
	\end{pmatrix}\label{band_basis}
\end{align}
The rotation angles $\phi_{e, \bm{k}}$, $\phi^{'}_{e,\bm{k}}$ and $\phi_{h,\bm{k}}$ depend on the parameters in $h_X$, $h_Y$, and $h_\Gamma$ as 
\bea
&&\tan{2 \phi_{e,\bm{k}}} = \frac{-4v k_x}{\epsilon_{1,Y} - \epsilon_{2,Y}}
 \notag \\
&& \tan{2 \phi^{'}_{e,\bm{k}}} = \frac{4v k_y}{\epsilon_{1,X} - \epsilon_{2,X}} \notag \\
&& \tan{2 \phi_{h,\bm{k}}} = \frac{2c k_x k_y}{b(k^2_x-k^2_y)}
\label{nnn_1}
\eea

Fermions labeled by $f_{1}$ and $f_{2}$ cross the Fermi level and form the electron pockets near $Y$ and $X$, respectively. Fermions labeled by $c$ and $d$
 form the two hole pockets near $\Gamma$. We call these fermions low-energy excitations.  Fermions labeled by $g_1$ and $g_2$ are gapped and do not cross the Fermi level. In principle, $g_{1,2}$ fermions have to be included into the parquet RG analysis as the gap in their excitation spectrum is of order $E_F$, which is the lower edge of
 parquet RG analysis.
  To avoid dealing with too many
   couplings, we assume that parameters are such that the
  gap in the spectra of $g_1$ and $g_2$ fermions is numerically much larger than $E_F$ and treat these fermions as high-energy, in which case they are not
  subjects of RG.

\begin{figure}[h]
	\centering{}\includegraphics[width=0.95\columnwidth]{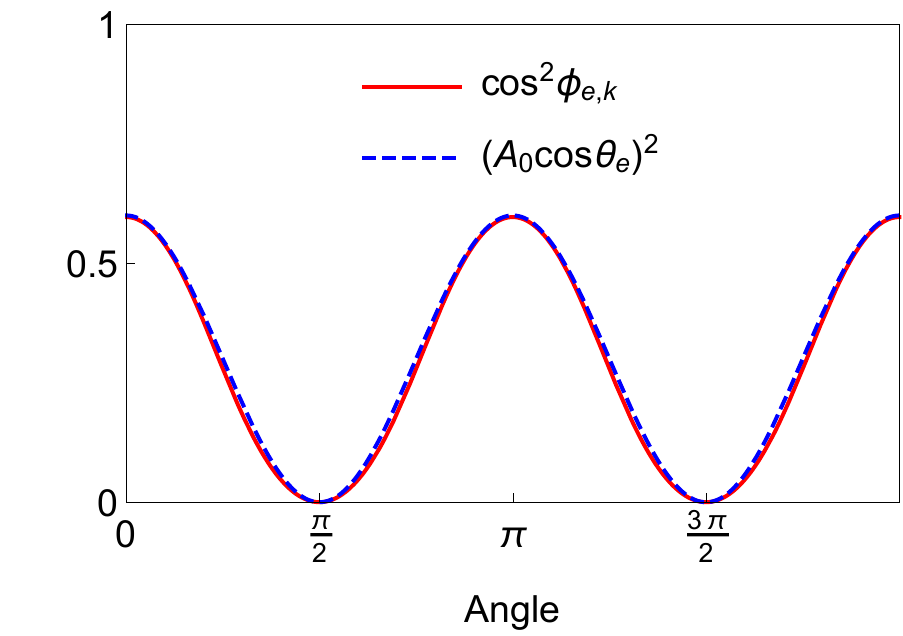}
	\protect\caption{Approximation for the transformation matrix between orbital and band basis.
 Solid line -- the actual  $\cos^{2}\phi_{e,k}$ from Eq. (\protect \ref{nnn_1}). Dashed line -- $(A_{0} \cos\theta_{e})^{2}$, where $\theta_e$ is the angle along an electron pocket.}
\label{fig:A0approx}
\end{figure}

 We make two additional assumptions to simplify the evaluation of the integrals below.  First, we assume that the hole pockets are circular rather than just $C_4$-symmetric.  This is the case when $c=b$ in Eq. (\ref{ch_1}). For circular hole pockets
 \begin{align}
\epsilon_{c,d}(\bm{k})=\mu_h - \frac{k^{2}}{2m_{c,d}},
\end{align}
where $\mu_h = \epsilon_3$, $m_c= m_3(1-2m_3 b), m_d= m_3(1 +2m_3 b)$,
  and the rotation angle  $\phi_{h,\bm{k}}$ in (\ref{band_basis}) coincides with the angle $\theta_h$ between $\bm{k}$ and $X$ axis along the hole Fermi surfaces.
  Second, on electron pockets we set
   $\cos\phi_{e,k} = A_{0} \cos\theta_{e}$,  $\sin\phi_{e,k} = \sqrt{1-A^2_{0} \cos^2\theta_{e}}$,  and
  $\cos\phi^{'}_{e,k} = -A_{0} \sin\theta_{e}$,  $\sin\phi^{'}_{e,k} = \sqrt{1-A^2_{0} \sin^2\theta_{e}}$,
  where $\theta_{e}$  is the angle between  $\bm{k}$ and $X$
  direction on both electron Fermi surfaces and $1/\sqrt{2} < A_0 < 1$.
    This parametrization is indeed an approximation,
    but it is consistent with symmetry, and we verified numerically (see Fig.~\ref{fig:A0approx}) that it matches quite accurately the actual $\phi_{e,\bm{k}}$ and $\phi^{'}_{e,\bm{k}}$ from Eq. (\ref{nnn_1}), at least
    for the parameters of the tight-binding dispersion listed in Ref. \cite{Cvetkovic2013}.
    The condition $ A_0 > 1/\sqrt{2}$
    follows from the fact that $\epsilon_{1,Y(X)}$ and $\epsilon_{2,Y(X)}$ must cross at some value of $\theta_e$ to
    ensure that over some range along each of the electron pockets the largest spectral weight comes from the $d_{xz} (d_{yz})$ orbital, while over the rest of the pockets the largest spectral weight comes from $d_{xy}$ orbital.
  (A larger value of $A_0$ would mean a larger total weight of the $d_{xz} (d_{yz})$ orbital 
   on an electron pocket).
    These two approximations simplify the evaluation of angular integrals later in the paper, but they do not affect the structure of RG equations and the interplay between susceptibilities in different channels.
 Expanding the dispersions of $f_1$ and $f_2$ fermions in Eq. (\ref{nnn_2}) in powers of momenta, we find that   electron pockets are elliptical, and the dispersions near these pockets are
\begin{align}
\epsilon_{f_{1},f_{2}}(\bm{k})=\frac{k_{x}^{2}}{2m_{x,y}}+\frac{k_{y}^{2}}{2m_{y,x}} -\mu_e\label{disp}.
\end{align}
The parameters $m_{x,y}$ and $\mu_e$ are determined by Eqn. (\ref{nnn_2}).

\subsection{The interactions}
\label{sec:interactions}

We now move to the interaction part of the Hamiltonian.
We can either derive the 4-fermion part of the Hamiltonian by using symmetry arguments, or just depart from the model with local Hubbard-Hund interactions.
In the notations of Ref. \cite{kemper}, we have
\begin{align}
H_{int}=& \frac{U}{2}\sum_{i,\mu}n_{i,\mu}n_{i,\mu}+\frac{U'}{2}\sum_{i,\mu\neq\mu'}n_{i\mu}n_{i\mu'}
\notag \\
& +\frac{J}{2}\sum_{i,\mu'\neq\mu}\sum_{\sigma\sigma'}\psi_{i\mu\sigma}^{\dag}\psi_{i\mu'\sigma'}^{\dag}\psi_{i\mu\sigma'}\psi_{i\mu'\sigma}+
\notag \\
& + \frac{J'}{2}\sum_{i,\mu'\neq\mu}\psi_{i\mu\sigma}^{\dag}\psi_{i\mu\sigma'}^{\dag}\psi_{i\mu'\sigma'}\psi_{i\mu'\sigma}\,.
\label{interaction_K_1}
\end{align}
Here $\psi_{i\mu\sigma}^{\dag}$ creates an electron  on iron site $\bm{R}_i$, in orbital state $\mu$, and  in spin state $\sigma$.
The operator
$n_{i,\mu}=\psi_{i,\mu}^{\dag}\psi_{i,\mu}$ is  the density operator at an orbital $\mu$ at site $i$.   $U$ and $U'$ are Hubbard intra-orbital and inter-orbital density interactions, $J$ is the Hund's exchange coupling, and  $J'$ is the amplitude of the inter-orbital pair hopping.
The Hamiltonian $H_{int}$ is invariant under SU(2) spin rotations.

The relation between the local operators, $\psi_{i\mu\sigma}$ and orbital operators $\psi_{a\sigma}(\bm{k})$ near $\Gamma$, $X$, and $Y$  is
\begin{align}\label{relation}
\psi_{i,xz,\sigma} &= \frac{1}{\sqrt{N}}
\sum_{\bm{k}} e^{i \bm{k} \bm{R}_{i}} \big[
\psi_{1\sigma}(\bm{k}) e^{i \bm{Q}_y \bm{R}_{i}} + \psi_{6\sigma}(\bm{k}) \big], \notag\\
\psi_{i,yz,\sigma} &= \frac{1}{\sqrt{N}}
\sum_{\bm{k}} e^{i \bm{k} \bm{R}_{i}} \big[
\psi_{3\sigma}(\bm{k}) e^{i \bm{Q}_x \bm{R}_{i}} + \psi_{5\sigma}(\bm{k}) \big], \notag\\
\psi_{i,xy,\sigma} &= \frac{1}{\sqrt{N}}
\sum_{\bm{k}} e^{i \bm{k} \bm{R}_{i}} \big[
\psi_{2\sigma}(\bm{k}) e^{i \bm{Q}_y \bm{R}_{i}} + \psi_{4\sigma}(\bm{k})e^{i \bm{Q}_x \bm{R}_{i}} \big]. \notag\\
\end{align}
Substituting Eq.~\eqref{relation} in Eq.~\eqref{interaction_K_1}, we obtain the interaction Hamiltonian in the orbital representation:
\begingroup
\allowdisplaybreaks
\begin{widetext}
\begin{align}\label{general_H}
H_{int}= & {U}_{1}\sum\nolimits'\left[\psi_{1\sigma}^{\dag}\psi_{1\sigma}\psi_{6\sigma'}^{\dag}\psi_{6\sigma'}+\psi_{3\sigma}^{\dag}\psi_{3\sigma}\psi_{5\sigma'}^{\dag}\psi_{5\sigma'}\right]+\bar{U}_{1}\sum\nolimits'\left[\psi_{1\sigma}^{\dag}\psi_{1\sigma}\psi_{5\sigma'}^{\dag}\psi_{5\sigma'}+\psi_{3\sigma}^{\dag}\psi_{3\sigma}\psi_{6\sigma'}^{\dag}\psi_{6\sigma'}\right]\notag\\
+&\tilde{U}_{1}\sum\nolimits'\left[\psi_{2\sigma}^{\dag}\psi_{2\sigma}\psi_{6\sigma'}^{\dag}\psi_{6\sigma'}+\psi_{4\sigma}^{\dag}\psi_{4\sigma}\psi_{5\sigma'}^{\dag}\psi_{5\sigma'}\right]+\tilde{\tilde{U}}_{1}\sum\nolimits'\left[\psi_{4\sigma}^{\dag}\psi_{4\sigma}\psi_{6\sigma'}^{\dag}\psi_{6\sigma'}+\psi_{2\sigma}^{\dag}\psi_{2\sigma}\psi_{5\sigma'}^{\dag}\psi_{5\sigma'}\right]\notag\\
+ & {U}_{2}\sum\nolimits'\left[\psi_{1\sigma}^{\dag}\psi_{6\sigma}\psi_{6\sigma'}^{\dag}\psi_{1\sigma'}+\psi_{3\sigma}^{\dag}\psi_{5\sigma}\psi_{5\sigma'}^{\dag}\psi_{3\sigma'}\right]+\bar{U}_{2}\sum\nolimits'\left[\psi_{1\sigma}^{\dag}\psi_{5\sigma}\psi_{5\sigma'}^{\dag}\psi_{1\sigma'}+\psi_{3\sigma}^{\dag}\psi_{6\sigma}\psi_{6\sigma'}^{\dag}\psi_{3\sigma'}\right]\notag\\
+ & \tilde{U}_{2}\sum\nolimits'\left[\psi_{2\sigma}^{\dag}\psi_{6\sigma}\psi_{6\sigma'}^{\dag}\psi_{2\sigma'}+\psi_{4\sigma}^{\dag}\psi_{5\sigma}\psi_{5\sigma'}^{\dag}\psi_{4\sigma'}\right]+\tilde{\tilde{U}}_{2}\sum\nolimits'\left[\psi_{2\sigma}^{\dag}\psi_{5\sigma}\psi_{5\sigma'}^{\dag}\psi_{2\sigma'}+\psi_{4\sigma}^{\dag}\psi_{6\sigma}\psi_{6\sigma'}^{\dag}\psi_{4\sigma'}\right]\notag\\
+ & \frac{{U}_{3}}{2}\sum\nolimits'\left[\psi_{1\sigma}^{\dag}\psi_{6\sigma}\psi_{1\sigma'}^{\dag}\psi_{6\sigma'}+\psi_{3\sigma}^{\dag}\psi_{5\sigma}\psi_{3\sigma'}^{\dag}\psi_{5\sigma'}+h.c.\right]
+\frac{\bar{U}_{3}}{2}\sum\nolimits'\left[\psi_{1\sigma}^{\dag}\psi_{5\sigma}\psi_{1\sigma'}^{\dag}\psi_{5\sigma'}+\psi_{3\sigma}^{\dag}\psi_{6\sigma}\psi_{3\sigma'}^{\dag}\psi_{6\sigma'}+h.c.\right]\notag\\
+ & \frac{\tilde{U}_{3}}{2}\sum\nolimits'\left[\psi_{2\sigma}^{\dag}\psi_{6\sigma}\psi_{2\sigma'}^{\dag}\psi_{6\sigma'}+\psi_{4\sigma}^{\dag}\psi_{5\sigma}\psi_{4\sigma'}^{\dag}\psi_{5\sigma'}+h.c.\right]
+\frac{\tilde{\tilde{U}}_{3}}{2}\sum\nolimits'\left[\psi_{2\sigma}^{\dag}\psi_{5\sigma}\psi_{2\sigma'}^{\dag}\psi_{5\sigma'}+\psi_{4\sigma}^{\dag}\psi_{6\sigma}\psi_{4\sigma'}^{\dag}\psi_{6\sigma'}+h.c.\right]\notag\\
+ & \frac{U_{4}}{2}\sum\nolimits'\left[\psi_{5\sigma}^{\dag}\psi_{5\sigma}\psi_{5\sigma'}^{\dag}\psi_{5\sigma'}+\psi_{6\sigma}^{\dag}\psi_{6\sigma}\psi_{6\sigma'}^{\dag}\psi_{6\sigma'}\right]+\frac{\bar{U}_{4}}{2}\sum\nolimits'\left[\psi_{5\sigma}^{\dag}\psi_{6\sigma}\psi_{5\sigma'}^{\dag}\psi_{6\sigma'}+\psi_{6\sigma}^{\dag}\psi_{5\sigma}\psi_{6\sigma'}^{\dag}\psi_{5\sigma'}\right]\notag\\
+ & \tilde{U}_{4}\sum\nolimits'\psi_{5\sigma}^{\dag}\psi_{5\sigma}\psi_{6\sigma'}^{\dag}\psi_{6\sigma'}+\tilde{\tilde{U}}_{4}\sum\nolimits'\psi_{5\sigma}^{\dag}\psi_{6\sigma}\psi_{6\sigma'}^{\dag}\psi_{5\sigma'}+ \frac{U_{5}}{2}\sum\nolimits'\left[\psi_{1\sigma}^{\dag}\psi_{1\sigma}\psi_{1\sigma'}^{\dag}\psi_{1\sigma'}+\psi_{3\sigma}^{\dag}\psi_{3\sigma}\psi_{3\sigma'}^{\dag}\psi_{3\sigma'}\right]\notag\\
+& \frac{\bar{U}_{5}}{2}\sum\nolimits'\left[\psi_{1\sigma}^{\dag}\psi_{3\sigma}\psi_{1\sigma'}^{\dag}\psi_{3\sigma'}+h.c.\right]+ \tilde{U}_{5}\sum\nolimits'\psi_{1\sigma}^{\dag}\psi_{1\sigma}\psi_{3\sigma'}^{\dag}\psi_{3\sigma'}+\tilde{\tilde{U}}_{5}\sum\nolimits'\psi_{1\sigma}^{\dag}\psi_{3\sigma}\psi_{3\sigma'}^{\dag}\psi_{1\sigma'}\notag\\
+& \frac{U_{6}}{2}\sum\nolimits'\left[\psi_{2\sigma}^{\dag}\psi_{2\sigma}\psi_{2\sigma'}^{\dag}\psi_{2\sigma'}+\psi_{4\sigma}^{\dag}\psi_{4\sigma}\psi_{4\sigma'}^{\dag}\psi_{4\sigma'}\right]
+ \frac{\bar{U}_{6}}{2}\sum\nolimits'\left[\psi_{2\sigma}^{\dag}\psi_{4\sigma}\psi_{2\sigma'}^{\dag}\psi_{4\sigma'}+h.c.\right]\notag\\
+& \tilde{U}_{6}\sum\nolimits'\psi_{2\sigma}^{\dag}\psi_{2\sigma}\psi_{4\sigma'}^{\dag}\psi_{4\sigma'}+\tilde{\tilde{U}}_{6}\sum\nolimits'\psi_{2\sigma}^{\dag}\psi_{4\sigma}\psi_{4\sigma'}^{\dag}\psi_{2\sigma'}\notag\\
+&\frac{\bar{U}_{7}}{2}\sum\nolimits'\left[\psi_{1\sigma}^{\dag}\psi_{2\sigma}\psi_{1\sigma'}^{\dag}\psi_{2\sigma'}+\psi_{3\sigma}^{\dag}\psi_{4\sigma}\psi_{3\sigma'}^{\dag}\psi_{4\sigma'}+h.c.\right]\notag\\
+& \tilde{U}_{7}\sum\nolimits'\left[\psi_{1\sigma}^{\dag}\psi_{1\sigma}\psi_{2\sigma'}^{\dag}\psi_{2\sigma'}+\psi_{3\sigma}^{\dag}\psi_{3\sigma}\psi_{4\sigma'}^{\dag}\psi_{4\sigma'}\right]+\tilde{\tilde{U}}_{7}\sum\nolimits'\left[\psi_{1\sigma}^{\dag}\psi_{2\sigma}\psi_{2\sigma'}^{\dag}\psi_{1\sigma'}+\psi_{3\sigma}^{\dag}\psi_{4\sigma}\psi_{4\sigma'}^{\dag}\psi_{3\sigma'}\right]\notag\\
+&\frac{\bar{U}_{8}}{2}\sum\nolimits'\left[\psi_{1\sigma}^{\dag}\psi_{4\sigma}\psi_{1\sigma'}^{\dag}\psi_{4\sigma'}+\psi_{2\sigma}^{\dag}\psi_{3\sigma}\psi_{2\sigma'}^{\dag}\psi_{3\sigma'}+h.c.\right]
+ \tilde{U}_{8}\sum\nolimits'\left[\psi_{1\sigma}^{\dag}\psi_{1\sigma}\psi_{4\sigma'}^{\dag}\psi_{4\sigma'}+\psi_{2\sigma}^{\dag}\psi_{2\sigma}\psi_{3\sigma'}^{\dag}\psi_{3\sigma'}\right]\notag\\
+&\tilde{\tilde{U}}_{8}\sum\nolimits'\left[\psi_{1\sigma}^{\dag}\psi_{4\sigma}\psi_{4\sigma'}^{\dag}\psi_{1\sigma'}+\psi_{2\sigma}^{\dag}\psi_{3\sigma}\psi_{3\sigma'}^{\dag}\psi_{2\sigma'}\right]\, ,
\end{align}
\end{widetext}
\endgroup
where
\begin{align}
{U}_{1} & ={U}_{2} ={U}_{3}=U_{4}=U_{5}=U_{6}=\bar{U}_{6}=\tilde{U}_{6}=\tilde{\tilde{U}}_{6}=U,\notag\\
\bar{U}_{1}&=\tilde{U}_{1}=\tilde{\tilde{U}}_{1}=\tilde{U}_{4}=\tilde{U}_{5}=\tilde{U}_{7}=\tilde{U}_{8} =U',\notag\\
\bar{U}_{2}&=\tilde{U}_{2}=\tilde{\tilde{U}}_{2}=\tilde{\tilde{U}}_{4}=\tilde{\tilde{U}}_{5}=\tilde{\tilde{U}}_{7}=\tilde{\tilde{U}}_{8} =J,\notag\\
\bar{U}_{3}& =\tilde{U}_{3}=\tilde{\tilde{U}}_{3}=\bar{U}_{4}=\bar{U}_{5}=\bar{U}_{7}=\bar{U}_{8}=J'.\label{Hubbard_relation}
\end{align}
In Eq.~\eqref{general_H} the momentum arguments of the fermion operators $\bm{k}_{i}$, $i =1,2,3,4$
are omitted and the summation is over spin indices $\sigma$, $\sigma'$ and momenta,  subject to the momentum conservation condition $\sum_{i=1}^4 \bm{k}_{i} = 0$.

The next step is to realize that Eq.~\eqref{general_H} with {\it arbitrary} prefactors is the most general form of the interaction for the 4-band, 3-orbital model,
 consistent with the $C_4$ lattice symmetry.
The $C_4$ symmetry implies that the four-fermion Hamiltonian must be invariant under the transformation $\psi_{1}\leftrightarrow\psi_{3}$, $\psi_{2}\leftrightarrow\psi_{4}$, $\psi_{5}\leftrightarrow\psi_{6}$.
 One can easily check that each term in (\ref{general_H}) is  $C_4$-symmetric on its own.
Then  the coupling constants do not have to be bound by the relations \eqref{Hubbard_relation}.

This reasoning implies that Eq.~\eqref{general_H} is the most generic form of fermion-fermion interaction  for a model with not necessarily local interactions.
The total number of different terms in Eq.~\eqref{general_H} is 30, hence there are 30 independent coupling constants.  This number was first reported in Ref.~\cite{Cvetkovic2013}. At a bare level, the couplings may be related, as in Eq.~\eqref{Hubbard_relation}. However, once we integrate out fermions with energies above a certain cutoff,  all 30  coupling constants
 renormalize differently.
As a result the conditions set by Eq.~\eqref{Hubbard_relation} do not hold for the running couplings.
 We also verified explicitly that no new interactions are generated by the RG flow, i.e.,
  the terms which we present in Eq.~\eqref{general_H} exhaust all possible symmetry allowed interactions between low-energy fermions.
In RG language this implies that the theory is renormalizable.

\section{Renormalization group (RG) analysis}
\label{sec:pRG}

Like we said, we will use the parquet RG technique to analyze the flow of the couplings.
The RG technique is generally applicable when interactions in some channels evolve logarithmically with the running energy.
Ladder RG is applicable when there is only one channel with logarithmic interactions (e.g,  the Cooper channel).
The parquet RG is applied when there is more than one channel, in which the renormalization of the coupling is logarithmic.

Parquet RG was first introduced in field-theory~\cite{sudakov}. In condensed matter it was successfully used to
  map the phase diagram of one-dimensional systems, where logarthmic renormalizations are present in both particle-particle and particle-hole channels~\cite{RG_1d},
  and was also applied to several 2D systems, e.g., to the 2D $\sigma$-model~\cite{RG_sigma}, fermions near a van-Hove singularity in the dispersion~\cite{RG_vh}, and bilayer graphene~\cite{graphene}.
The leading logarithmic contributions at each order of perturbation are represented by the so-called parquet diagrams.
The RG technique allows one to express infinite series of logarithmic renormalizations by differential equations for fully renormalized vertices.

The application of parquet RG to FeSCs has been discussed before~\cite{earlier_RG}.
Like we said, the new key element of our analysis is the
 inclusion
  of the orbital content of the excitations around the Fermi pockets.  We refer to earlier literature for details and here just
 state the two main  reasons to use parquet RG for FeSCs. First, the very fact that hole and electron dispersions have opposite signs implies that
  the renormalizations in the particle-hole channel at momenta separating hole and electron pockets are logarithmic at energies between the
   bandwidth and the Fermi energy.  Nesting does not play a crucial role here as the two dispersions have opposite sign with or without nesting.
   Nesting (the near equivalence between hole and electronic dispersions, up to a sign) extends the logarithmic renormalizations in the particle-hole channel to
    energies smaller than $E_F$
    (down to energies of order $\abs{\mu_h-\mu_e}$),
    but at such low energies parquet RG is already not applicable  as particle-particle and particle-hole channels no longer "talk" to each other.
    Second, the logarithm in the particle-particle channel is not the Cooper logarithm (which comes from fermions in the near vicinity of the Fermi surfaces),
     but the one
     associated with the renormalization of the scattering amplitude in 2D (Ref.~\cite{2d_scat_ampl}).
       In this respect, the pairing instability  within the parquet RG is actually towards a bound state formation of  two particles.   In a one-band 2D system, the actual superconducting $T_c$ would be much smaller than this temperature~\cite{mohit_one_band}  However,
      in our case, when hole and electron bands are both present, the temperature, at which bound pairs develop, and the actual superconducting $T_c$ are of the same order~\cite{cee_1}. By this reason we will not distinguish between a bound state formation
        (which  develops within parquet RG)  and a true superconductivity.

   In the calculations below we assume that the bare values of the  interactions are small compared to the bandwidth and restrict the analysis to one loop parquet RG.
    We  show that some interactions grow in the process of the RG flow, i.e., the system flows towards strong coupling.  If  we set the bare interactions to be larger,  the system will more rapidly flow towards strong coupling, and the temperature of the leading instability will increase.  In FeSe the leading instability is at $T_s = 85 K \sim 8$ meV. This  temperature is two orders of magnitude smaller than the bandwidth $W \sim 1$ eV.  We consider the smallness of $T_s/W$ as at least partial justification  to apply the RG procedure.  At the same time we caution that our approximation of the dispersions of the bands which cross the Fermi level by parabolas is not well justified, as other bands hybridize with $d_{xz}, d_{yz}$, and $d_{xy}$ bands already at energies  below $W$.
    These additional bands, however, affect only the value of the upper cutoff for parquet RG, but not the outcome of the RG flow.\\

\subsection{RG equations for the interactions}\label{sec:RG_equtions_flow}

In this Section we derive and solve the set of parquet RG equations for  the interactions.
The derivation of 30 coupled RG equations is a cumbersome but straightforward procedure.
As we said, solving one-loop parquet RG equations  is equivalent to summing up all leading logarithmic contributions originating from both
particle-particle and particle-hole channels.  Like in BCS theory,  logarithms come from internal energies larger, in logarithmic sense,  than the external ones.
 To logarithmic accuracy we set all external frequencies to be of the same order $E$ and set external momenta $k_{ext}$ to be of order $(2mE)^{1/2}$.
We obtain the interactions  $U_i (k_{ext})$  by integrating first over internal frequency and then over internal momentum between
  $\sqrt{2mW}$ and $(2mE)^{1/2}$.

	\begin{figure*}[ht]
		\centering{\includegraphics[width=2.05\columnwidth]{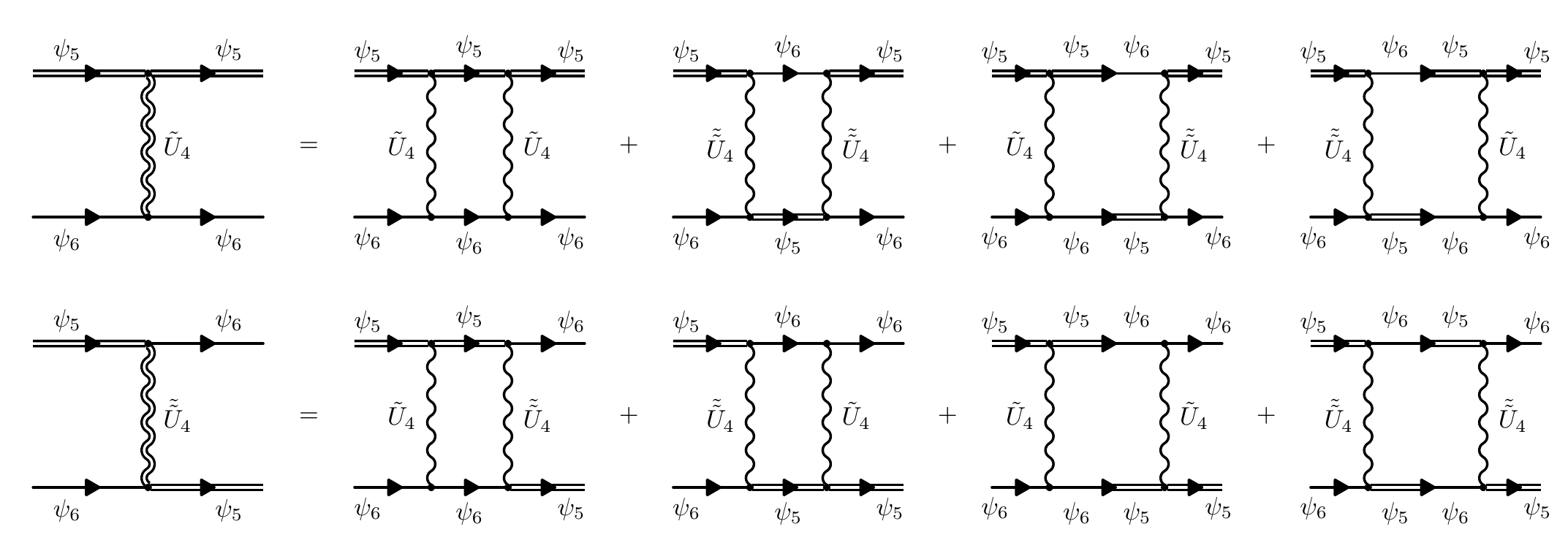}}
		 \protect\caption{The diagrams for the renormalizations of $\tilde{U}_{4}$ and $\tilde{\tilde{U}}_{4}$.
			The propagators are identified by their label. Note that there are contributions involving Green's functions which are non-diagonal in the orbital index.}
		\label{fig:pRG_example}
	\end{figure*}

As an example, we derive the RG equations for the interactions $\tilde{U}_{4}$ and $\tilde{\tilde{U}}_{4}$.
The renormalizations of these couplings are given by diagrams shown in Fig.~\ref{fig:pRG_example}. Evaluating the diagrams, we obtain
\begingroup
\allowdisplaybreaks
\begin{align}\label{RG_U4}
\tilde{U}_{4}&(k_{ext})=  -\int_{k_{ext}}^{\sqrt{2mW}}\frac{d^{2}\bm{k}}{4\pi^{2}}(\tilde{U}_{4}^{2}(\bm{k})+\tilde{\tilde{U}}_{4}^{2}(\bm{k}))
\notag\\ & \times
\int\frac{d\epsilon}{2\pi}G_{\psi_{5};\psi_{5}}(i\epsilon,\bm{k})G_{\psi_{6};\psi_{6}}(-i\epsilon,-\bm{k})\notag\\
 & -\int_{k_{ext}}^{\sqrt{2mW}}\frac{d^{2}\bm{k}}{4\pi^{2}}2\tilde{U}_{4}(\bm{k})\tilde{\tilde{U}}_{4}(\bm{k})
 \notag \\
& \times \int\frac{d\epsilon}{2\pi}G_{\psi_{5};\psi_{6}}(i\epsilon,\bm{k})G_{\psi_{6};\psi_{5}}(-i\epsilon,-\bm{k}),
\end{align}
\endgroup
\begingroup
\allowdisplaybreaks
\begin{align}\label{RG_U4_a}
\tilde{\tilde{U}}_{4}&(k_{ext})= - \int_{k_{ext}}^{\sqrt{2mW}}\frac{d^{2}\bm{k}}{4\pi^{2}}2\tilde{U}_{4}(\bm{k})\tilde{\tilde{U}}_{4}(\bm{k})
\notag\\ & \times
\int\frac{d\epsilon}{2\pi}G_{\psi_{5};\psi_{5}}(i\epsilon,\bm{k})G_{\psi_{6};\psi_{6}}(-i\epsilon,-\bm{k})\notag\\
 & -\int_{k_{ext}}^{\sqrt{2mW}}\frac{d^{2}\bm{k}}{4\pi^{2}}(\tilde{U}_{4}^{2}(\bm{k})+\tilde{\tilde{U}}_{4}^{2}(\bm{k}))
\notag \\
& \times
 \int\frac{d\epsilon}{2\pi}G_{\psi_{5};\psi_{6}}(i\epsilon,\bm{k})G_{\psi_{6};\psi_{5}}(-i\epsilon,-\bm{k}).
\end{align}
\endgroup
where $G_{\psi_{i};\psi_{j}}(i\epsilon,\bm{k})=-i\langle T\psi_{i}(\epsilon,\bm{k}) \psi_{j}^{\dag}(\epsilon,\bm{k})\rangle$.
To select the logarithms, we use Eq.~\eqref{band_basis}
and  re-express $G_{\psi_{i};\psi_{j}}(i\epsilon,\bm{k})$ in terms of Green's functions of band operators
 (in this case operators  c and d),  $G_{c,d} (i\epsilon, \bm{k}) = 1/(i\epsilon - \epsilon_{c,d}(\bm{k}))$.
Integrating over frequency, introducing the logarithmic variable
\begin{align}
L=\log\frac{W }{k_{ext}^{2}/2m}=\log\frac{W}{E}\,
\label{L_definition}
\end{align}
and combing the equations (\ref{RG_U4}) and (\ref{RG_U4_a}),
we obtain
\begin{align} \label{U4_t3}
\tilde{U}_{4}&(L) \pm  \tilde{\tilde{U}}_{4}(L)=  -A_{\pm} \int_{0}^{L}dL'\frac{(\tilde{U}_{4}(L')\pm\tilde{\tilde{U}}_{4}(L'))^{2}}{4\pi}
\end{align}
where
\begin{align}
A_{\pm} = \left[\frac{1}{8}(m_{c}+m_{d})+\frac{3}{8}\frac{4m_{c}m_{d}}{m_{c}+m_{d}}\pm\frac{1}{8}\frac{(m_{c}-m_{d})^{2}}{m_{c}+m_{d}}\right]\,.
\end{align}
Note that $A_{\pm} >0$.
Differentiating in (\ref{U4_t3}) over the upper limit, we obtain
\begin{align}
4 & \pi  \frac{d(\tilde{U}_{4}\pm\tilde{\tilde{U}}_{4})}{dL}=  - A_{\pm} (\tilde{U}_{4}\pm\tilde{\tilde{U}}_{4})^{2}
\label{U4_t3a}
\end{align}
Solving the equations for $\tilde{U}_{4} +\tilde{\tilde{U}}_{4}$ and $\tilde{U}_{4} -\tilde{\tilde{U}}_{4}$, we find that
both interactions  flow to zero under RG, provided that at the
bare level $\tilde{U}_{4}>\tilde{\tilde{U}}_{4}$.  Using Eq.  (\ref{Hubbard_relation}) for the bare couplings, we see that this holds if
 $U'>J$.  We assume in this paper that this condition is satisfied. If it is not satisfied, the conclusions will be different~\cite{ch_va}.

Because both  $\tilde{U}_{4}$ and $\tilde{\tilde{U}}_{4}$ vanish under RG for any $A_+$ and $A_-$, i.e., for any ratio of $m_c/m_d$, as long as both masses are non-zero, below we  set  $m_{c}=m_{d}=m_{h}$ to reduce the number of parameters  in the RG equations.
 By the same reason we also set $m_{x}=m_{y}=m_{e}$, i.e., approximate electron pockets as circular.
 Keeping $m_c \neq m_d$ and $m_x \neq m_y$ only complicates the formulas but does not lead to any novel system behavior.
With these approximations,  Eq.~\eqref{U4_t3a} simplifies to
\begin{align}
4\pi\frac{d(\tilde{U}_{4}\pm\tilde{\tilde{U}}_{4})}{dL}= & -A_{h}(\tilde{U}_{4}\pm\tilde{\tilde{U}}_{4})^{2}
\label{U4_t32}
\end{align}
where $A_h=m_h$.

\begin{figure}[h]
	\centering{}\includegraphics[width=0.95\columnwidth]{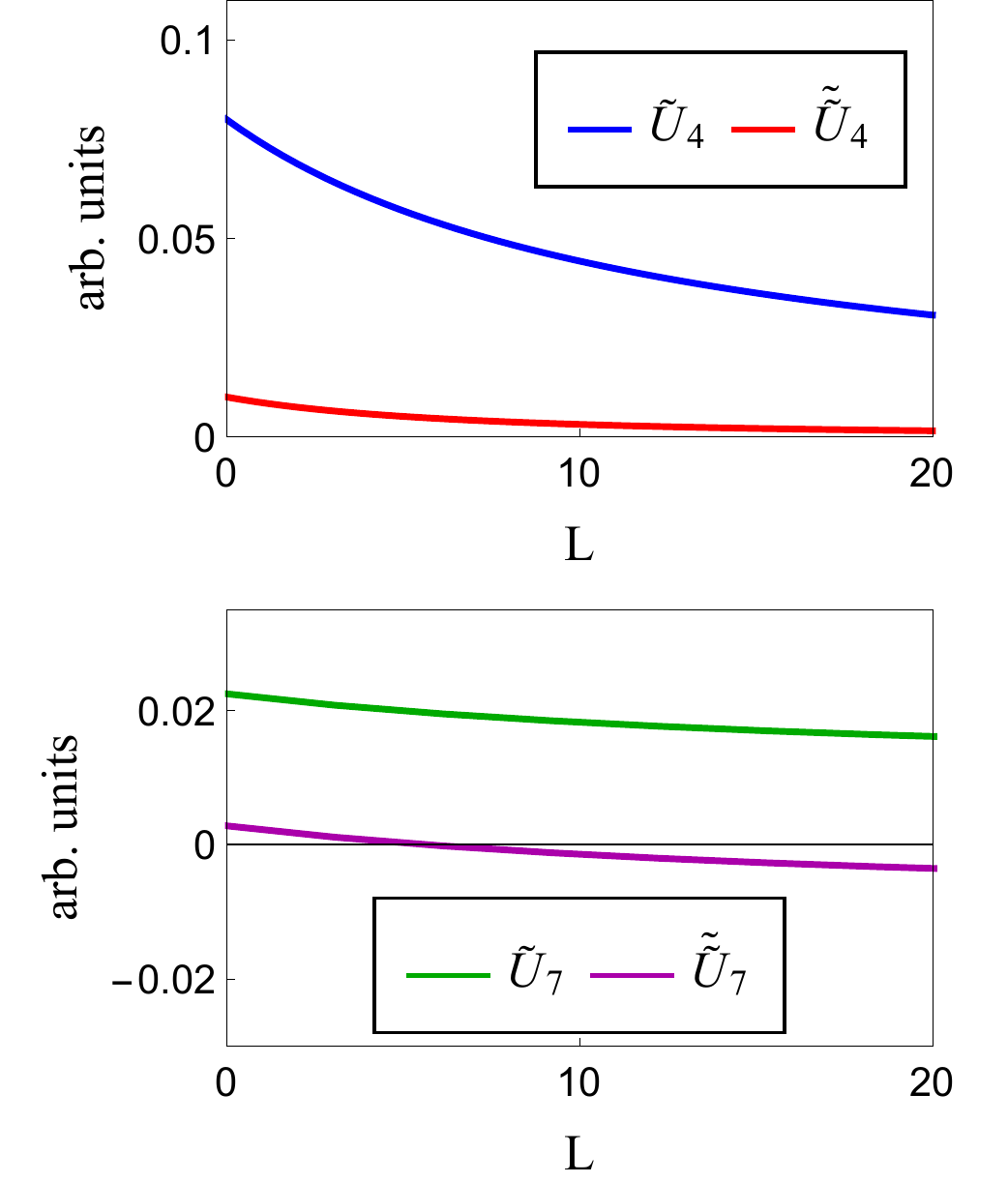}
	\protect\caption{Representatrive RG flow of some of  10 decoupled interactions.
		The upper panel shows the flow of $\tilde{U}_4$ and $\tilde{\tilde{U}}_4$. Both flow to zero under RG.  The flow of  $\tilde{U}_{5}$ and $\tilde{\tilde{U}}_{5}$, $\tilde{U}_{6}$ and $\tilde{\tilde{U}}_{6}$, and $\tilde{U}_{8}$ and $\tilde{\tilde{U}}_{8}$ are similar. The lower panel shows the flow of $\tilde{U}_7$ and $\tilde{\tilde{U}}_7$.  Both flow to small but finite values under RG.}\label{fig:10flow}
\end{figure}

Using the same reasoning, we  obtain eight similar-looking RG equations
\begingroup
\allowdisplaybreaks
\begin{align}
4\pi\frac{d\tilde{U}_{5}}{dL}= & -A^{'}_{e}(\tilde{U}^{2}_{5}+\tilde{\tilde{U}}^{2}_{5})\notag\\
4\pi\frac{d\tilde{\tilde{U}}_{5}}{dL}&=  -2A^{'}_{e}\tilde{U}_{5}\tilde{\tilde{U}}_{5}\notag\\
4\pi\frac{d\tilde{U}_{6}}{dL}= & -A^{''}_{e}(\tilde{U}^{2}_{6}+\tilde{\tilde{U}}^{2}_{6})\notag\\
4\pi\frac{d\tilde{\tilde{U}}_{6}}{dL}&=  -2A^{''}_{e}\tilde{U}_{6}\tilde{\tilde{U}}_{6}\notag\\
4\pi\frac{d\tilde{U}_{7}}{dL}= & 4\pi\frac{d\tilde{\tilde{U}}_{7}}{dL}= -\tilde{A}_{e}(\tilde{U}_{7}+\tilde{\tilde{U}}_{7})^{2}, \notag\\
4\pi\frac{d\tilde{U}_{8}}{dL}= & -A^{'''}_{e}(\tilde{U}^{2}_{8}+\tilde{\tilde{U}}^{2}_{8})\notag\\
4\pi\frac{d\tilde{\tilde{U}}_{8}}{dL}&=  -2A^{'''}_{e}\tilde{U}_{8}\tilde{\tilde{U}}_{8},
\label{10_RG_equations}
\end{align}
\endgroup
where
\begingroup
\allowdisplaybreaks
\begin{align}\label{A1}
\tilde{A}_{e}&=m_{e}\int \frac{d\theta_e}{2\pi}\cos^{2}\phi_{e,k}\sin^{2}\phi_{e,k}\notag\\
A^{'}_{e}&=m_{e}\int \frac{d\theta_e}{2\pi}\cos^{2}\phi_{e,k}\cos^{2}\phi^{'}_{e,k}\notag\\
A^{''}_{e}&=m_{e}\int \frac{d\theta_e}{2\pi}\sin^{2}\phi_{e,k}\sin^{2}\phi^{'}_{e,k}\notag\\
A^{'''}_{e}&=m_{e}\int \frac{d\theta_e}{2\pi}\cos^{2}\phi_{e,k}\sin^{2}\phi^{'}_{e,k}\, .
\end{align}
\endgroup
 and we remind that we set   $\cos\phi_{e,k} = A_{0} \cos\theta_{e}$,  $\sin\phi_{e,k} = \sqrt{1-A^2_{0} \cos^2\theta_{e}}$,  and
  $\cos\phi^{'}_{e,k} = -A_{0} \sin\theta_{e}$,  $\sin\phi^{'}_{e,k} = \sqrt{1-A^2_{0} \sin^2\theta_{e}}$, where $1/\sqrt{2} < A_0<1$.
The different A's in Eq.~(\ref{A1}) all scale as $m_e$ and are functions of $A_0$.
One can easily see from Eq.~\eqref{10_RG_equations} that  the couplings  $\tilde{U}_{j}$ and $\tilde{\tilde{U}}_{j}$ with $j =5,6,8$ flow to zero
 (upper panel in Fig.~\ref{fig:10flow}) if
 the bare values  $\tilde{U}_{j}$ and $\tilde{\tilde{U}}_{j}$ are positive and bare $\tilde{U}_{j} \geq \tilde{\tilde{U}}_{j}$,  which is the case when $U' >J$.
 Like we said, we assume that this holds.

\begin{figure}[h]
	\centering{}\includegraphics[width=1.0\columnwidth]{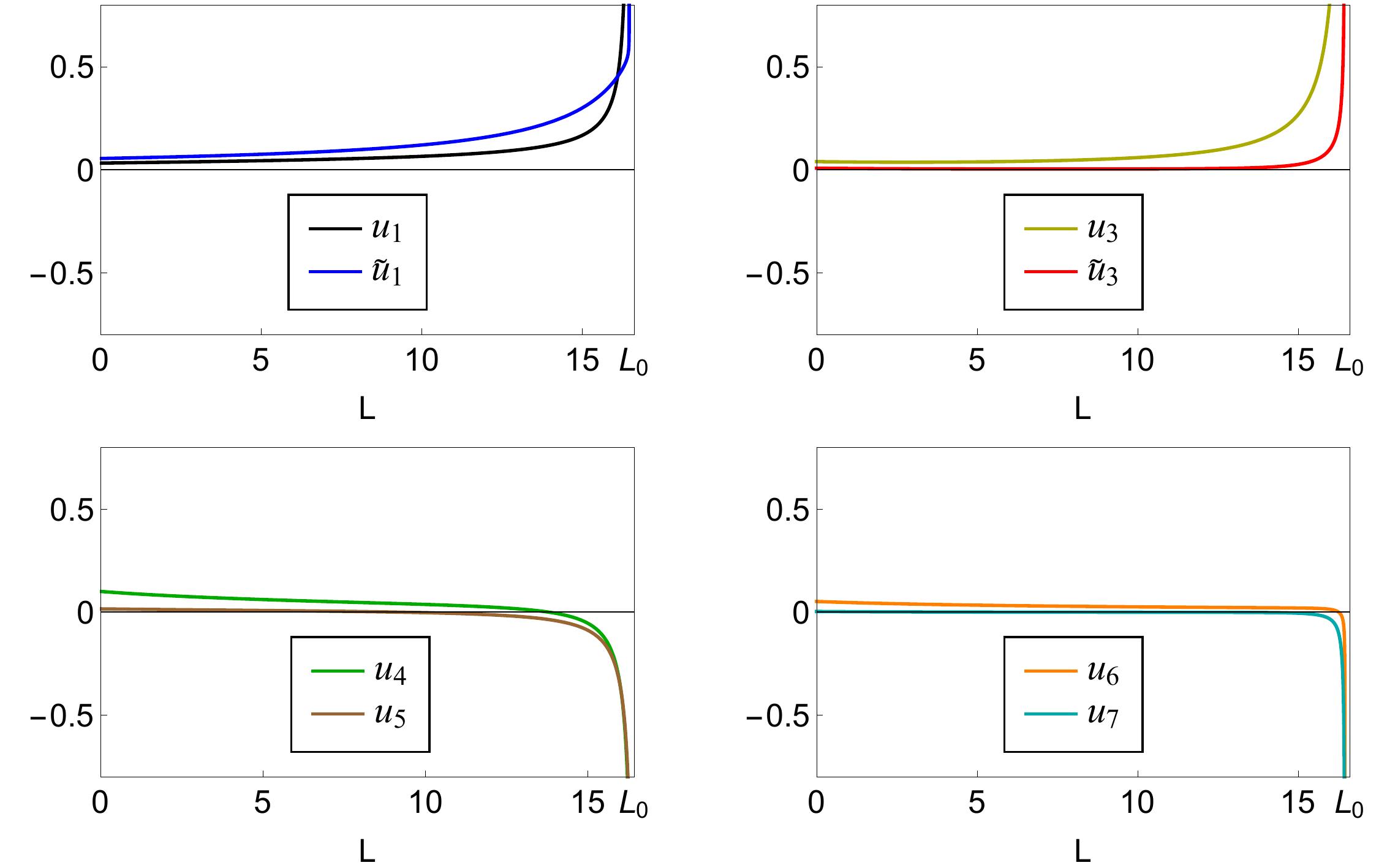}
	\caption{Representative RG flows of some of the 20 coupled interactions. The flow of eight  couplings is  shown.
 All couplings diverge at $L = L_0$. In the particular case  we show here $L_0=16.47$.}
		\label{fig:20flow}
\end{figure}

 The RG equations for ${\tilde U}_7$ and $\tilde{ \tilde U}_7$ are somewhat different compared to the other six equations in Eq.~(\ref{10_RG_equations}).
The reason is that the couplings ${\tilde U}_7$ and $\tilde{ \tilde U}_7$ are density-density and exchange couplings for $d_{xz}$ and $d_{xy}$ (or $d_{yz}$ and $d_{xy}$)
 orbital components on the same pockets. Because only one combination of these orbitals forms the band which crosses the Fermi level, the difference
 ${\tilde U}_7 - \tilde{ \tilde U}_7$ does not flow under RG. [The situation is similar to the case of RG flow of ${\tilde U}_4$ and $\tilde{ \tilde U}_4$ when one of the masses vanishes and $A_-$ becomes equal to zero].  Solving the RG equations for $\tilde{U}_{7}$ and $\tilde{\tilde{U}}_{7}$, we find that
these two couplings tend to finite values under RG, $\tilde{U}_{7} = -\tilde{\tilde{U}}_{7} =
const$.  We will see that the other couplings increase under RG, and in comparison the couplings $\tilde{U}_{7}$ and $\tilde{\tilde{U}}_{7}$ become negligible
(compare the lower panel in Fig.~\ref{fig:10flow} and Fig.~\ref{fig:20flow}).

Performing an analogous diagrammatic analysis for the remaining 20 couplings, we obtain 20 coupled RG equations.
We  write these equations below for dimensionless couplings, which we introduce as follows:

\allowdisplaybreaks
\begin{align}\label{u}
u_{1,2} & =\frac{A}{4\pi}U_{1,2}\,, \bar{u}_{1,2} =\frac{\bar{A}}{4\pi}\bar{U}_{1,2}\,,  \notag \\
\tilde{u}_{1,2}&=\frac{\tilde{A}}{4\pi}\tilde{U}_{1,2}\,, \tilde{\tilde{u}}_{1,2}=\frac{\tilde{A}}{4\pi}\tilde{\tilde{U}}_{1,2}\,, \notag\\
u_{3} & =\frac{AC}{4\pi}U_{3}\,, \bar{u}_{3}=\frac{\bar{A}\bar{C}}{4\pi}\bar{U}_{3}\,, \tilde{u}_{3}=\frac{\tilde{A}\tilde{C}}{4\pi}\tilde{U}_{3}\,, \tilde{\tilde{u}}_{3}=\frac{\tilde{A}\tilde{C}}{4\pi}\tilde{\tilde{U}}_{3}\,, \notag\\
u_{4} & =\frac{A_{h}}{4\pi}U_{4}\,, \bar{u}_{4}=\frac{A_{h}}{4\pi}\bar{U}_{4}\,,  u_{5}=\frac{A_{e}}{4\pi}U_{5}\,, \bar{u}_{5}=\frac{A_{e}}{4\pi}\bar{U}_{5}\,, \notag\\
u_{6} & =\frac{\bar{A}_{e}}{4\pi}U_{6}\,, \bar{u}_{6}=\frac{\bar{A}_{e}}{4\pi}\bar{U}_{6}\,,
\notag \\
\bar{u}_{7} & =\frac{\sqrt{{A}_{e}\bar{A}_{e}}}{4\pi}\bar{U}_{7}\,,  \bar{u}_{8}=\frac{\sqrt{{A}_{e}\bar{A}_{e}}}{4\pi}\bar{U}_{8}\,,
\end{align}
where 
\begin{align}\label{A2}
A=\bar{A}&=\frac{2m_{e}m_{h}}{m_{e}+m_{h}}\int \frac{d\theta}{2\pi}cos^{2}\phi_{e,k}\notag\\
\tilde{A}&=\frac{2m_{e}m_{h}}{m_{e}+m_{h}}\int \frac{d\theta}{2\pi}sin^{2}\phi_{e,k}\notag\\
{A}_{e}&=m_{e}\int \frac{d\theta}{2\pi}\cos^{4}\phi_{e,k}\notag\\
\bar{A}_{e}&=m_{e}\int \frac{d\theta}{2\pi}\sin^{4}\phi_{e,k}\notag\\
\end{align}
and
\begin{align}\label{CE}
C=\bar{C}&= \frac{\sqrt{A_{h}A_{e}}}{A}=\frac{m_{e}+m_{h}}{2\sqrt{m_{e}m_{h}}}\frac{\sqrt{\int \frac{d\theta}{2\pi}\cos^{4}\phi_{e,k}}}{\int \frac{d\theta}{2\pi}\cos^{2}\phi_{e,k}}\notag\\
\tilde{C}&= \frac{\sqrt{A_{h}\bar{A}_{e}}}{\tilde{A}}=\frac{m_{e}+m_{h}}{2\sqrt{m_{e}m_{h}}}\frac{\sqrt{\int \frac{d\theta}{2\pi}\sin^{4}\phi_{e,k}}}{\int \frac{d\theta}{2\pi}\sin^{2}\phi_{e,k}}
\end{align}
  We also introduce the ratio
\begin{align}\label{CE_1}
E&=\frac{\tilde{A}_{e}}{\sqrt{{A}_{e}\bar{A}_{e}}}=\frac{\int \frac{d\theta}{2\pi}\cos^{2}\phi_{e,k}\sin^{2}\phi_{e,k}}{\sqrt{\int \frac{d\theta}{2\pi}\cos^{4}\phi_{e,k}\int \frac{d\theta}{2\pi}\sin^{4}\phi_{e,k}}}\, .
\end{align}
With these notations the 20 coupled RG equations read
\begin{align}
\dot{u}_{1}&={u}_{1}^{2}+{u}_{3}^{2}/C^{2}\notag\\
\dot{\bar{u}}_{1} & =\bar{u}_{1}^{2}+\bar{u}_{3}^{2}/{C}^{2}\notag\\
\dot{\tilde{u}}_{1} & =\tilde{u}_{1}^{2}+\tilde{u}_{3}^{2}/\tilde{C}^{2}\notag\\
\dot{\tilde{\tilde{u}}}_{1} & =\tilde{\tilde{u}}_{1}^{2}+\tilde{\tilde{u}}_{3}^{2}/\tilde{C}^{2}\notag\\
\dot{u}_{2} & =2u_{1}u_{2}-2u_{2}^{2}\notag\\
\dot{\bar{u}}_{2} & =2{\bar{u}}_{1}{\bar{u}}_{2}-2{\bar{u}}_{2}^{2}\notag\\
\dot{\tilde{u}}_{2} & =2{\tilde{u}}_{1}{\tilde{u}}_{2}-2{\tilde{u}}_{2}^{2}\notag\\
\dot{\tilde{\tilde{u}}}_{2} &=2\tilde{\tilde{u}}_{1}\tilde{\tilde{u}}_{2}-2\tilde{\tilde{u}}_{2}^{2}\notag\\
\dot{u}_{3}&=-\bar{u}_{3}\bar{u}_{5}-u_{3}u_{5}-\tilde{u}_{3}\bar{u}_{7}-\tilde{\tilde{u}}_{3}\bar{u}_{8}
\notag \\ &
-E(\tilde{u}_{3}u_{5}+\tilde{\tilde{u}}_{3}\bar{u}_{5}+u_{3}\bar{u}_{7}+\bar{u}_{3}\bar{u}_{8})\notag\\
&-u_{3}u_{4}-\bar{u}_{3}\bar{u}_{4}+4u_{1}u_{3}-2u_{2}u_{3}\notag\\
\dot{\bar{u}}_{3}&=-\bar{u}_{3}u_{5}-u_{3}\bar{u}_{5}-\tilde{\tilde{u}}_{3}\bar{u}_{7}-\tilde{u}_{3}\bar{u}_{8}
\notag\\ &
-E(\tilde{u}_{3}\bar{u}_{5}+\tilde{\tilde{u}}_{3}u_{5}+\bar{u}_{3}\bar{u}_{7}+u_{3}\bar{u}_{8})\notag\\
&-u_{3}\bar{u}_{4}-\bar{u}_{3}u_{4}+4\bar{u}_{1}\bar{u}_{3}-2\bar{u}_{2}\bar{u}_{3}\notag\\
\dot{\tilde{u}}_{3}&=-{u}_{3}\bar{u}_{7}-\bar{u}_{3}\bar{u}_{8}-\tilde{u}_{3}{u}_{6}-\tilde{\tilde{u}}_{3}\bar{u}_{6}
\notag\\ &
-E({u}_{3}{u}_{6}+\bar{u}_{3}\bar{u}_{6}+\tilde{u}_{3}\bar{u}_{7}+\tilde{\tilde{u}}_{3}\bar{u}_{8})\notag\\
&-\tilde{u}_{3}\bar{u}_{4}-\tilde{\tilde{u}}_{3}u_{4}+4\tilde{u}_{1}\tilde{u}_{3}-2\tilde{u}_{2}\tilde{u}_{3}\notag\\
\dot{\tilde{\tilde{u}}}_{3}&=-\bar{u}_{3}\bar{u}_{7}-{u}_{3}\bar{u}_{8}-\tilde{\tilde{u}}_{3}{u}_{6}-\tilde{u}_{3}\bar{u}_{6}
\notag\\ &
-E(\bar{u}_{3}{u}_{6}+{u}_{3}\bar{u}_{6}+\tilde{\tilde{u}}_{3}\bar{u}_{7}+\tilde{u}_{3}\bar{u}_{8})\notag\\
&-\tilde{u}_{3}{u}_{4}-\tilde{\tilde{u}}_{3}\bar{u}_{4}+4\tilde{\tilde{u}}_{1}\tilde{\tilde{u}}_{3}-2\tilde{\tilde{u}}_{2}\tilde{\tilde{u}}_{3}\notag\\
\dot{u}_{4}&=-{u}_{3}^{2}-\bar{u}_{3}^{2}-\tilde{u}_{3}^{2}-\tilde{\tilde{u}}_{3}^{2}-E(2u_{3}\tilde{u}_{3}+2\bar{u}_{3}\tilde{\tilde{u}}_{3})-{u}_{4}^{2}-\bar{u}_{4}^{2}\notag\\
\dot{\bar{u}}_{4}&=-2u_{3}\bar{u}_{3}-2\tilde{u}_{3}\tilde{\tilde{u}}_{3}-E(2u_{3}\tilde{\tilde{u}}_{3}+2\bar{u}_{3}\tilde{u}_{3})-2u_{4}\bar{u}_{4}\notag\\
\dot{u}_{5}&=-u_{5}^{2}-\bar{u}_{5}^{2}-\bar{u}_{7}^{2}-\bar{u}_{8}^{2}-E(2u_{5}\bar{u}_{7}+2\bar{u}_{5}\bar{u}_{8})-u_{3}^{2}-\bar{u}_{3}^{2}\notag\\
\dot{\bar{u}}_{5}&=-2u_{5}\bar{u}_{5}-2\bar{u}_{7}\bar{u}_{8}-E(2\bar{u}_{5}\bar{u}_{7}+2u_{5}\bar{u}_{8})-2u_{3}\bar{u}_{3}\notag\\
\dot{u}_{6}&=-\bar{u}_{7}^{2}-\bar{u}_{8}^{2}-{u}_{6}^{2}-\bar{u}_{6}^{2}-E(2u_{6}\bar{u}_{7}+2\bar{u}_{6}\bar{u}_{8})-\tilde{u}_{3}^{2}-\tilde{\tilde{u}}_{3}^{2}\notag\\
\dot{\bar{u}}_{6}&=-2\bar{u}_{7}\bar{u}_{8}-2u_{6}\bar{u}_{6}- E(2\bar{u}_{6}\bar{u}_{7}+2u_{6}\bar{u}_{8})-2\tilde{u}_{3}\tilde{\tilde{u}}_{3}\notag\\
\dot{\bar{u}}_{7}&=-u_{5}\bar{u}_{7}-\bar{u}_{5}\bar{u}_{8}-\bar{u}_{6}\bar{u}_{8}-u_{6}\bar{u}_{7} \nonumber \\
&-E(u_{5}u_{6}+\bar{u}_{5}\bar{u}_{6}+\bar{u}_{7}^{2}+\bar{u}_{8}^{2}) -u_{3}\tilde{u}_{3}-\bar{u}_{3}\tilde{\tilde{u}}_{3}\notag\\
\dot{\bar{u}}_{8}&=-u_{5}\bar{u}_{8}-\bar{u}_{5}\bar{u}_{7}-\bar{u}_{6}\bar{u}_{7}-u_{6}\bar{u}_{8} \notag \\
&-E(u_{5}\bar{u}_{6}+\bar{u}_{5}{u}_{6}+2\bar{u}_{7}\bar{u}_{8}) -u_{3}\tilde{\tilde{u}}_{3}-\bar{u}_{3}\tilde{u}_{3}.\label{20_RG_equations}
\end{align}
where  $\dot{u}=\frac{du}{dL}$.  The three parameters in this RG set, $C$, ${\tilde C}$, and $E$ depend on the ratio of hole and electron masses $m_h/m_e$ and on $A_0$.
 We remind that $A_0$ determines over which portion of the electron Fermi surface the
 $d_{xy}$ orbital component is stronger than the $d_{xz}$ ($d_{yz}$) component.

The analysis of the set shows that 
 couplings grow under the RG and
 diverge at a finite critical $L =L_0$ 
  (see Fig.~\ref{fig:20flow}). Physically this scale can be seen as a temperature of order $W e^{-L_0}$.
This signals an instability of the normal state.
The symmetry that is actually broken at $L_0$ has to be determined by comparing the susceptibilities in different channels. We will do this
 after we analyze the flow of the couplings.

\subsection{Fixed trajectories  of the RG equations}
\label{sec:traj}

A stable fixed trajectory is  the solution for $u_i$, to which the system flows from all directions as $L$ tends to the critical value $L_0$. Each fixed trajectory has a basin of attraction in the space of bare interactions.
A fixed trajectory is universal in the sense that the system behavior on this trajectory does  not depend on the initial conditions.
 The latter only determine how fast the system approaches  a given fixed trajectory.  We will show that in our case there are two stable fixed trajectories.

 An unstable fixed trajectory is  approached from some directions,  but along other directions the system moves away from it (i.e.,
 the stability analysis for deviations from an unstable fixed trajectory yields at least one positive exponent).
  Unstable fixed trajectories are located in between stable fixed trajectories and in general are  irrelevant for the RG analysis because the RG flow moves the system
  away from these trajectories towards the stable ones.  In our case, however, we show that
   there is one weakly unstable fixed trajectory with just one positive exponent, whose value is small. In this situation, a weakly unstable fixed trajectory behaves, for practical purposes,  as a stable one because deviations from it become relevant only near the end of the RG flow, when the hierarchy of susceptibilities is already established. By this reason, below  we treat the two stable and one weakly unstable fixed trajectories on equal footing.

 As a first step, we verified, both analytically and numerically, that along stable and weakly unstable trajectories
\begin{align}\label{anz}
u_{1}&=\bar{u}_{1},
\tilde{u}_{1}=\tilde{\tilde{u}}_{1},
u_{2}=\bar{u}_{2},
\tilde{u}_{2}=\tilde{\tilde{u}}_{2},
\notag \\
u_{3}&=\bar{u}_{3},
\tilde{u}_{3}=\tilde{\tilde{u}}_{3},
u_{4}=\bar{u}_{4},
u_{5}=\bar{u}_{5},
\notag \\
u_{6}&=\bar{u}_{6},
\bar{u}_{7}=\bar{u}_{8}\equiv{u}_{7}.
\end{align}
Specifically, we verified that  if we set initially $u_{1} \neq \bar{u}_{1}$, the difference between the two running couplings will decrease in the process of the RG flow.

 Eq.~\eqref{anz} allows one to reduce the 20 RG equations from Eq.~\eqref{20_RG_equations} to 10 equations:
\begingroup
\allowdisplaybreaks
\begin{align}
\dot{u}_{1}&={u}_{1}^{2}+{u}_{3}^{2}/C^{2}\quad
\dot{\tilde{u}}_{1}  =\tilde{u}_{1}^{2}+\tilde{u}_{3}^{2}/\tilde{C}^{2}\notag\\
\dot{u}_{2} & =2u_{1}u_{2}-2u_{2}^{2}\quad \quad
\dot{\tilde{u}}_{2}  =2{\tilde{u}}_{1}{\tilde{u}}_{2}-2{\tilde{u}}_{2}^{2}\notag\\
\dot{u}_{3}&=-2u_{3}u_{5}-2\tilde{u}_{3}{u}_{7}-E(2\tilde{u}_{3}u_{5}+2u_{3}{u}_{7})
\notag \\
&-2u_{3}u_{4}+4u_{1}u_{3}-2u_{2}u_{3}\notag\\
\dot{\tilde{u}}_{3}&=-2{u}_{3}{u}_{7}-2\tilde{u}_{3}{u}_{6}-E(2{u}_{3}{u}_{6}+2\tilde{u}_{3}{u}_{7})
\notag\\
&-2\tilde{u}_{3}{u}_{4}+4\tilde{u}_{1}\tilde{u}_{3}-2\tilde{u}_{2}\tilde{u}_{3}\notag\\
\dot{u}_{4}&=-2{u}_{3}^{2}-2\tilde{u}_{3}^{2}-E(4u_{3}\tilde{u}_{3})-2{u}_{4}^{2}
\notag \\
\dot{u}_{5}&=-2u_{5}^{2}-2{u}_{7}^{2}-E(4u_{5}{u}_{7})-2u_{3}^{2}\notag\\
\dot{u}_{6}&=-2{u}_{7}^{2}-2{u}_{6}^{2}-E(4u_{6}\bar{u})-2\tilde{u}_{3}^{2}\notag\\
\dot{u}_{7}&=-2u_{5}{u}_{7}-2u_{6}{u}_{7}-E(2u_{5}u_{6}+2{u}_{7}^{2})-2u_{3}\tilde{u}_{3}.
\end{align}
\endgroup

Along the fixed trajectories, the couplings grow (and diverge at $L = L_0$), but their ratios tend to universal  constant values.
We introduce such ratios by selecting, say, $u_1$, and writing $u_{i}=\gamma_{i}u_{1}$, $\tilde{u}_{i}=\tilde{\gamma}_{i}u_{1}$, etc.
We then set $\gamma_i$ to be constants and solve the algebraic equations for $\gamma_i$.
 The solutions yield non-divergent $\gamma_i$ if $u_1$ is  one 
 of the
 most strongly divergent couplings.  If some $\gamma_i$ come out infinite, we select another
  coupling as the primary one and repeat the procedure until all $\gamma_i$ are non-divergent.
We then check the stability of the solution by expanding around it to linear order and solving for the deviations $\delta u_j$. The deviations behave as
$\delta \gamma_j = \sum_m A_{mj} (\frac{1}{L_0-L})^{\beta_m}$ ($m=1,2,...,10$).
If all $\beta_m$ are negative, the trajectory is fully stable.  If one or more $\beta_m >0$, the trajectory is unstable.
 Like we said, we call a trajectory weakly unstable if only one $\beta_m >0$ and its  value is numerically small.

Carrying out this procedure, we obtain two stable fixed trajectories and one weakly unstable trajectory.
 We present technical details of our analysis in  the Appendix (see Sec.~\ref{sec:fixed_point_solution_detail}).
 For the two stable fixed trajectories, we find
\begin{align}
\label{FT_1}
&u_{1}(L)=\frac{1}{1+\gamma_{3}^{2}/C^{2}}\frac{1}{L_{0}-L},\notag\\
&\tilde{\gamma}_{1}={\gamma}_{2}=\tilde{\gamma}_{2}=\tilde{\gamma}_{3}=\gamma_{6}=\gamma_{7}=0,\notag\\
&\gamma_{3}=+C \sqrt{-1 + 8 C^2 + 4 \sqrt{1 - C^2 + 4 C^4}},\notag\\
&\gamma_{4}=\gamma_{5}=1-2C^2-\sqrt{1-C^2+4C^4}
\end{align}
for the one  and
\begin{align}
\label{FT_2}
&\tilde{u}_{1}(L)=\frac{1}{1+\tilde{\gamma}_{3}^{2}/\tilde{C}^{2}}\frac{1}{L_{0}-L},\notag\\
&\gamma_{1}={\gamma}_{2}=\tilde{\gamma}_{2}=\gamma_{3}=\gamma_{5}=\gamma_{7}=0,\notag\\
&\tilde{\gamma}_{3}=+\tilde{C} \sqrt{-1 + 8 \tilde{C}^2 + 4 \sqrt{1 - \tilde{C}^2 + 4 \tilde{C}^4}},\notag\\
&\gamma_{4}=\gamma_{6}=1-2\tilde{C}^2-\sqrt{1-\tilde{C}^2+4\tilde{C}^4}.
\notag\\
\end{align}
for the other.

Along the first fixed trajectory, the interactions $\tilde{U}_{1}$, $U_{2}$, $\tilde{U}_{2}$, $\tilde{U}_{3}$, $U_{6}$, and $U_{7}$ become negligible compared to interactions
$U_{1}$, $U_3$, $U_4$, and $U_5$. Going back to Eq.~(\ref{general_H}), we find that this separation between the couplings implies that the interactions
 involving $d_{xy}$ components on electron pockets vanish compared to interactions involving $d_{xz}$ or $d_{yz}$ components.  In other words, the electron pockets can be effectively approximated as pure $d_{xz}$ (the $Y$ pocket) and pure $d_{yz}$ (the $X$ pocket).
 Along the second fixed trajectory, the situation is opposite --  the interactions
 involving $d_{xz}$ or $d_{yz}$ orbital components on electron pockets vanish compared to interactions involving $d_{xy}$ components.  In this case,
  both electron pockets can be effectively approximated as pure $d_{xy}$.
  These two situations correspond to the two approximate models, considered in  Ref. \cite{CKF2016} -- model I and II, respectively.

Additionally, we found a new fixed trajectory not present in the approximate models.
This new fixed trajectory is formally an unstable one, but it is weakly unstable, with only one unstable direction. Furthermore the
 corresponding positive exponent is numerically small, e.g.,  $\beta_1 = 0.10 $ for $m_h/m_e =1$ and $A_0 =0.8$.
On this fixed trajectory the couplings behave as
\begin{align}
\label{FT_3}
&\tilde{u}_{1}=u_1=\frac{1}{1+\gamma_{3}^{2}/C^{2}}\frac{1}{L_{0}-L}>0,\notag\\
&{\gamma}_{2}=\tilde{\gamma}_{2}=0\notag \\
&\gamma_{3}/C=\tilde{\gamma}_{3}/\tilde{C},\notag\\
&\gamma_{4},\gamma_{5},\gamma_{6},\gamma_{7}<0
\end{align}
The values of the couplings $\gamma_3$, $\gamma_4$, $\gamma_5$, $\gamma_6$  and $\gamma_7$ depend on the three parameters  $C$, ${\tilde C}$, and $E$, defined in Eqs. (\ref{CE}) and (\ref{CE_1}), which in turn
 depend on the ratios of hole and electron masses and on $A_0$.  The analytical formulas
for ${\gamma}_i$ are somewhat involved and we present them in the Appendix (see Sec.~\ref{sec:fixed_point_solution_detail}).
  For $m_h/m_e =1$ and $A_0 =0.8$, the numbers are $\gamma_3 = 9.62$, $\gamma_4 = -14.69$, $\gamma_5 = -5.50$, $\gamma_6 = -3.74$, $\gamma_7 = -4.56$.

The key feature of this fixed trajectory is that now
 interactions involving $d_{xz}/d_{yz}$ and $d_{xy}$ orbital components of the electron pockets remain of the same order and both grow under RG.  We will see below that,
  as a consequence, an instability towards orbital order leads to simultaneous appearance of three order parameters, two involving $d_{xz}/d_{yz}$ orbitals on hole pockets and on electron pockets, and one involving $d_{xy}$ orbitals on the electron pockets.  We argue below that all three   orbital order parameters  are required  to explain recent ARPES data on FeSe~\cite{borisenko}.

\section{Scaling of susceptibilities and the hierarchy of phase transitions}
 \label{sec:inst}

In this section we analyze the hierarchy of instabilities, which break different symmetries. For this we introduce auxiliary  order parameter fields in different channels.
We obtain the RG equations for the vertices, which couple the corresponding auxiliary fields to fermions, and solve them using the running couplings as inputs. We then express the running susceptibilities $\chi_i (L)$ in terms of running vertices and obtain the expressions for $\chi_i (L)$ in different channels.
Similar procedure was applied to other problems~\cite{graphene, suscept_RG}.
The divergence of the susceptibility in a particular channel signals an instability towards developing a long-range order in this channel.  We will see that not all susceptibilities diverge as $L$ approaches $L_0$. For divergent susceptibilities, we compare the exponents and select the channel, in which the
 the  exponent is the largest, as the one where the leading instability occurs.

Below we consider SDW, charge-density-wave (CDW), superconducting, and orbital channels. The interplay between the susceptibilities in these channels on the
the two stable fixed trajectories is the same as in the two approximate models considered in Ref. \cite{CKF2016}. We will not
  repeat the analysis here and focus on the system behavior along the weakly unstable fixed trajectory.

 \subsection{SDW and CDW order parameters}
 \label{sec:SDW0pi}

The SDW order introduces a spatial modulation at wave-vectors $X=(\pi,0)$ and/or $Y=(0,\pi)$
and breaks spin SU(2) symmetry.
If SDW order develops at a single wave-vector, $X$ or $Y$, it in addition breaks the $C_4$ lattice rotational symmetry (the stripe order).
If the modulations at $X$ and $Y$ wave-vectors coexist, the resulting checkerboard SDW order preserves the $C_4$ lattice symmetry.
In the RG approach we perform a linear stability analysis of the paramagnetic state, i.e. we analyze the behavior of susceptibilities at temperatures   above the one for the  leading instability. By symmetry, SDW susceptibilities  at $X$ and $Y$ are  equivalent in the paramagnetic (non-nematic) phase.  To distinguish between stripe and checkerboard orders one has to include  non-linear couplings between the $X$ and $Y$  SDW order parameters~\cite{rafael}.
This analysis is  beyond the scope of our RG analysis.

In a multi-orbital system, the orbital content must be included in the  classification of  different order parameters in terms of irreducible representations of the
  symmetry group of the lattice.
Specifically, for  a tetragonal lattice,  the SDW order parameters must come in degenerate pairs because an SDW order parameter at the wave-vector $X$ transforms into an SDW order parameter at the wave-vector $Y$ under a rotation by
$\pi/2$.
In addition, SDW order parameters split into two distinct groups, depending on whether the order parameter is diagonal or off-diagonal in the orbital index~\cite{Cvetkovic2013}.
 The SDW order in the first group gives rise to a finite magnetization on Fe ions, while for the order in the second group the magnetization vanishes on Fe
sites,
but is finite on
 pnictogen or chalcogen ions (on Se ions in FeSe).  In our case the first group contains two elements (SDW involving $d_{xz}$ or $d_{yz}$ orbitals) and the second group contains six elements (between $d_{xz}$ at $\Gamma$ and $d_{yz}$  at $X$, or $d_{xy}$ at $X$ or $d_{xy}$ at $Y$, and  between  $d_{yz}$ at $\Gamma$ and $d_{xz}$  at $Y$, or $d_{xy}$ at $Y$ or $d_{xy}$ at $X$).
 Accordingly, we introduce eight auxiliary fields $\bm{s}^{(0)}_{i(i')}$, $i =1,\ldots, 4$,
 choose them to be along the $z$ direction for definiteness,
  and couple them to fermions as
 \begin{align}
 	& H_{SDW}=\label{extra_ac_1}\notag\\
 	&\sum_{k}[\bm{s}_{1}^{(0)}\cdot\psi^{\dag}_{1,\alpha}(k)\bm{\sigma}_{\alpha,\beta}\psi_{6,\beta}(k)+
 \bm{s}_{1'}^{(0)}\cdot\psi^{\dag}_{1,\alpha}(k)\bm{\sigma}_{\alpha,\beta}\psi_{5,\beta}(k)\notag\\
 	&+\bm{s}_{2}^{(0)}\cdot\psi^{\dag}_{2,\alpha}(k)\bm{\sigma}_{\alpha,\beta}\psi_{5,\beta}(k)+\bm{s}_{2'}^{(0)}\cdot\psi^{\dag}_{2,\alpha}(k)\bm{\sigma}_{\alpha,\beta}\psi_{6,\beta}(k)\notag\\
 	&+\bm{s}_{3}^{(0)}\cdot\psi^{\dag}_{3,\alpha}(k)\bm{\sigma}_{\alpha,\beta}\psi_{5,\beta}(k)+\bm{s}_{3'}^{(0)}\cdot\psi^{\dag}_{3,\alpha}(k\bm{\sigma}_{\alpha,\beta}\psi_{6,\beta}(k)\notag\\
 	&+\bm{s}_{4}^{(0)}\cdot\psi^{\dag}_{4,\alpha}(k)\bm{\sigma}_{\alpha,\beta}\psi_{6,\beta}(k)+
 \bm{s}_{4'}^{(0)}\cdot\psi^{\dag}_{4,\alpha}(k)\bm{\sigma}_{\alpha,\beta}\psi_{5,\beta}(k)\notag \\
 &+h.c.]
 \end{align}
We recall that in our notations $\psi_{1}({\bm{k}})$ and $\psi_{6}({\bm{k}})$ are Bloch states of pure $d_{xz}$ character at $Y$ and at $\Gamma$,
 $\psi_{3}({\bm{k}})$ and $\psi_{5}({\bm{k}})$ are the Bloch states of pure $d_{yz}$ character at $X$ and at $\Gamma$, and  $\psi_{2}({\bm{k}})$ and $\psi_{4}({\bm{k}})$ are the Bloch states of pure $d_{xy}$ character at $Y$ and at $X$, respectively.
 The field  $\bm{s}_{1}^{(0)}$ ($\bm{s}_{3}^{(0)}$) couples to intra-orbital SDW order parameters on  $d_{xz}$ ($d_{yz}$) Fe orbitals,
The intra-orbital  SDW at $Y$ and at $X$  are related by $C_4$ lattice rotation, and the susceptibilities with respect to
$\bm{s}_{1}^{(0)}$  and $\bm{s}_{3}^{(0)}$ must be
 equal by symmetry.  The other four auxiliary fields couple to
 inter-orbital
 SDW. By symmetry, the susceptibilities with respect to
 $\bm{s}_{1'}^{(0)}$ and $\bm{s}_{3'}^{(0)}$ and
 with respect to $\bm{s}_{2(2')}^{(0)}$ and $\bm{s}_{4(4')}^{(0)}$ must coincide.

The RG equations for the flow of the running auxiliary fields $\bm{s}_{i(i')}$ away from the bare values $\bm{s}^{(0)}_{i(i')}$ from Eq. (\ref{extra_ac_1}) are obtained in the same way as the flow of the couplings -- by analyzing diagrams with renormalizations of $\bm{s}^{(0)}_{i(i')}$ due to the interactions.
 In general, $\bm{s}^{(0)}_{i(i')}$ are complex fields, whose real part Re $\bm{s}_{i(i')}$ describes the actual SDW, and whose imaginary part Im $\bm{s}_{i(i')}$ describes spin currents.
 We analyzed RG flows for both Re $\bm{s}_{i(i')}$ and Im $\bm{s}_{i(i')}$ and found that the RG equations for the two decouple, and in the process of the RG flow,
 Re $\bm{s}_{i(i')}$ becomes larger than Im $\bm{s}_{i(i')}$, even if the bare values of the two are comparable.
 For brevity, we then only consider the
 flow
 of Re $\bm{s}_{i(i')}$ and skip ``Re'' in the formulas below.
 \begin{figure}[h]
 	\centering{}
 	\includegraphics[width=1.05\columnwidth]{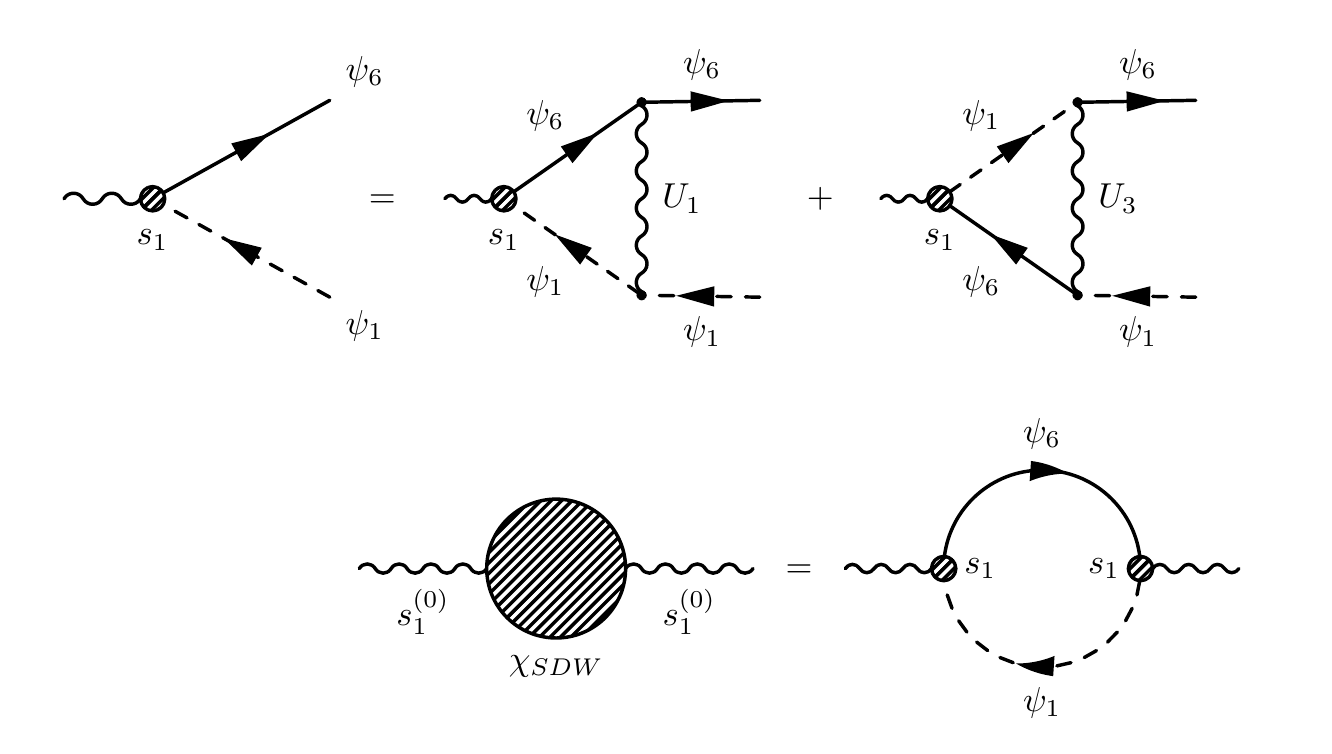} \protect\caption{The diagrams for the renormalization of the SDW vertex(upper panel) and the
 susceptibility(lower panel).}
 		\label{fig:FD_SDW}
 \end{figure}

The
RG
equations describing the renormalization of $\bm{s}_{1}$, $\bm{s}_{1'}$, $\bm{s}_{2}$, and $\bm{s}_{2'}$ are (see Fig.~\ref{fig:FD_SDW})
 \begin{align}
 	\frac{d {s}_{1}}{dL}&={s}_{1}\left(u_1+\frac{u_3}{C}\right)\notag\\
 	\frac{d {s}_{2}}{dL}&={s}_{2}\left(\tilde{u}_1+\frac{\tilde{u}_3}{\tilde{C}}\right) \notag\\
 \frac{d {s}_{1'}}{dL}&={s}_{1'}\left({\bar u}_1+\frac{{\bar u}_3}{C}\right)\notag\\
 	\frac{d {s}_{2'}}{dL}&={s}_{2'}\left(\tilde{{\tilde u}}_1+\frac{\tilde{{\tilde u}}_3}{\tilde{C}}\right)\, . \label{pRG_SDW}
 \end{align}
On the fixed trajectories, the running  couplings satisfy  Eq.~\eqref{anz}, and
 the RG equations for ${s}_{i}$ and ${s}_{i'}$ become identical.
We emphasize that this  equivalence is the property of the fixed trajectories of the RG flow of the couplings
 rather than the consequence of tetragonal symmetry.
The latter only guarantees that the vertices $s_{3(3')}$ and $s_{4(4')}$ satisfy the same parquet RG equations as $s_{1(1')}$ and $s_{2(2')}$, respectively.

On the weakly unstable fixed trajectory the couplings are related by Eq.~\eqref{FT_3}.
 Expressing $u_3 (L)$ in terms of $u_1 (L)$ and using $u_1 (L) = \frac{1}{1 + \gamma^2_3/C^2} \frac{1}{L_0 -L}$ we obtain from the first equation in
~\eqref{pRG_SDW}
\begin{align}\label{s1L}
s_{1}(L)& \propto s_{1}^{(0)} (\frac{1}{L_0-L})^\frac{1+\gamma_{3}/C}{1+\gamma_{3}^{2}/{C^2}}.
\end{align}
The running susceptibility $\chi_{SDW,1} (L)$ is represented within RG  by the bubble diagram with fully renormalized side vertices. We emphasize that both vertices should be
  treated as the running ones (see Fig.~\ref{fig:FD_SDW}).
  The RG equation for $\chi_{SDW,1} (L)$  is then
 \begin{align}\label{chi_SDW}
 (s_{1}^{(0)})^2\frac{d\chi_{SDW,1}}{dL}= s_{1}^2\, .
 \end{align}
Solving this equation we obtain
 \begin{align}
 \chi_{SDW,1}(L)&=
(\frac{1}{L_0-L})^{\alpha_{SDW,1}}\, ,
\label{spin_susceptibility}
 \end{align}
 where the scaling exponent $\alpha_{SDW,1}$ is given by
 \begin{align}
 \alpha_{SDW,1} = 2\frac{1+\gamma_{3}/C}{1+\gamma_{3}^{2}/{C^2}}-1.
 \end{align}
 Performing the same calculations for $\chi_{SDW,2}(L)$ we obtain
 \begin{align}
 \chi_{SDW,2}(L) \propto (\frac{1}{L_0-L})^{\alpha_{SDW,2}}
 \end{align}
 with
\begin{align}\label{SDW_1}
\alpha_{SDW,2}=2\frac{1+\tilde{\gamma}_{3}/\tilde{C}}{1+\tilde{\gamma}_{3}^{2}/{\tilde{C}^2}}-1
 \end{align}
Using the fact that on the weakly unstable fixed trajectory $\gamma_3/C = \tilde{\gamma}_3/\tilde{C}$ (see Eq. \eqref{FT_3}), we find
\begin{align}\label{SDW_1_1}
\alpha_{SDW,2}= \alpha_{SDW,1} \equiv \alpha_{SDW}.
 \end{align}
 By $C_4$ symmetry,  the other susceptibilities $\chi_{SDW,i (i')}(L)$ have the same exponent
 $\alpha_{SDW,1'} =  \alpha_{SDW,3}=  \alpha_{SDW,3'} =  \alpha_{SDW,2'}=  \alpha_{SDW,4}=  \alpha_{SDW,4'} = \alpha_{SDW}$.
 We plot $\alpha_{SDW}$ along with the exponents in other channels in Fig.~\ref{fig:exponents_case3}.

We also analyzed the susceptibility in the CDW channel at the same momentum $X$ and $Y$. We found (see Sec.~\ref{sec:CDW0pi} in the Appendix) that along the weakly unstable fixed trajectory different CDW components are degenerate and the susceptibility exponent in the CDW channel is the same as in the SDW channel, i.e.,
\begin{align}\label{SDW_1_1_1}
\alpha_{CDW} =  \alpha_{SDW}\, .
 \end{align}

 \subsection{Superconducting order parameters}
 \label{sec:SC}
To study superconductivity we analyze the response to the auxiliary intra-orbital  pairing fields $\Delta_{i}^{(0)}$, $i=1,\ldots,6$ 
and inter-orbital pairing fields $\Delta_{1,2}^{(0)}$, $\Delta_{3,4}^{(0)}$, $\Delta_{5,6}^{(0)}$.
These 12 auxiliary fields couple to 12 distinct singlet superconducting order parameters which one can construct out of $d_{xz}, d_{yz}$, and $d_{xy}$ orbitals:
\begingroup
 \begin{align}
 \notag\\
 & H^{(0)}_{SC}=\notag\\
 &\sum_{k\sigma}[\Delta_{1}^{(0)}\psi^{\dag}_{1,\sigma}(k)\psi^{\dag}_{1,-\sigma}(-k)+\Delta_{2}^{(0)}\psi^{\dag}_{2,\sigma}(k)\psi^{\dag}_{2,-\sigma}(-k)\notag\\
 &+\Delta_{3}^{(0)}\psi^{\dag}_{3,\sigma}(k)\psi^{\dag}_{3,-\sigma}(-k)+\Delta_{4}^{(0)}\psi^{\dag}_{4,\sigma}(k)\psi^{\dag}_{4,-\sigma}(-k)\notag\\
 &+\Delta_{5}^{(0)}\psi^{\dag}_{5,\sigma}(k)\psi^{\dag}_{5,-\sigma}(-k)+\Delta_{6}^{(0)}\psi^{\dag}_{6,\sigma}(k)\psi^{\dag}_{6,-\sigma}(-k)\notag\\
 &+\Delta_{1,2}^{(0)}\psi^{\dag}_{1,\sigma}(k)\psi^{\dag}_{2,-\sigma}(-k)+\Delta_{3,4}^{(0)}\psi^{\dag}_{3,\sigma}(k)\psi^{\dag}_{4,-\sigma}(-k)\notag\\
 &+\Delta_{5,6}^{(0)}\psi^{\dag}_{5,\sigma}(k)\psi^{\dag}_{6,-\sigma}(-k)+h.c.]\label{test_SC}
 \end{align}
 \endgroup
The parameters $\Delta_{i}^{(0)}$ and $\Delta_{ij}^{(0)}$ play the role of bare superconducting vertices with zero total momenta in the particle-particle channel.  We label the full vertices  as $\Delta_{i}$ and $\Delta_{i,j}$, without the superscript.

 Like we did in the SDW case, we first obtain and solve the RG equations for the vertices and then use the results to obtain the exponents for superconducting susceptibilities.
  For symmetry analyses it is useful to introduce the symmetrized combinations $\Delta_{1}\pm\Delta_{3}$, $\Delta_{2}\pm\Delta_{4}$, and $\Delta_{5}\pm\Delta_{6}$. These combinations  transform  as $A_{1g}$ ($B_{1g}$) representations of the $D_{4h}$ point group of a tetragonal lattice.  	
Similarly, the combinations $\Delta^{re(im)}_{1,2}= \Delta_{1,2}\pm\Delta_{1,2}^*$, $\Delta^{re(im)}_{3,4}= \Delta_{3,4}\pm\Delta_{3,4}^*$ and
$\Delta^{re(im)}_{5,6}= \Delta_{5,6}\pm\Delta_{5,6}^*$ transform as $A_{2g}$ ($B_{2g}$)  representations of the $D_{4h}$ group.
 Because the interaction is a $D_{4h}$ scalar, renormalizations do not mix vertices from different representations.
As a result, all  symmetrized combinations, belonging to different representations,  decouple and flow separately under RG.

\begin{figure*}[ht]
	\centering{}
	\includegraphics[width=1.0\columnwidth]{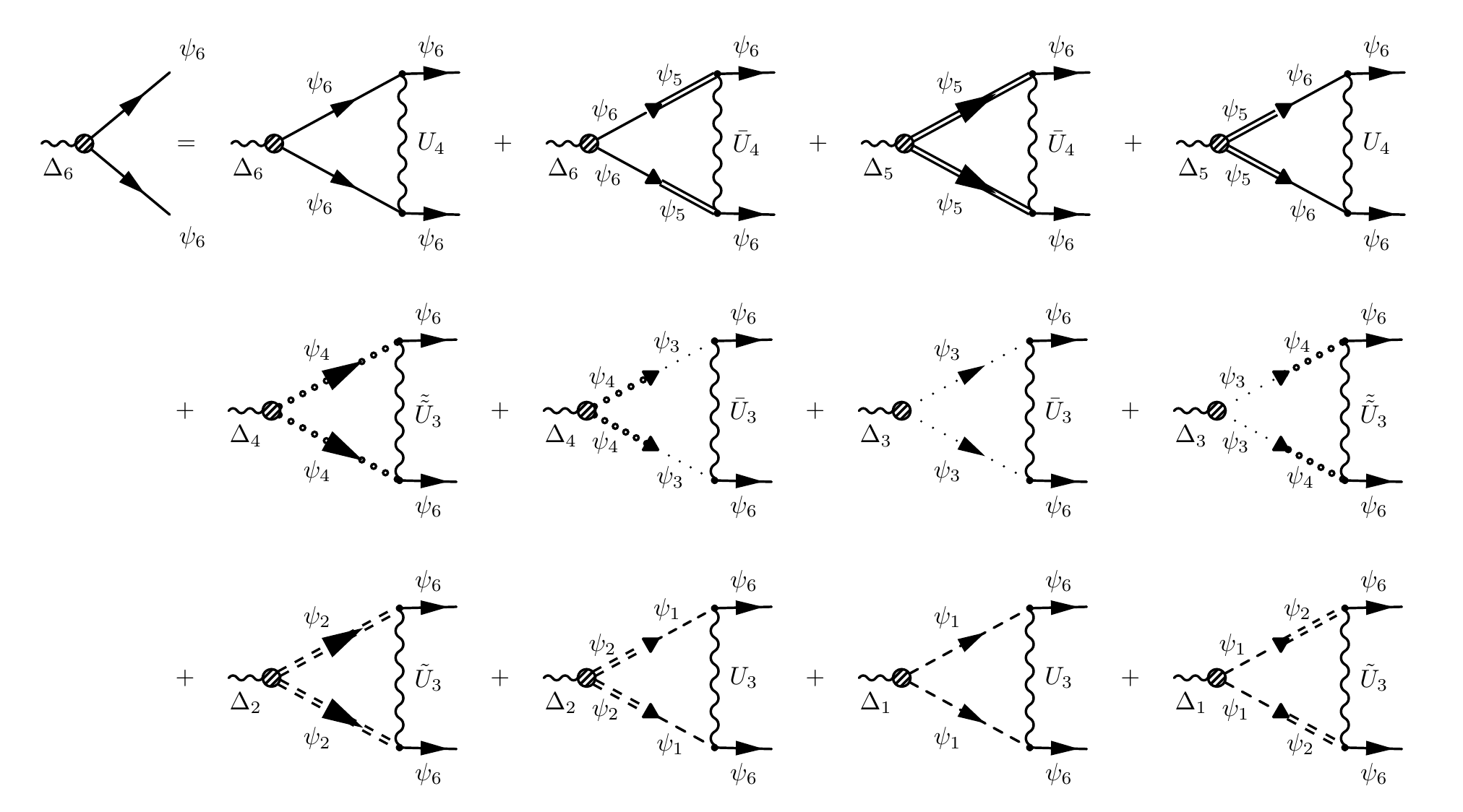} \protect\caption{The  diagraamatic representation of the  renormalizations of the vertices in SC channel.}
	\label{fig:FD_SC}
\end{figure*}

The derivation of RG equations proceeds in the same way as for SDW vertices.
We show the corresponding diagrams in Fig.~\ref{fig:FD_SC}.
The RG equations for particle-particle vertices in the $A_{1g}$ channel on the weakly unstable fixed trajectory are
\begin{widetext}
  \begin{align}\label{SC_matrix}
  \frac{d}{dL}
  \begin{pmatrix}
  \Delta_{1}+\Delta_{3}        \\
  \Delta_{2}+\Delta_{4}           \\
  \Delta_{5}+\Delta_{6}
  \end{pmatrix}
  =
  \begin{pmatrix}
  -2(u_{5}+E u_{7}) & -2(\sqrt{\frac{\overline{A}_{e}}{A_e}}{u}_{7}+\sqrt{\frac{\overline{A}_{e}}{A_e}}Eu_5) &-2(\sqrt{\frac{A_h}{A_e}}u_{3})  \\
  -2(\sqrt{\frac{A_e}{\overline{A}_{e}}}u_{7}+\sqrt{\frac{A_e}{\overline{A}_{e}}}Eu_6) & -2(u_{6}+E u_{7}) & -2(\sqrt{\frac{A_h}{\overline{A}_{e}}}\tilde{u}_{3}) \\
  -2(\sqrt{\frac{{A}_{e}}{A_h}}u_{3}+\sqrt{\frac{{A}_{e}}{A_h}}E\tilde{u}_{3}) & -2(\sqrt{\frac{\bar{A}_{e}}{A_h}}\tilde{u}_{3}+\sqrt{\frac{\bar{A}_{e}}{A_h}}E{u}_{3}) & -2u_4
  \end{pmatrix}
  \begin{pmatrix}
  \Delta_{1}+\Delta_{3}        \\
  \Delta_{2}+\Delta_{4}           \\
  \Delta_{5}+\Delta_{6}
  \end{pmatrix}\, ,
  \end{align}
 \end{widetext}
We solve the RG equations by
taking the  interactions $u_i$  to be on the weakly unstable fixed trajectory, \eqref{FT_3}. The diagonalization of the $3 \times 3$  matrix
 yields three independent combinations of $A_{1g}$ vertices, corresponding to three eigenvalues: $e_{\alpha} u_1$, $e_{\beta} u_1$ and $e_{\gamma} u_1$.
Solving the three independent RG equations we obtain
  \begin{align}
  \Delta_{\mu}(L)& \propto
  \left(\frac{1}{L_{0}-L}\right)^{\frac{e_\mu}{1+\gamma_3^{2}/C^2}}\, ,
  \end{align}
where   $\mu=\alpha,\beta,\gamma$.
We choose the solution with the largest exponent, which corresponds to the largest eigenvalue $e_{SC} =\max\{e_{\alpha},e_{\beta},e_{\gamma}\}$.
For the corresponding susceptibility we then obtain $\chi_{SC} (L) \propto 1/(L_0 -L)^{\alpha_{SC}}$, where
 \begin{align}\label{alpha_SC_f}
 \alpha_{SC}=2\frac{e_{SC}}{1+\gamma_3^{2}/C^2}-1\, .
 \end{align}
The values $e_{\alpha,\beta,\gamma}$  depend on our three parameters $C, {\tilde C}$, and $E$, 
which, we recall,  depend on $A_0$ and $m_h/m_e$.
 We obtain the largest exponent $\alpha_{SC}$ numerically and show it in Fig.~\ref{fig:exponents_case3} as a function of
 $A_0$ along with the exponents in the other channels.
The corresponding $A_{1g}$ order parameter is plotted in Fig.~\ref{fig:SCangular_dependence_case3} as a function of the angle along the Fermi surfaces. The order parameter has opposite signs on electron and hole Fermi surfaces, i.e. the gap structure is  $s^{+-}$.

We also analyzed the three other superconducting channels
$B_{1g}$, $A_{2g}$ and $B_{2g}$, and found that the vertices and susceptibilities in these channels do not diverge.
In the $B_{1g}$ channel, the analog of the $3 \times 3$ matrix for $\Delta_1-\Delta_3$, $\Delta_2 - \Delta_4$, and $\Delta_5 - \Delta_6$ vanishes
 on the weakly unstable fixed trajectory, because the corresponding couplings vanish (see Eq.~\eqref{anz}).
 The vertices in $A_{2g}$ and $B_{2g}$ channels  describe inter-orbital pairing. These vertices are  renormalized
 via the interactions $\tilde{U}_4$, $\tilde{\tilde{U}}_4$, $\tilde{U}_7$ and $\tilde{\tilde{U}}_7$.
For our choice $U' >J$,  these interactions renormalize to zero under RG, hence the corresponding vertices do not increase.

\subsection{Orbital order parameters}
\label{sec:Pom}
\begin{figure}[h]
   	\centering{}\includegraphics[width=0.98\columnwidth]{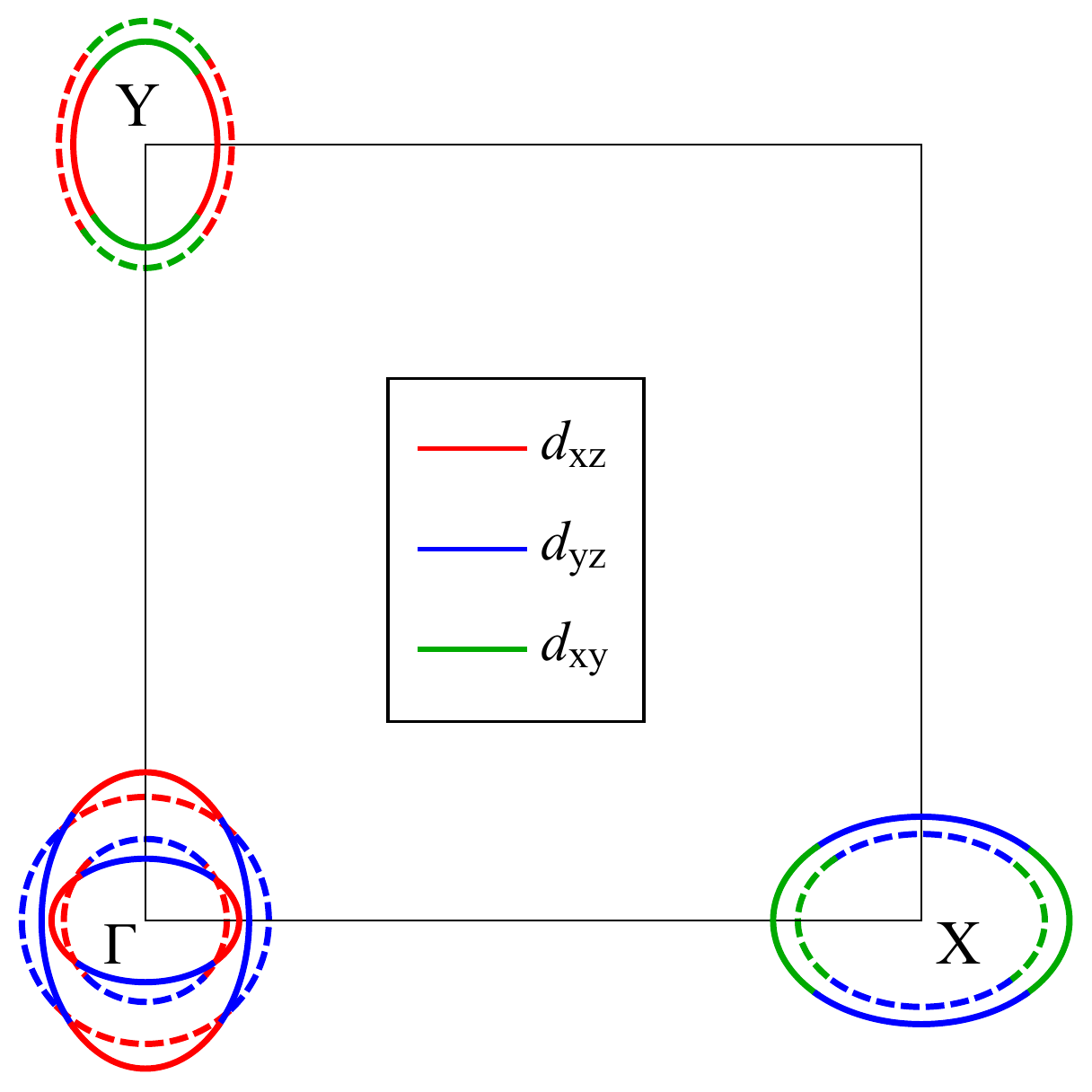} \caption{Electron structure in the nematic state above and below the onset of $B_{1g}$ Pomeranchuk instability. The two originally circular hole pockets (dashed lines)
   		are distorted into ellipses with orthogonal directions of elongation (solid lines).  The electron pockets at $X$ and $Y$, originally of the same size and form (dashed lines),   become inequivalent in the presence of a nematic order (solid lines).}
   		\label{fig:Pomeranchuk_sketch}
   \end{figure}

We consider orbital order parameters with zero transferred momentum.  In the band basis, an instability  leading to condensation of
 any of such orbital order parameters is a
Pomeranchuk instability.
A non-s-wave Pomeranchuk order breaks the rotational symmetry of the lattice but does not break the translational invariance.
The  reconstruction of the Fermi surfaces for a d-wave 
($B_{1g}$) Pomeranchuk order is shown in  Fig.~\ref{fig:Pomeranchuk_sketch}.

To analyze the susceptibilities in the orbital channel  we again introduce auxiliary fields, this time real charge fields $\Gamma_{i}^{(0)}$ and complex charge fields $\Gamma_{i,j}^{(0)}$. The coupling of auxiliary fields to fermions is described by
\begin{align}
H_{Pom}=&\sum_{k\sigma}\Big[\Gamma_{1}^{(0)}\psi^{\dag}_{1,\sigma}(k)\psi_{1,\sigma}(k)+\Gamma_{2}^{(0)}\psi^{\dag}_{2,\sigma}(k)\psi_{2,\sigma}(k)
\notag \\
+& \Gamma_{3}^{(0)}\psi^{\dag}_{3,\sigma}(k)\psi_{3,\sigma}(k)
+\Gamma_{4}^{(0)}\psi^{\dag}_{4,\sigma}(k)\psi_{4,\sigma}(k)
\notag \\
+& \Gamma_{5}^{(0)}\psi^{\dag}_{5,\sigma}(k)\psi^{\dag}_{5,-\sigma}(k)+\Gamma_{6}^{(0)}\psi^{\dag}_{6,\sigma}(k)\psi_{6,\sigma}(k)
\notag \\
+&
 (\Gamma_{1,2}^{(0)}\psi^{\dag}_{1,\sigma}(k)\psi_{2,\sigma}(k)+\Gamma_{3,4}^{(0)}\psi^{\dag}_{3,\sigma}(k)\psi_{4,\sigma}(k)
\notag \\
+& \Gamma_{5,6}^{(0)}\psi^{\dag}_{5,\sigma}(k)\psi_{6,\sigma}(k)+h.c.)\Big]\, .
\label{test_Pom}
\end{align}
The coefficients $\Gamma_{i}^{(0)}$ and $\Gamma_{i,j}^{(0)}$ are bare vertices with zero momentum in the particle-hole charge channel.
 We label dressed vertices by the same $\Gamma_{i}$ and $\Gamma_{i,j}$, but  without the superscript.
 We  introduce symmetrized combinations $\Gamma_{1}\pm\Gamma_{3}$, $\Gamma_{2}\pm\Gamma_{4}$, and $\Gamma_{5}\pm\Gamma_{6}$, which
 transform as $A_{1g}$ ($B_{1g}$) representations of the $D_{4h}$ group, and the combinations $\Gamma^{re(im)}_{1,2}= \Gamma_{1,2}\pm\Gamma_{1,2}^*$, $\Gamma^{re(im)}_{3,4}= \Gamma_{3,4}\pm\Gamma_{3,4}^*$ and
$\Gamma^{re(im)}_{5,6}= \Gamma_{5,6}\pm\Gamma_{5,6}^*$, which  transform as $A_{2g}$ ($B_{2g}$).
Combinations belonging to different representations again decouple in the RG equations.
	\begin{figure*}[ht]
		\centering{}
		\includegraphics[width=2.1\columnwidth]{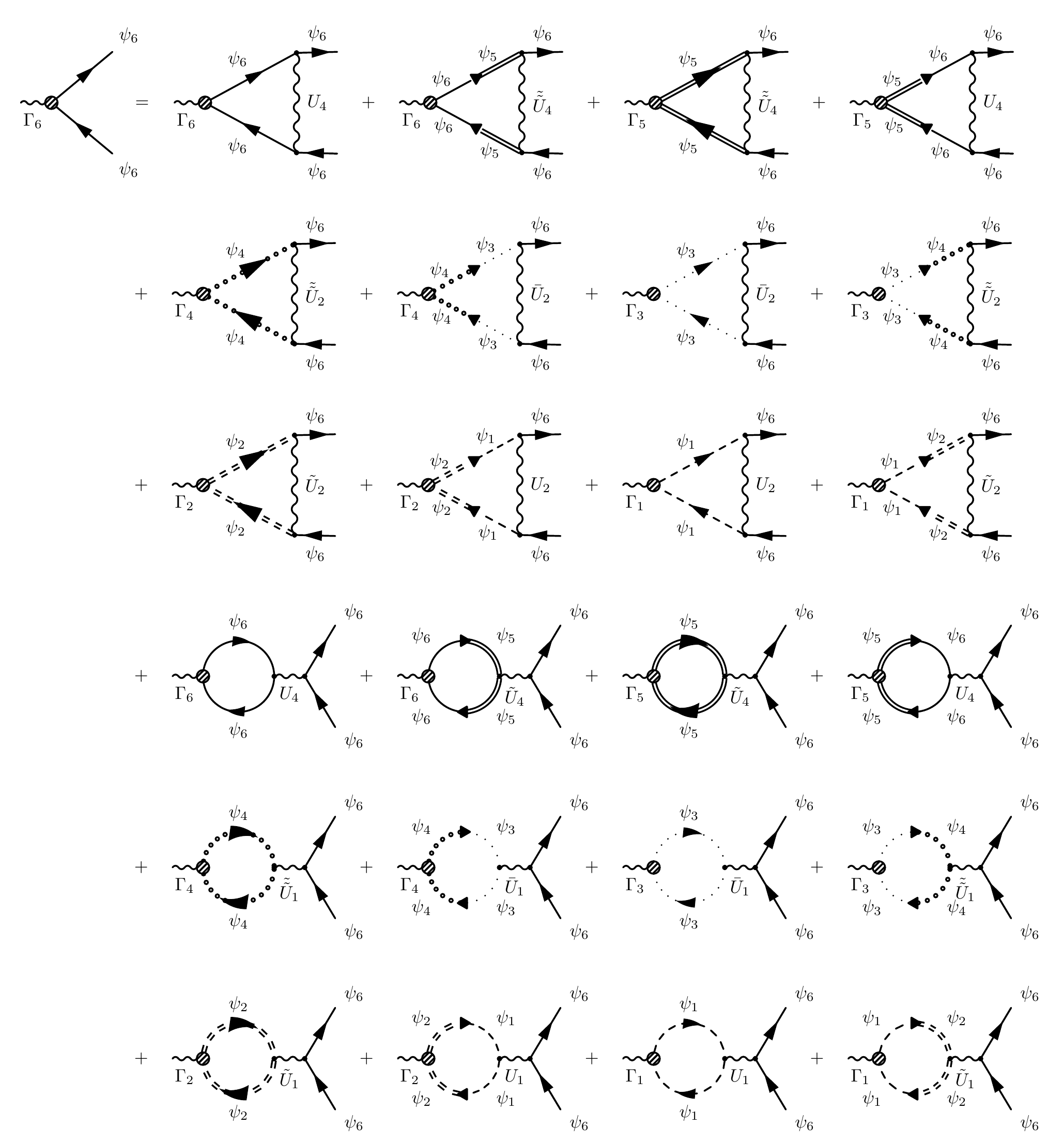} \protect\caption{The diagramatic representation of the renormalization  of the vertices in the Pomeranchuk channel.}
		\label{fig:FD_Pom}
	\end{figure*}

The renormalization of the vertices in the orbital (Pomeranchuk) channels is different from the ones in SDW and superconducting channels, because
the  particle-hole susceptibility at zero momentum transfer is non-logarithmical.
 Still, the renormalization involves the running couplings $u_i (L)$.  The key point, discussed in detail in Ref. ~\cite{kh_ch_last}, is that
 the renormalization of $\Gamma_i (L)$ and $\Gamma_{ij} (L)$ comes from internal energies comparable to $L$.  As a consequence,  the vertices at a scale $L$ are
  expressed in terms of interactions $u_i$ {\it at the same} scale  $L$.

We now need to select diagrammatic series for $\Gamma_i$. In SDW and superconducting channels the  RG flow of the vertices is given by series of ladder diagrams.
The selection of these diagrams is rigorously justified within the one-loop RG. For the Pomeranchuk vertices, there are no logarithms and hence no parameter to select a particular set of diagrams. We choose, without proof, the same set of ladder diagrams as for SDW and superconducting channels
(see Fig.~\ref{fig:FD_Pom}).
 Within this approximation, the equations for the dressed Pomeranchuk vertices are
\begin{align}\label{RPA_Pom1}
\begin{pmatrix}
\Gamma_{1}-\Gamma_{3}        \\
\Gamma_{2}-\Gamma_{4}         \\
\Gamma_{5}-\Gamma_{6}	     		     	
\end{pmatrix}
\!=\!
\bm{M}_{B_{1g}}
\begin{pmatrix}
\Gamma_{1}-\Gamma_{3}        \\
\Gamma_{2}-\Gamma_{4}         \\
\Gamma_{5}-\Gamma_{6}	     	    	
\end{pmatrix}
\!+\!\begin{pmatrix}\!
\Gamma_{1}^{(0)}-\Gamma_{3}^{(0)}        \\
\Gamma_{2}^{(0)}-\Gamma_{4}^{(0)}         \\
\Gamma_{5}^{(0)}-\Gamma_{6}^{(0)}	     	\!     	        	\end{pmatrix}\, ,
\end{align}
\begin{align}\label{RPA_Pom2}
\begin{pmatrix}
\Gamma_{1}+\Gamma_{3}        \\
\Gamma_{2}+\Gamma_{4}         \\
\Gamma_{5}+\Gamma_{6}	     	
\end{pmatrix}
\!=\!
\bm{M}_{A_{1g}} \!
\begin{pmatrix}
\Gamma_{1}+\Gamma_{3}        \\
\Gamma_{2}+\Gamma_{4}         \\
\Gamma_{5}+\Gamma_{6}	     	
\end{pmatrix}
\!+\!\begin{pmatrix}\!
\Gamma_{1}^{(0)}+\Gamma_{3}^{(0)}        \\
\Gamma_{2}^{(0)}+\Gamma_{4}^{(0)}         \\
\Gamma_{5}^{(0)}+\Gamma_{6}^{(0)}	\!     	        	\end{pmatrix}\, ,
\end{align}
where
\begin{align}
\bm{M}_{B_{1g}}&\!=\!
\setlength{\arraycolsep}{1pt}
\begin{pmatrix}
-2u_{5} & -2\frac{\tilde{A}_{e}}{A_{e}}u_{5} & -2\frac{A_{h}}{A}\epsilon_1 \\
-2\frac{\tilde{A}_{e}}{\overline{A}_{e}}u_6 &-2u_6& -2\frac{A_{h}}{\tilde{A}}\epsilon_2   \\
-2\frac{A_{e}}{A}\epsilon_1-2\frac{\tilde{A}_{e}}{\tilde{A}}\epsilon_2  &
-2\frac{\tilde{A}_{e}}{A}\epsilon_1-2\frac{\bar{A}_{e}}{\tilde{A}}\epsilon_2 & -2u_4 \\
\end{pmatrix}
\label{M_B1}	
\end{align}	
and
\begin{align}
\bm{M}_{A_{1g}}\!\!=\!
\setlength{\arraycolsep}{1pt}
\begin{pmatrix}
-2u_{5} &-2\frac{\tilde{A}_{e}}{A_{e}}u_{5} & -8\frac{A_h}{A}u_1 \\
-2\frac{\tilde{A}_{e}}{\overline{A}_{e}}u_6 &-2u_6&-8\frac{A_h}{{\tilde A}}\tilde{u}_1 \\
-8(\frac{A_{e}}{A}u_1+\frac{\tilde{A}_{e}}{\tilde{A}}\tilde{u}_1) & -8(\frac{A_{e}}{A}u_1+\frac{\bar{A}_{e}}{\tilde{A}}\tilde{u}_1) &-2u_{4} \\
\end{pmatrix}\!.
\label{M_A1}
\end{align}
In Eqs.~\eqref{M_B1} and \eqref{M_A1}
we introduced
\begin{align}\label{e_12}
\epsilon_1&=u_2-\bar{u}_2-2(u_1-\bar{u}_1)\notag\\
\epsilon_2&=\tilde{u}_2-\tilde{\tilde{u}}_2-2(\tilde{u}_1-\tilde{\tilde{u}}_1)\, .
\end{align}
The ratios $A_e/A$, etc. are functions of $A_0$ and $m_h/m_e$.

 In what follows, we focus on the instability in the $B_{1g}$ channel, which gives rise to a true $C_4$ breaking orbital order.
 Solving the $3\times 3$ matrix equation for the three order parameters $\Gamma_{1,e}= \Gamma_1-\Gamma_3$, $\Gamma_{2,e} = \Gamma_2-\Gamma_4$, $\Gamma_{1,h} = -(\Gamma_5-\Gamma_6)$,
  we obtain
 \begin{align}\label{Gamma_eh}
\Gamma_{1,e}, \Gamma_{2,e}, \Gamma_{1,h} \propto 1/(1-\lambda u_1) \propto (L_{B_{1g}}-L)^{-1}\, ,
\end{align}
where $\lambda u_1$ is the largest eigenvalue of ${\bf M}_{B_{1g}}$ and
\begin{align}
L_{B_{1g}}&=L_0-\frac{\lambda}{1+\gamma_3^{2}/C^2}\, .
\end{align}
We verified that the largest eigenvalue of ${\bf M}_{B_{1g}}$ is positive. Then $L_{B_{1g}} <L_0$, i.e., the instability in the orbital channel
 occurs at a larger $T$ than the one in the superconducting channel.  On the other hand, the correction to $L_0$ is non-logarithmical, i.e., strictly speaking,
  the difference between $L_0$ and $L_{B_{1g}}$ is outside of the applicability of the RG analysis.  Still, however,
    the exponent for the $B_{1g}$ Pomeranchuk vertices is $\beta_{Pom} =1$.
 Evaluating then the susceptibility in the Pomeranchuk channel we obtain that, even if we neglect the difference between
 $L_0$ and $L_{B_{1g}}$, we have
\begin{align}\label{Pom_chi_B1}
\chi_{B_{1g}}\propto (L_0-L)^{-1},
\end{align}
i.e. the susceptibility  exponent  $\alpha_{Pom}=1$.

For completeness, we also considered A$_{1g}$, $A_{2g}$ and $B_{2g}$ Pomeranchuk channels.
The divergence of the Pomeranchuk susceptibility in the $A_{1g}$ channel gives rise to a shift of the chemical potential, with different magnitudes on hole and electron pockets~\cite{CKF2016}.  It does not, however,  give rise to a true symmetry breaking as $A_{1g}$ symmetry is the same as the symmetry of the  tetragonal phase. In practice it implies that the divergence of the $A_{1g}$ Pomeranchuk susceptibility is very likely cut by terms beyond RG.
 The order parameters $\Gamma_{ij}$  in  $A_{2g}$ and $B_{2g}$ channels are inter-pocket ones and do not break symmetry between x and y directions. We discuss these orders in
   Sec. \ref{sec:Pom_A2g} in the Appendix.

 \subsection{Comparative analysis of susceptibilities}
 \label{sec:resultsusc}
\begin{figure}[h]
	\centering{}\includegraphics[width=0.95\columnwidth]{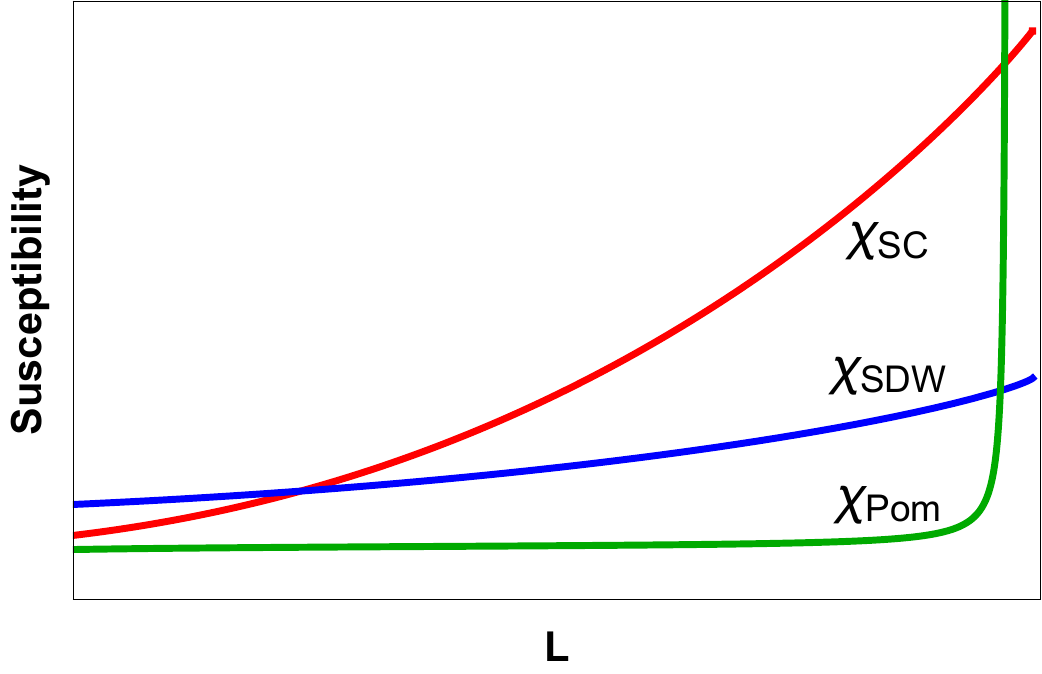} \protect\caption{The RG flow of susceptibilities as functions of the RG parameter L.
 The susceptibilities in the
 superconducting and Pomeranchuk channels diverges, while the one in the SDW channel initially increases but remains finite as $L$ approaches $L_0$,
 which is slightly to the right from the right boundary of the figure.
	\label{fig:susceptibility}}
\end{figure}

The explicit results for the RG flow of susceptibilities in SC, SDW, and d-wave 
($B_{1g}$) Pomeranchuk channels are presented in Fig.~\ref{fig:susceptibility}.
 We obtained this flow by selecting a particular set of initial conditions,
 for which the RG flow moves the system towards the weakly unstable fixed trajectory,
 solving for $u_i (L)$, and using these running couplings to obtain $L$ dependencies first
of the vertices and then of the susceptibilities.   We see the same behavior as we obtained by analyzing the fixed trajectories.  Namely, the susceptibility in the
Pomeranchuk channel becomes the largest at $L \approx L_0$.  The susceptibility in the SC channel increases but not as fast as the Pomeranchuk susceptibility, and the susceptibility in SDW channel does not diverge at $L=L_0$.

\begin{figure}[h]
\includegraphics[width=0.95\columnwidth]{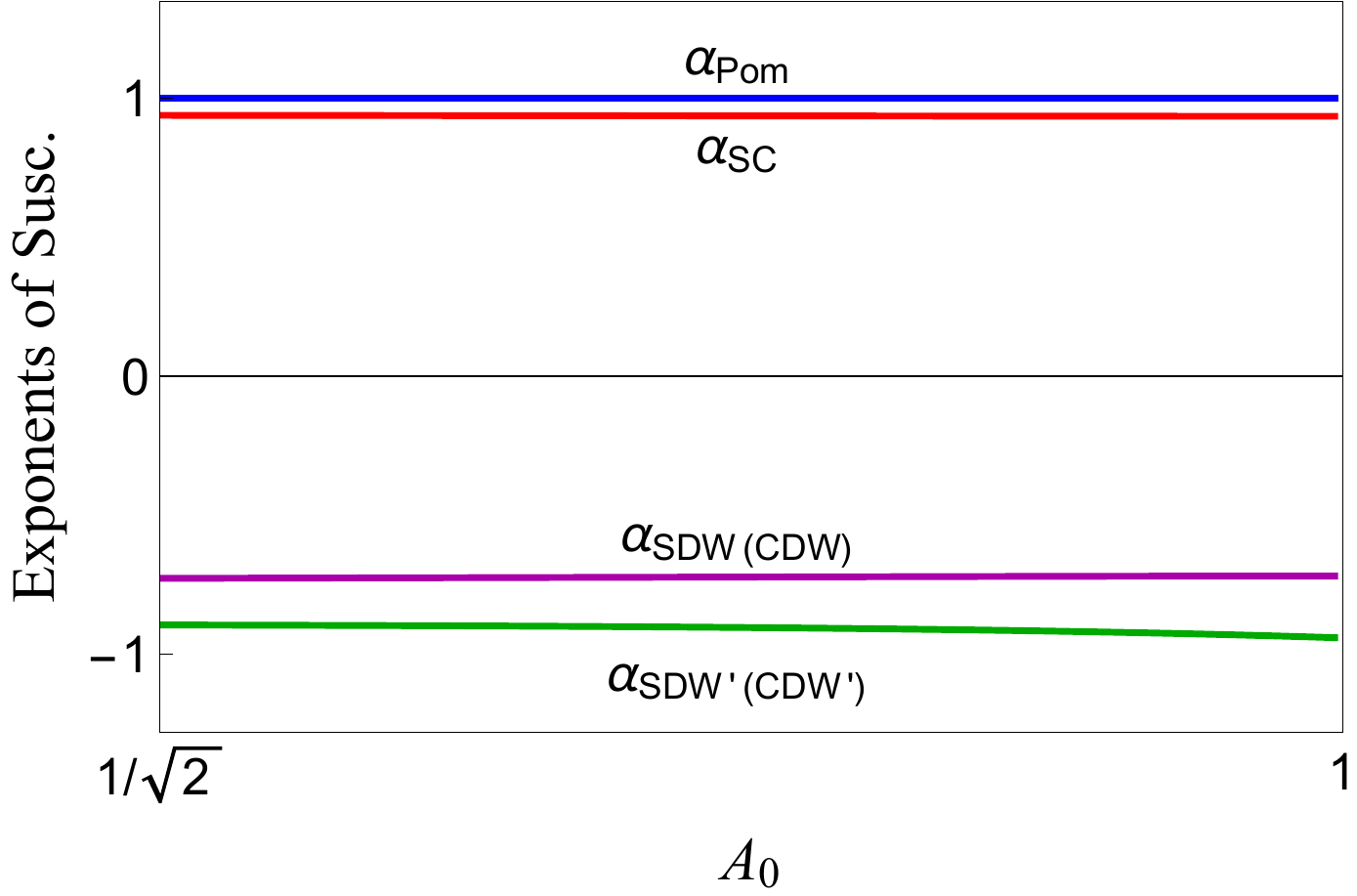}\\
\includegraphics[width=0.95\columnwidth]{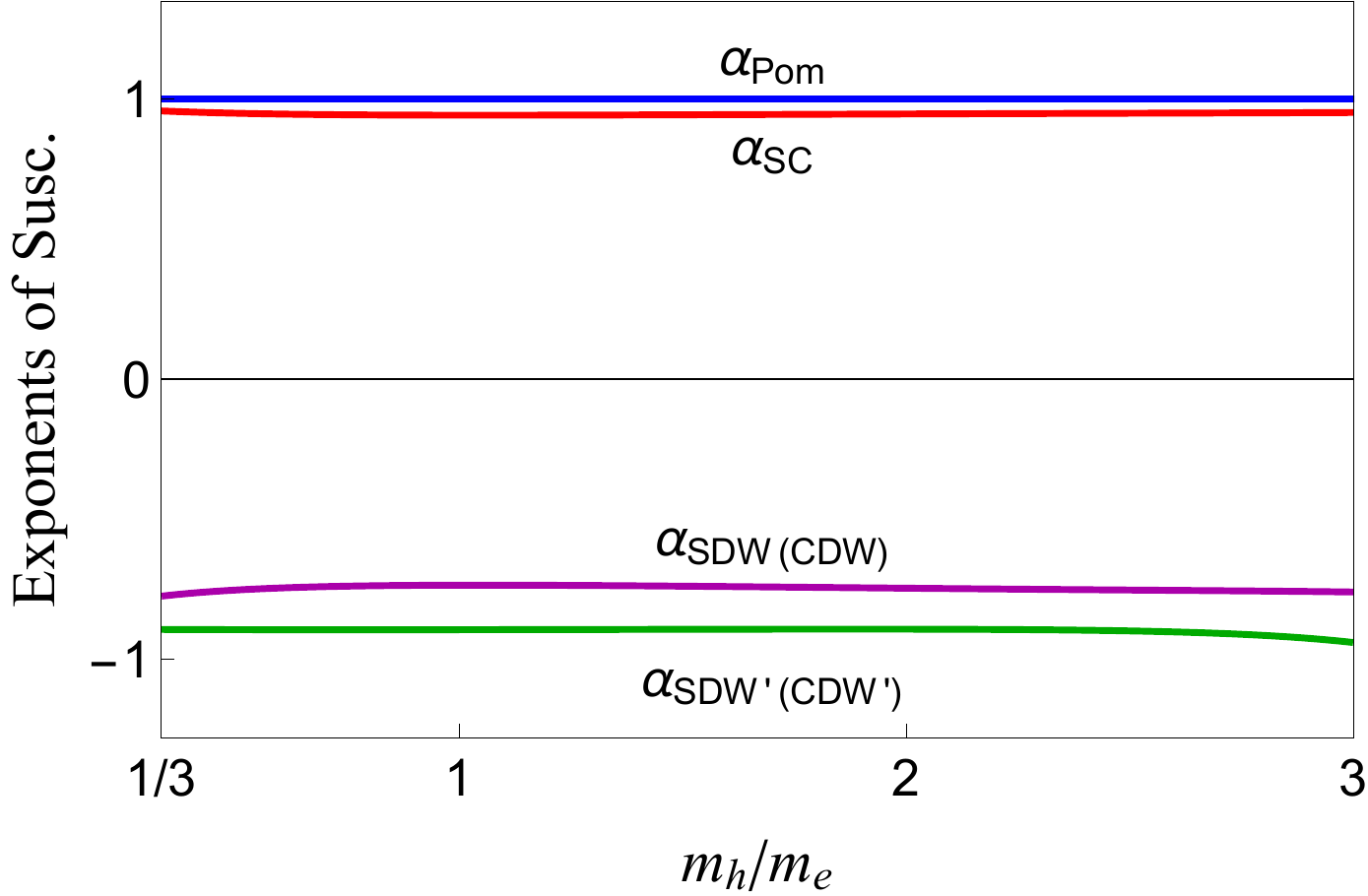}
\protect\caption{The exponents $\alpha_i$ for  susceptibilities $\chi_0 \propto 1/(L_0-L)^{\alpha_i}$  in SDW, SC, and d-wave Pomeranchuk channels for interactions on the weakly unstable fixed trajectory, Eq. \protect\eqref{FT_3}. Upper panel -- $\alpha_i$ as functions of $A_0$ at fixed $m_h/m_e = 1$.  Lower panel -- $\alpha_i$ as functions of $m_h/m_e$ at fixed $A_0 =0.8$. A larger exponent means a faster divergence of the susceptibility. We recall that $A_{0}^{2}$  determines a relative weight of $d_{xz}(d_{yz})$ and $d_{xy}$ orbitals along the electron pockets. }\label{fig:exponents_case3}
 \end{figure}

To see more accurately the scaling behavior of various susceptibilities
we compared the exponents $\alpha$ in $\chi (L) \propto 1/(L_0 - L)^\alpha$ in SDW, SC, and Pomeranchuk channels for $u_i$ on the weakly unstable fixed trajectory.
 We plot $\alpha_i$ in Fig. \ref{fig:exponents_case3} as function of $A_0$ at fixed $m_h/m_e$ and as function of $m_h/m_e$ at fixed $A_0$.
  The value of $\alpha_{Pom}=1$ is independent on the $m_h/m_e$ mass ratio and the parameter $A_0$. The values of $\alpha_{SDW} = \alpha_{CDW}$ and $\alpha_{SC}$ weakly
 depend on on $A_0$ (and on $m_h/m_e$). We see that $\alpha_{SC}$ is positive, but smaller than one, and $\alpha_{SDW} = \alpha_{CDW}$ is negative.
   This is fully consistent with Fig. \ref{fig:susceptibility}.  The conclusion from both figures is then that,  upon increasing $L$ (i.e., lowering the temperature),
 the first instability occurs in the  $B_{1g}$ Pomeranchuk channel and leads to a spontaneous orbital order which breaks $C_4$ rotational symmetry.  Superconducting order develops at a lower temperature (which will be further reduced due to a negative feedback from the orbital order), and SDW and CDW orders do not develop down to $T=0$.

We caution that this result  only applies to systems for which $L_0 \leq L_F = \log{W/E_F}$. When $L_F < L_0$, the one-loop parquet RG runs up to $L = L_F$,
  and the system generally develops an instability in the channel in which the susceptibility is the largest at $L = L_F$ (see Refs.~\cite{Chubukov_rev,CKF2016, kh_ch_last}).

\section{The structure of superconducting and orbital order parameters and implications for the experiments}
 \label{sec:implication_exp}

The susceptibility analysis reveals  instabilities towards superconducting and orbital order. In this section we determine the structure of the corresponding order parameters and discuss the implications of the orbital order for the band structure.

The magnitudes of different order parameters at $T \to 0$ can only be obtained by solving the full set of non-linear gap equations, which include a non-linear coupling between orbital and superconducting orders. This accounts for the fact that, once orbital order develops first, it tends to suppress the onset of superconducting order.
   This analysis is beyond the scope of our RG study, in which we approach the instabilities from the disordered state at higher $T$.
   Nevertheless, the RG analysis allows one to detect the symmetry of superconducting and orbital orders, and also find the ratios between  different components of superconducting and orbital order parameters near their onsets, i.e.~between superconducting gaps on hole and electron pockets and between various $xy$ and $xz/yz$ components of the orbital order parameter.  We assume that the vertices $\Delta_{i}$ for superconductivity and $\Gamma_i$ for orbital order ($i =1-6$) in the equations Eqs.~(\ref{SC_matrix}), (\ref{RPA_Pom1}), and (\ref{RPA_Pom2}), are proportional to the corresponding condensates in the ordered phases.
   Furthermore we assume that, with one exception, which we discuss below, the ratios between the components  of $\Delta_i$ and $\Gamma_i$  do not change between the onset of the order and  lower $T$, at which they are measured by ARPES and other techniques.

   We  consider the superconducting channel first and then analyze the orbital channel.

\subsection{Superconducting order parameter}

\begin{figure}[h]
	\centering{}\includegraphics[width=1.0\columnwidth]{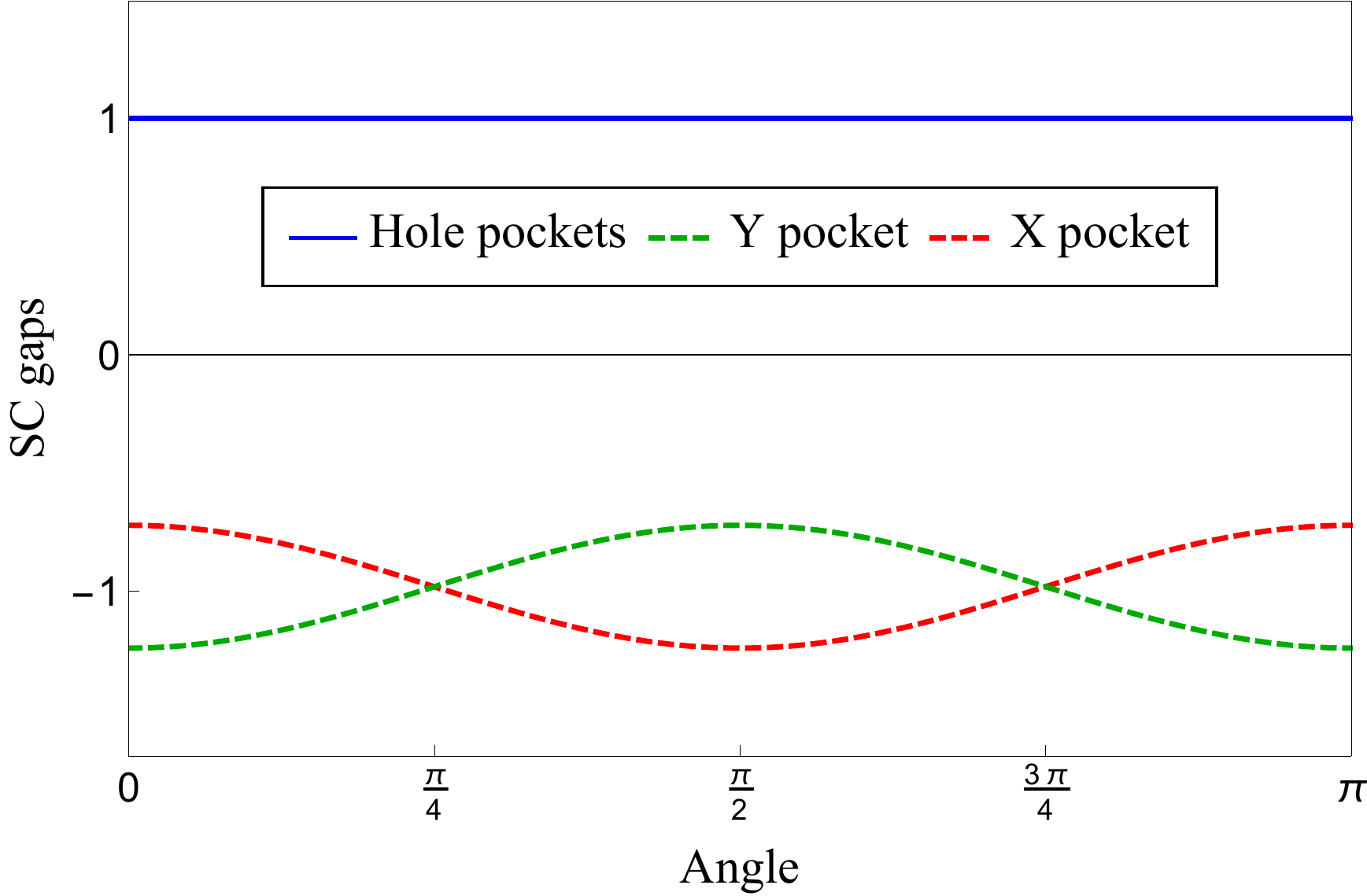} \protect\caption{Superconducting  gaps along the Fermi surfaces for the interactions
  on the weakly unstable fixed trajectory.
   The solid blue line is the gap $\Delta_h$ on  the two  hole pockets, the dashed lines are the gaps on the  electron pockets --  the green one is the gap on the  Y pocket and red one is on the X pocket. The angle is counted  anti-clockwise from $k_x$ direction.  We set $m_h/m_e=1$, $A_0=0.8$.}
		\label{fig:SCangular_dependence_case3}
\end{figure}

   The vertices $\Delta_i$ represent fermionic bilinears in the particle-particle channel  in the orbital basis. To obtain the superconducting order parameters on different pockets,
     we need to convert these $\Delta_i$ into band basis using the orbital-to-band transformation from Eq.~(\ref{band_basis}). For the SC order parameter on hole pockets
      we obtain
\begin{align}
\langle c_{k}c_{-k} \rangle &=\langle \cos^{2}\theta_{h}\psi_5(k)\psi_5(-k)+\sin^{2}\theta_{h}\psi_6(k)\psi_6(-k)\notag\\
&+\frac{1}{2}\sin2\theta_{h}(\psi_5(k)\psi_6(-k)+\psi_6(k)\psi_5(-k))\rangle\notag\\
&= \cos^{2}\theta_{h}\Delta_{5}+\sin^{2}\theta_{h}\Delta_{6} + \frac{1}{2}\sin2\theta_{h} \left(\Delta_{5,6} + \Delta_{6,5}\right) \notag\\
&=\Delta_{5} \equiv \Delta_h\label{SCgap_hole}
\end{align}
and
\begin{align}
\langle d_{k}d_{-k} \rangle &=\langle \sin^{2}\theta_{h}\psi_5(k)\psi_5(-k)+\cos^{2}\theta_{h}\psi_6(k)\psi_6(-k)\notag\\
&+\frac{1}{2}\sin2\theta_{h}(\psi_5(k)\psi_6(-k)+\psi_6(k)\psi_5(-k))\rangle\notag\\
&=\sin^{2}\theta_{h}\Delta_{5}+\cos^{2}\theta_{h}\Delta_{6} + \frac{1}{2}\sin2\theta_{h} \left(\Delta_{5,6} + \Delta_{6,5}\right) \notag\\
&=\Delta_{5} \equiv \Delta_h\label{SCgap_hole_1}
\end{align}
The $A_{1g}$ SC order parameter on the electron pocket near $Y$ is
\begin{align}
\langle f_{1,k} & f_{1,-k} \rangle =\langle \cos^{2}\phi_{e,k}\psi_1(k)\psi_1(-k)+\sin^{2}\phi_{e,k}\psi_2(k)\psi_2(-k)\notag\\
&+\frac{1}{2}\sin2\phi_{e,k}(\psi_1(k)\psi_2(-k)+\psi_2(k)\psi_1(-k))\rangle\notag\\
& = \cos^{2}\phi_{e,k}\Delta_{1}+\sin^{2}\phi_{e,k}\Delta_{2}\notag\\
&=A_{0}^{2}\cos^{2}\theta_{e}\Delta_{1}+(1-A_{0}^{2}\cos^{2}\theta_{e})\Delta_{2}\notag\\
&=(\frac{A_{0}^{2}}{2}\Delta_{1}+(1-\frac{A_{0}^{2}}{2})\Delta_{2})+A_{0}^{2}\frac{\Delta_{1}-\Delta_{2}}{2}\cos2\theta_{e}\notag\\
&= \Delta_{a,e} +\Delta_{b,e}  \cos2\theta_{e},\label{SCgap_Y}
\end{align}
where $\Delta_{a,e}=\Delta_1A_0^2/2+\Delta_{2}(1-A_0^2/2)$ and $\Delta_{b,e}=(\Delta_{1}-\Delta_{2})A_0^2/2$.
 The order parameter on the electron pocket near $X$ is obtained from (\ref{SCgap_Y}) by $\pi/2$ rotation:
\begin{align}
\langle f_{2,k}f_{2,-k} \rangle = \Delta_{a,e} -\Delta_{b,e}  \cos2\theta_{e},
\label{SCgap_X}
\end{align}
The ratios of $\Delta_h$, $\Delta_{a,e}$, and $\Delta_{b,e}$ are determined by extracting the components $\Delta_1= \Delta_3, \Delta_2=\Delta_4$, and $\Delta_5=\Delta_6$ from the matrix equation \eqref{SC_matrix}, i.e. from the solution which corresponds to the largest eigenvalue of this matrix.
 We show the results for SC gaps on hole and electron pockets on the  weakly unstable fixed trajectory in Fig.~\ref{fig:SCangular_dependence_case3}.
 We see that  all three components $\Delta_h, \Delta_{a,e}$, and $\Delta_{b,e}$ are non-zero and of the same order.
  The  two angle-independent components $\Delta_{h}$ and $\Delta_{a,e}$  have opposite signs, i.e., the $A_{1g}$ order parameter has $s^{+-}$  structure, as expected.
   We also see that
  $\Delta_{a,e} > \Delta_{b,e}$, i.e.,  there are no accidental nodes on the electron pockets.  We note by passing that on  the stable fixed trajectories the angular dependence of the gaps on the electron pockets becomes more pronounced.

\subsection{Orbital order parameter}

Long-range orbital order in our RG analysis emerges as a $d-$wave Pomeranchuk order.
 Such an order leads to unequal occupations of $d_{xz}$ and $d_{yz}$ orbital states near hole and electron pockets, and also to unequal occupations
  of $d_{xy}$ orbital states near $X$ and $Y$ electron pockets.  The three $B_{1g}$ order parameters in the orbital basis are
 $\Gamma_{1,e} = n_{xz}^Y - n_{yz}^X$, $\Gamma_{1,h} =  n_{xz}^\Gamma -  n_{yz}^\Gamma$, and $\Gamma_{2,e} = n_{xy}^Y - n_{xy}^X$.  In our notations,
 $\Gamma_{1,e} = \Gamma_1 - \Gamma_3 = 2\Gamma_1,  \Gamma_{2,e} = \Gamma_2-\Gamma_4 = 2\Gamma_2$, and $\Gamma_{1,h} = \Gamma_6 - \Gamma_5 =2\Gamma_6$.
 Transforming from orbital to band basis, we obtain for electron and hole densities
 \begin{figure}[h]
 	\centering{}\includegraphics[width=0.99\columnwidth]{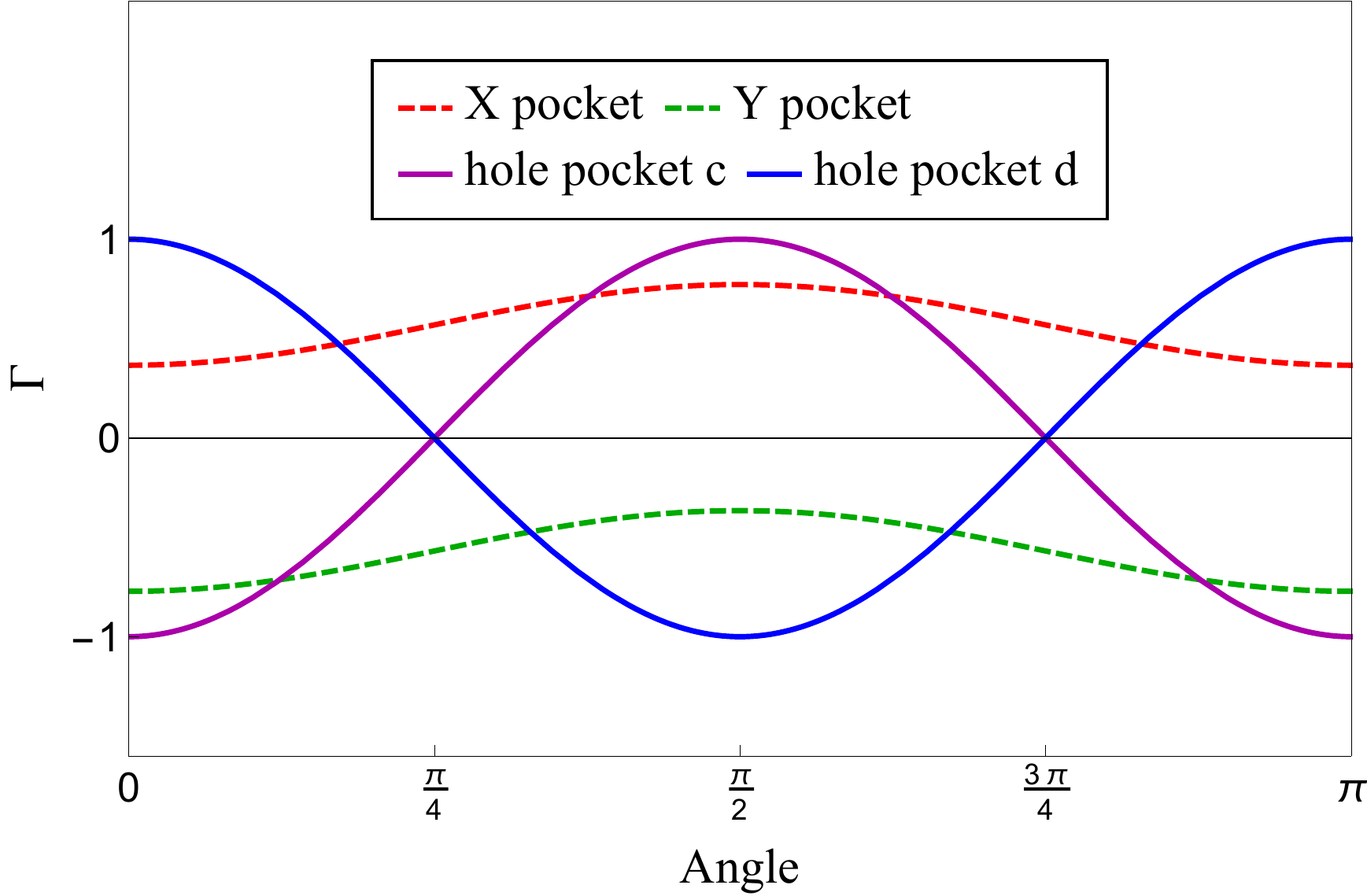} \caption{$d-$wave Pomeranchuk order parameters (Eq. \protect \ref{mon_1})
  for interactions on the weakly unstable fixed trajectory.  The order parameters on the hole pockets are shown by  solid lines, and the ones on the  electron pockets by
  dashed lines.  The $\cos{2\theta}$ form of order parameters on the hole pockets deform $C_4$-symmetric hole pockets into ellipses, with long axis along
   orthogonal directions on the two pockets.  Almost constant order parameters of opposite sign on the two electron pockets make one pocket larger and the other smaller in the nematic phase (see Fig.~\protect\ref{fig:Pomeranchuk_sketch}). The angle $\theta$ is counted anti-clockwise from $k_x$ direction. We use $m_h/m_e=1$, $A_0=0.8$ to determine the
    order parameters on the electron pockets.  The overall magnitude of the order parameters on the hole pockets was adjusted to be comparable to that on the electron pockets.}
 		\label{fig:Pomangular_dependence_case3}
 \end{figure}

{\allowdisplaybreaks
	\allowdisplaybreaks
\begin{align}
\langle f_{1,k}^{\dag}f_{1,k} \rangle &\propto(\frac{A_{0}^{2}}{2}\Gamma_{1}+(1-\frac{A_{0}^{2}}{2})\Gamma_{2})+A_{0}^{2}\frac{\Gamma_{1}-\Gamma_{2}}{2}\cos2\theta_{e}\notag\\
&\equiv \Gamma_{a,e}+ \Gamma_{b,e}  \cos2\theta_{e}\notag\\
\langle f_{2,k}^{dag}f_{2,k} \rangle &\propto(\frac{A_{0}^{2}}{2}\Gamma_{3}+(1-\frac{A_{0}^{2}}{2})\Gamma_{4})+(A_{0}^{2}\frac{\Gamma_{3}-\Gamma_{4}}{2})cos2\theta_{e}\notag\\
&=-(\frac{A_{0}^{2}}{2}\Gamma_{1}+(1-\frac{A_{0}^{2}}{2})\Gamma_{2})+A_{0}^{2}\frac{\Gamma_{1}-\Gamma_{2}}{2}\cos2\theta_{e}\notag\\
&= -\Gamma_{a,e}+ \Gamma_{b,e} \cos2\theta_{e}\notag\\
\langle c_{k}^{\dag}c_{k} \rangle &\propto \cos^{2}\theta_{h}\Gamma_{5}+\sin^{2}\theta_{h}\Gamma_{6}
=-\cos2\theta_{h}\Gamma_{6}\notag\\
\langle d_{k}^{\dag}d_{k} \rangle &\propto \cos^{2}\theta_{h}\Gamma_{6}+\sin^{2}\theta_{h}\Gamma_{5}
=\cos2\theta_{h}\Gamma_{6},
\label{mon_1}
\end{align}}
where $\Gamma_{a,e}=\Gamma_1 A_0^2/2+\Gamma_{2}(1-A_0^2/2)$ and $\Gamma_{b,e}=(\Gamma_{1}-\Gamma_{2})A_0^2/2$.
We show the order parameters on hole and electron pockets in Fig.~\ref{fig:Pomangular_dependence_case3}.
Like in the superconducting case, we extract the relations between $\Gamma_1$, $\Gamma_2$, and $\Gamma_6$ from the matrix equation \eqref{M_B1}, i.e.~from the solution which corresponds to the largest eigenvalue.
\begin{figure}[h]
	\centering{}\includegraphics[width=0.95\columnwidth]{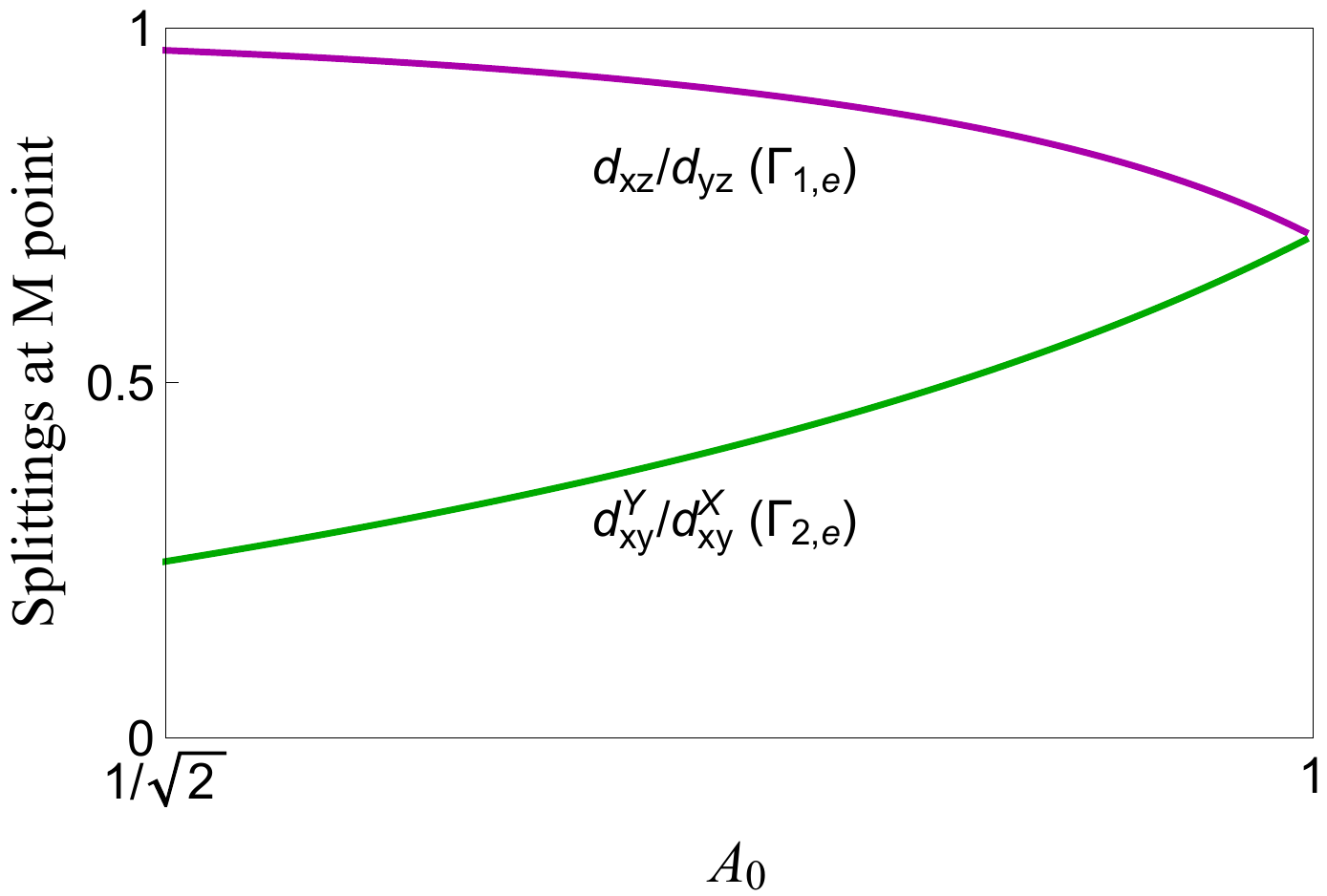} \caption{The orbital order parameters on the electron pockets
$\Gamma_{1,e} = n_{xz}^Y - n_{yz}^X = 2\Gamma_1$ and $\Gamma_{2,e} = n_{xy}^Y - n_{xy}^X =2\Gamma_2$ as functions of $A_0$ at fixed $m_h/m_e =1$
 for interactions on the weakly unstable fixed trajectory.  Each order parameter determines the splitting of the corresponding bands at $M$ point in the 2Fe BZ.}
		\label{fig:gamma1_gamma2}
\end{figure}

Exactly on the weakly unstable fixed trajectory the $3 \times 3$ matrix $M_{B_{1g}}$  decouples into the $2 \times 2$  set for $\Gamma_1$ and $\Gamma_2$ and a single equation
 for $\Gamma_6$, because $\epsilon_1$ and $\epsilon_2$ in  \eqref{M_B1} vanish.  From the $2 \times 2$ set one can obtain the ratio $\Gamma_1/\Gamma_2 = \Gamma_{1,e}/\Gamma_{2,e}$.
  The ratio
  does depend on $A_0$, but is generally close to one
  (see Fig.~\ref{fig:gamma1_gamma2}).
  This result implies that the $d-$wave order parameters made out of  $d_{xz}/d_{yz}$ orbitals and $d_{xy}$ orbitals at X and at Y pockets are nearly equal.  This is very different from the behavior on the two stable fixed trajectories, where either $\Gamma_1$ or $\Gamma_2$ vanishes.

To obtain the ratios of the order parameters on hole and on electron pockets, e.g. $\Gamma_6/\Gamma_1 = \Gamma_{1,h}/\Gamma_{1,e}$, one needs to include the fact that in reality the
 system approaches a fixed trajectory in the process of the RG flow, but is never strictly on the fixed trajectory, i.e., $\epsilon_1$ and $\epsilon_2$ are small but non-zero.
   Analyzing the RG flow towards the weakly unstable fixed trajectory, we find that $\Gamma_{1,h}/\Gamma_{1,e}$ is {\it negative} and its magnitude is large.

\subsection{Implications for experiments}

We now compare our theoretical results with the experiments on FeSe, where at ambient pressure a nematic order has been observed below $85K$, and superconductivity has been observed below $8K$.
 This sequence of transitions is consistent with the outcome of our RG analysis.  We identify the nematic order with a spontaneous $d-$wave orbital order.

\begin{figure}[h]
   	\centering{}\includegraphics[width=0.98\columnwidth]{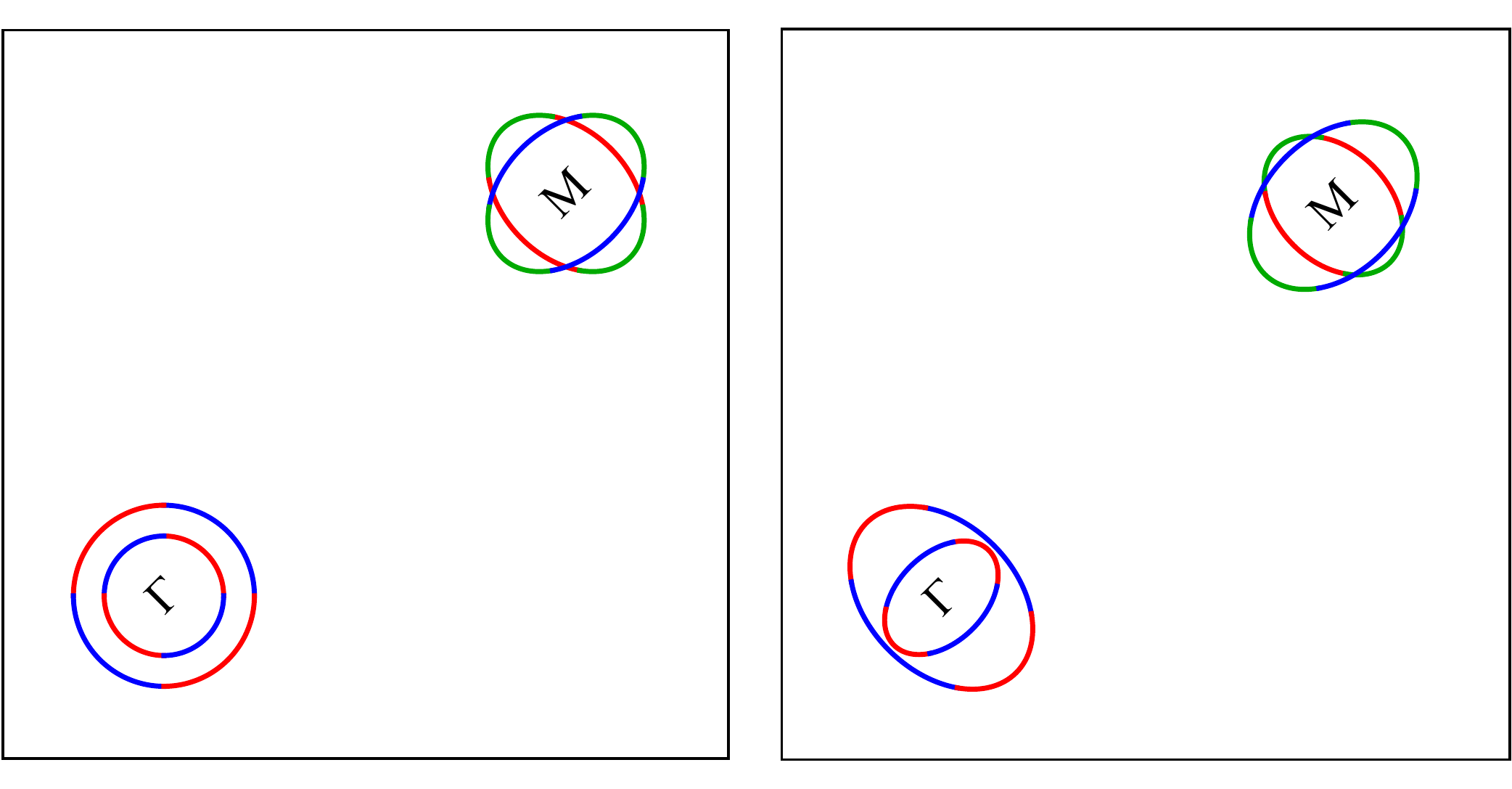} \caption{Left panel -- Fermi surfaces in 2Fe Brillouin zone above the nematic transition.
    Each of the two hole pockets is $C_4$ symmetric.  The two electron pockets are centered at $M = (\pi,\pi)$ and form an inner and outer pockets.
     The inner pocket predominantly consists of $d_{xz}$ and $d_{yz}$ orbital states, the outer pocket is predominantly made out out of $d_{xy}$ orbital states.
    These pockets
      touch each other along ${\tilde k}_x = \pi$  and ${\tilde k}_y = \pi$ directions (${\tilde {\bf k}}$ is the momentum in 2Fe BZ).
       Within our model, the location of the pockets in 2Fe BZ and their dispersion can be obtained by a simple folding, i.e, by changing momentum components $k_x$ and $k_y$   in the 1Fe BZ to ${\tilde k}_x = k_x + k_y$ and ${\tilde k}_y = k_y-k_x$.  Spin-orbit interaction, however,  splits the inner and the outer pockets.
      Right panel -- the structure of hole and electron pockets in the nematic phase in the 2Fe BZ.}
   		\label{fig:M_pocket}
   \end{figure}

 The most generic feature of $d-$wave orbital order is the elongation of the pockets due to breaking of $C_4$ lattice rotational symmetry down to $C_2$.
 This effect is particularly pronounced for the two hole pockets, which in the absence of orbital order are $C_4$-symmetric.  Below the nematic transition, the pockets become elongated. In the 2Fe Brillouin zone, where ARPES experiments are performed, one pocket becomes elongated along one BZ diagonal and the other along the other zone diagonal (see Fig. \ref{fig:M_pocket}).   Such an elongation has been observed in ARPES experiments on FeSe by several groups~\cite{borisenko,zhang2015,coldea}.
  In addition, there is an elongation of electron pockets as well. In the 2Fe BZ  the X and the Y electron pockets are centered at the same M point (Fig.~\ref{fig:M_pocket}). The two form an inner and outer pocket that touch each other in the absence of spin-orbit coupling, but split in the presence of such a coupling. The inner pocket predominantly consists of $d_{xz}$ and $d_{yz}$ orbital states, the outer pocket is predominantly made out out of $d_{xy}$ orbital states.  Above the nematic transition
   both inner and outer pockets are $C_4$-symmetric, but in the presence of orbital order  each pocket is elongated along the diagonal directions
   (Fig.~\ref{fig:M_pocket}).

\begin{figure}[h]
	\centering{}\includegraphics[width=1.0\columnwidth]{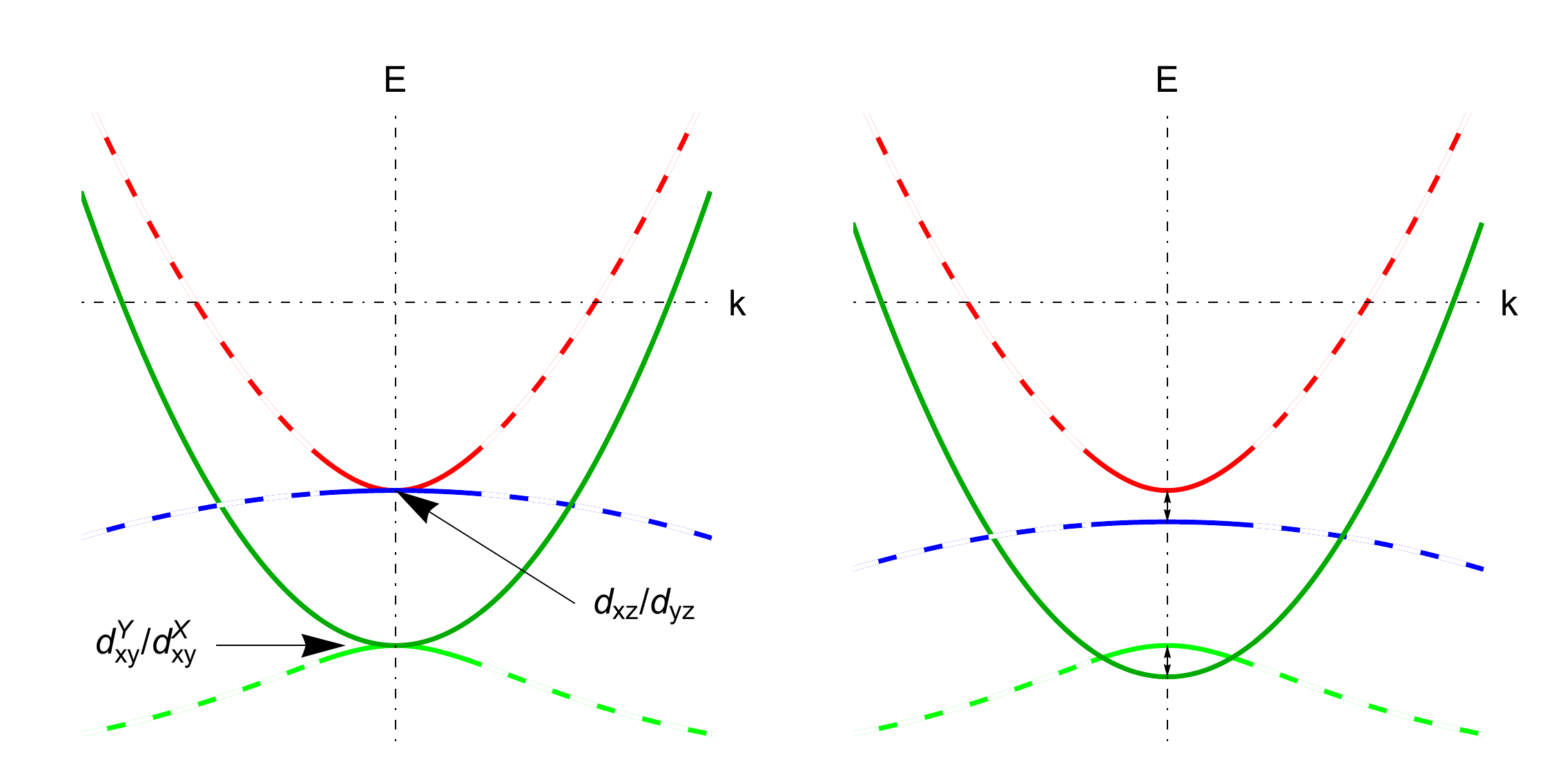} \caption{The splittings in the band dispersions near the M point in the 2Fe BZ for interaction on the weakly unstable fixed trajectory. The M point is taken as the origin of the coordinates and the cut is along $M-\Gamma$ ($k_{x}=k_{y} =k$). Left panel -- above the nematic transition. Right panel -- in the nematic phase.  Solid and dashed lines describe excitations with near-pure and mixed orbital content, respectively.}
		\label{fig:M_splitting}
\end{figure}

    The orbital order also affects the states away from the Fermi surface, in particular the hierarchy of electronic states at high-symmetry $\Gamma$ and $M$ points in the 2Fe BZ.
  In the absence of orbital order, the states at $M$ are doubly degenerate even in the presence of spin-orbit interaction~\cite{fern_vaf}
   (left panel in Fig.~\ref{fig:M_splitting}).
   One degeneracy is between $d_{xz}$ and $d_{yz}$ states, another is between two $d_{xy}$ states. In the 1Fe Brillouin zone one of the states in each subset comes from the pocket at X, another from the pocket at Y.  In the presence of orbital order, these degenerate states split. The splitting of $d_{xz}/d_{yz}$ states is $2 \Gamma_{1,e} (=4\Gamma_1)$, the splitting of $d_{xy}$ states is $2\Gamma_{2,e} (=4\Gamma_2)$. Assuming that one can extend the results of the RG analysis to the high-symmetry points, one can compare the ratios of the two splittings between theory and experiment. In the RG analysis, either $\Gamma_{1,e}$ or $\Gamma_{2,e}$ vanish
    on the stable fixed trajectories, but the ratio of the two is close to one on the weakly unstable fixed trajectory (see Fig.~\ref{fig:gamma1_gamma2}).

\begin{figure}[h]
	\centering{}\includegraphics[width=1.0\columnwidth]{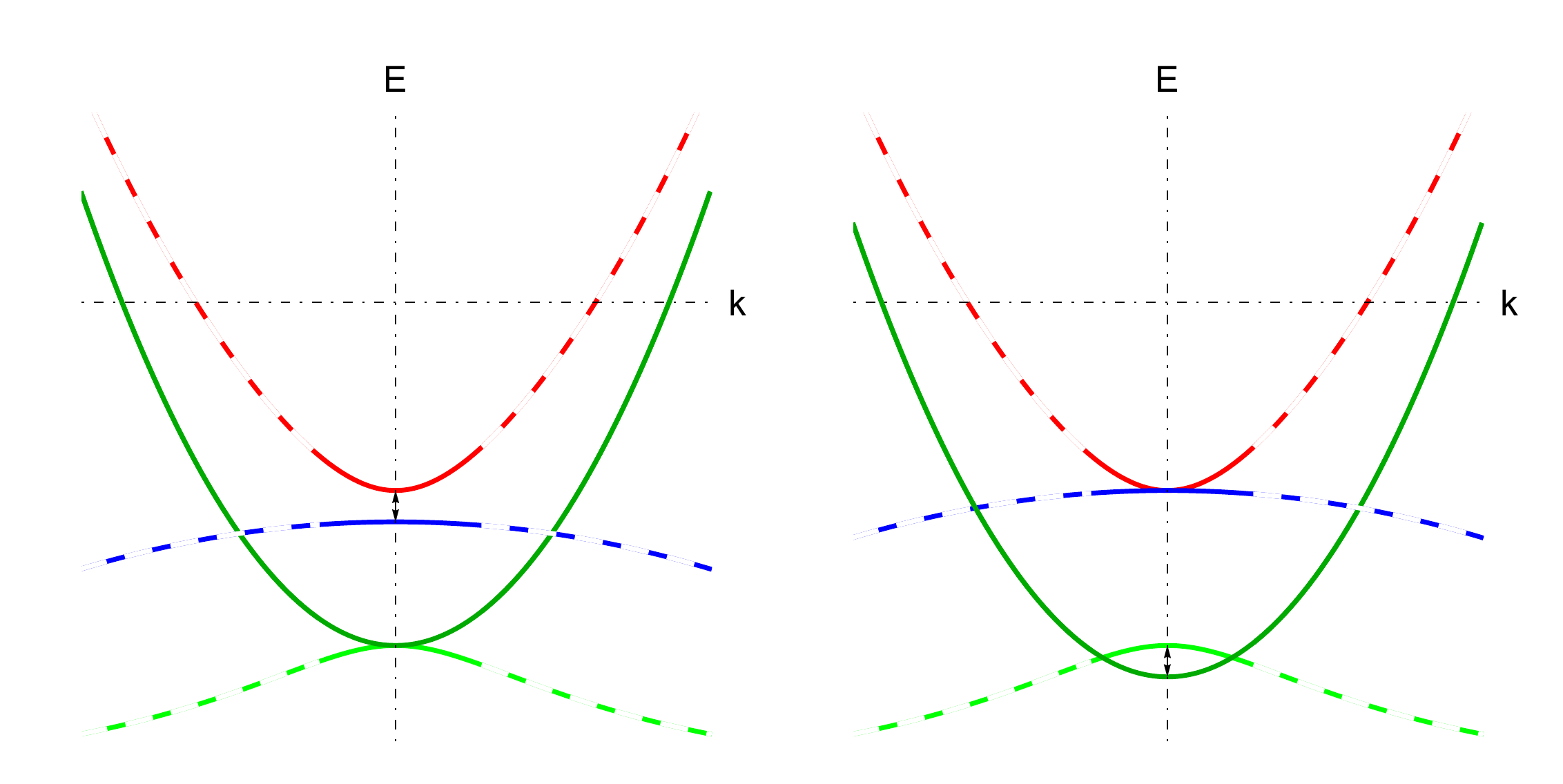} \caption{The same as 
in Fig. \protect \ref{fig:M_splitting} but for
 interactions on one of the two  stable fixed trajectories.  In this situation either the splitting between $d_{xz}/d_{yz}$ bands or the splitting between the two $d_{xy}$ bands vanishes.}
		\label{fig:M_splitting2}
\end{figure}

  ARPES data for $\Gamma_{1,e}/\Gamma_{2,e}$ from different groups~\cite{borisenko,coldea,brouer,zhang2015} are similar but not identical. We will use recent ARPES data from Ref.~\cite{borisenko} for comparison.  These authors have found that
 the magnitudes of the splittings within $d_{xz}/d_{yz}$ and $d_{xy}$ subsets are close to each other -- each is about 15 meV.
    In our notations, this implies that $\Gamma_{1,e} \approx \Gamma_{2,e} \approx 7.5$ meV.  Such near-equal splitting is not reproduced on the two stable fixed trajectories, but it is
     well reproduced on the weakly unstable fixed trajectory. We illustrate this in
  Figs.~\ref{fig:M_splitting} and ~\ref{fig:M_splitting2}.
   We argue, based on this comparison, that the RG analysis does agree with the ARPES data on the electron pockets, if, indeed,
  the  parameters for
  FeSe are such that the system is in the basin of attraction of the weakly unstable fixed trajectory.

 The comparison with orbital order on the hole pockets requires more care. On one hand, Suzuki et al reported~\cite{suzuki}, based on their ARPES data, that the
  signs of the $d_{xz}/d_{yz}$ order parameters on hole and electron pockets are opposite. This is consistent with the RG result that on the weakly unstable fixed trajectory, as we found that
  $\Gamma_{1,h}$ and $\Gamma_{1,e}$ have different signs [the same sign difference between $\Gamma_{1,h}$ and $\Gamma_{1,e}$  holds on the two stable fixed trajectories~\cite{CKF2016}].  On the other hand, our RG analysis yields a larger magnitude of $\Gamma_{1,h}$ than that of $\Gamma_{1,e}$, and, hence, a larger splitting at the $\Gamma$ point than that at the $M$ point. The authors of Ref. \cite{borisenko}, meanwhile, argued that the splitting at $\Gamma$ is comparable to that at $M$.
   However, when comparing our RG result for $\Gamma_{1,h}$
   with the measured splitting at $\Gamma$, one has to bear in mind that our $\Gamma_{1,h}$ was obtained without spin-orbit coupling.
    Meanwhile,  Ref.~\cite{borisenko} found
    that  the splitting at $\Gamma$ largely survives above the nematic transition and hence is predominantly due to spin-orbit coupling, which is known to split the bands at $\Gamma$ already in the absence of an orbital order~\cite{fern_vaf}.  The $\Gamma_{1,h} \sim 15$ meV was extracted from the ARPES data in  Ref. \cite{borisenko}  by detecting an additional splitting in the nematic phase  at low $T$.  Because of this,
        a meaningful comparison of the magnitude of $\Gamma_{1,h}$ between experiment and theory is only possible  after the inclusion of spin-orbit interaction into the theoretical analysis.


\section{Conclusion}
\label{sec:concl}

In this communication we reported the results of the parquet RG study of competing instabilities in the full 2D four pocket, three orbital low-energy model
 for FeSCs.
  Our four-pocket model consists of two $\Gamma$- centered hole pockets, made out of $d_{xz}$ and $d_{yz}$ orbitals,
  and two symmetry-related electron pockets centered at $X = (\pi,0)$ and $Y= (0,\pi)$ points in the 1Fe BZ
  and  made out of $d_{yz}/d_{xy}$  and $d_{xz}/d_{xy}$ orbitals, respectively.
   We derived and analyzed the RG flow of 30 couplings, which describe all symmetry-allowed interactions between low-energy fermions. Despite that the number of couplings is large, we argued that there are only two stable fixed trajectories of the RG flow
     and one weakly unstable trajectory with a single unstable direction.
      On  one stable trajectory the interactions involving $d_{xz}/d_{yz}$ orbital components on electron pockets  vanish relative to interactions involving $d_{xy}$ components, on the other interactions involving $d_{xy}$ orbital components vanish relative to $d_{xz}/d_{yz}$ components.  On the weakly unstable  trajectory, interactions involving $d_{xz}/d_{yz}$ and $d_{xy}$ orbital states on electron pockets remain comparable. The behavior along the two stable fixed trajectories has been analyzed in Ref. \cite{CKF2016}.
       In this work we  analyzed the system behavior along the weakly unstable trajectory.
     We argued, based on the analysis of susceptibilities along this trajectory,
    that the leading instability upon lowering the temperature is towards a {\it three-component} d-wave orbital nematic order. Two orbital components are the differences between fermionic densities
       on $d_{xz}$ and $d_{yz}$ orbitals on hole pockets and on electron pockets, $\Gamma_{1,h} = n^\Gamma_{xz} - n^\Gamma_{yz}$, $\Gamma_{1,e} = n^Y_{xz} - n^X_{yz}$, the third one is
   the difference between the densities of $d_{xy}$ orbitals on $X$ and $Y$ pockets, $\Gamma_{2,e} = n^Y_{xy} - n^X_{xy}$.  In our RG analysis, the magnitudes
    of $\Gamma_{1,e}$ and $\Gamma_{2,e}$ turn out to be nearly equal, and the sign of $\Gamma_{1,h}$ is opposite to that of $\Gamma_{1,e}$.  We applied the results to FeSe and found both qualitative and quantitative agreement with ARPES data~\cite{borisenko,coldea,zhang2015,suzuki}, specifically on the ratio of $\Gamma_{1,e}/\Gamma_{2,e}$.     We argue, based on this agreement and the fact that Fermi surfaces  in FeSe are all small,
     that the nematicity, observed in FeSe below $85K$ is likely the result of a spontaneous
      orbital order, which is captured by RG.   The situation in other Fe-pnictides, where either hole or/and electron pockets are larger, is different, and there the nematic order is likely due to softening of composite spin fluctuations. This last scenario gives rise to a two-step magnetic transition into the stripe SDW state, with an intermediate Ising-nematic phase, in which $C_4$ symmetry is broken, but $O(3)$ spin-rotation symmetry remains intact.

   \section{Acknowledgments}

We thank L. Basones, L. Benfatto, G. Blumberg, A. Boehmer, S. Borisenko,
A. Coldea, L. Fanfarino, R. Fernandes, I. Fisher, P. Hirschfeld, C. Honerkamp, I.
Eremin, S. Kivelson, H. Kontani, D-H Lee, I. Mazin, C. Meingast, J.
Schmalian, R. Thomale, V. K. Thorsm{\o}lle, O. Vafek, R. Valenti, A.
Yaresko, Fa Wang, and Y. Wang for useful discussions. R.X. and A.V.C.  are
supported by the Office of Basic Energy Sciences, U.S. Department
of Energy, under award DE-SC0014402. L.C. thanks the School of Physics and Astronomy of the University of Minnesota for hospitality during this work and acknowledges funding by the Studienstiftung des deutschen Volkes and the HGSFP at Heidelberg University.
MK is supported by the Israel Science Foundation, Grant No. 1287/15
and NSF DMR-1506668.

\section{Appendix}

\subsection{Details of RG  analysis on fixed trajectories}\label{sec:fixed_point_solution_detail}

As we wrote in Sec.~\ref{sec:traj}, the solution of the parquet RG equations leads to a divergence of various couplings, which occurs in a universal way in the sense that the coupling ratios tend to constants. These constants characterize the different solutions - the fixed trajectories - of the flow. In the following we present the detailed solution of the parquet RG equations and the stability analysis of the resulting fixed trajectories.

\subsubsection{Stable fixed trajectories}

For the first fixed trajectory we rewrite all couplings in terms of the ratios $\gamma_i,\tilde\gamma_i$ as  $u_{i}=\gamma_{i}u_{1}$, $\tilde{u}_{i}=\tilde{\gamma}_{i}u_{1}$. This leads to flow equations for the ratios 
 $u_1\frac{d\gamma_i}{dL}=\frac{d}{dL}u_i-\gamma_i\frac{d}{dL}u_1$ and analogously for $\tilde\gamma_i$. A fixed trajectory is set by conditions $\frac{d\gamma_i}{dL}=0$.  In our case
\begin{align}
\dot{u}_{1}&={u}_{1}^{2}+{u}_{3}^{2}/C^{2}\notag\\
{\tilde{\gamma}}_{1}(1+{\gamma}_{3}^{2}/C^{2}) & =\tilde{\gamma}_{1}^{2}+\tilde{\gamma}_{3}^{2}/\tilde{C}^{2}\notag\\
\gamma_{2}(1+{\gamma}_{3}^{2}/C^{2}) & =2\gamma_{2}-2\gamma_{2}^{2}\notag\\
{\tilde{\gamma}}_{2}(1+{\gamma}_{3}^{2}/C^{2}) & =2{\tilde{\gamma}}_{1}{\tilde{\gamma}}_{2}-2{\tilde{\gamma}}_{2}^{2}\notag\\
{\gamma}_{3}(1+{\gamma}_{3}^{2}/C^{2})&=-2\gamma_{3}\gamma_{5}-2\tilde{\gamma}_{3}{\gamma}_{7}-E(2\tilde{\gamma}_{3}\gamma_{5}\notag\\
&+2\gamma_{3}{\gamma}_{7})-2\gamma_{3}\gamma_{4}+4\gamma_{3}-2\gamma_{2}\gamma_{3}\notag\\
{\tilde{\gamma}}_{3}(1+{\gamma}_{3}^{2}/C^{2})&=-2{\gamma}_{3}{\gamma}_{7}-2\tilde{\gamma}_{3}{\gamma}_{6}-E(2{\gamma}_{3}{\gamma}_{6}\notag\\
&+2\tilde{\gamma}_{3}{\gamma}_{7})-2\tilde{\gamma}_{3}{\gamma}_{4}+4\tilde{\gamma}_{1}\tilde{\gamma}_{3}-2\tilde{\gamma}_{2}\tilde{\gamma}_{3}\notag\\
{\gamma}_{4}(1+{\gamma}_{3}^{2}/C^{2})&=-2{\gamma}_{3}^{2}-2\tilde{\gamma}_{3}^{2}-E(4\gamma_{3}\tilde{\gamma}_{3})-2{\gamma}_{4}^{2}\notag\\
{\gamma}_{5}(1+{\gamma}_{3}^{2}/C^{2})&=-2\gamma_{5}^{2}-2{\gamma}_{7}^{2}-E(4\gamma_{5}{\gamma}_{7})-2\gamma_{3}^{2}\notag\\
{\gamma}_{6}(1+{\gamma}_{3}^{2}/C^{2})&=-2{\gamma}_{7}^{2}-2{\gamma}_{6}^{2}-E(4\gamma_{6}{\gamma}_{7})-2\tilde{\gamma}_{3}^{2}\notag\\
{\gamma}_{7}(1+{\gamma}_{3}^{2}/C^{2})&=-2\gamma_{5}{\gamma}_{7}-2\gamma_{6}{\gamma}_{7}-E(2\gamma_{5}\gamma_{6}+2{\gamma}_{7}^{2})\notag\\
&-2\gamma_{3}\tilde{\gamma}_{3}.\label{9_fixed_point_equations}
\end{align}
The solutions for $\gamma_i$  in  Ref.~\cite{CKF2016} are reproduced by setting $\tilde{\gamma}_{1}=\gamma_{2}=\tilde{\gamma}_{2}=\tilde{\gamma}_{3}=\gamma_{6}=\gamma_{7}=0$ to obtain
\begin{align}
{\gamma}_{3}(1+{\gamma}_{3}^{2}/C^{2})&=-2\gamma_{3}\gamma_{5}-2\gamma_{3}\gamma_{4}+4\gamma_{3}-2\gamma_{2}\gamma_{3}\notag\\
{\gamma}_{4}(1+{\gamma}_{3}^{2}/C^{2})&=-2{\gamma}_{3}^{2}-2{\gamma}_{4}^{2}\notag\\
{\gamma}_{5}(1+{\gamma}_{3}^{2}/C^{2})&=-2\gamma_{5}^{2}-2\gamma_{3}^{2}\label{case_1}
\end{align}
This leads to the following five solutions
\begin{enumerate}[(1)]

\item
$\gamma_{3}= \pm C \sqrt{-1 + 8 C^2 + 4 \sqrt{1 - C^2 + 4 C^4}}$\\
$\gamma_{4}=\gamma_{5}= 1 - 2 C^2 \pm \sqrt{1 - C^2 + 4 C^4}$

\item
$\gamma_{3}=0, \gamma_{4}=-\frac{1}{2}, \gamma_{5}=-\frac{1}{2}$

\item
$\gamma_{3}=0, \gamma_{4}=-\frac{1}{2}, \gamma_{5}=0$

\item
$\gamma_{3}=0, \gamma_{4}=0, \gamma_{5}=-\frac{1}{2}$

\item
$\gamma_{3}=0, \gamma_{4}=0, \gamma_{5}=0$
\end{enumerate}
Analyzing the stability as explained in the next section, we find that solutions (2)-(5) are unstable with more than one unstable direction. Additionally the negative sign in the expression for $\gamma_4$ and $\gamma_5$ in solution (1) also leads to several unstable directions.
Regarding the remaining solution in (1), we anticipate that it is stable and that $\gamma_{3}$  retains a positive sign, because its initial value is positive for repulsive interactions. Therefore we obtain as a first stable fixed trajectory
\begin{align}
u_{1}(L)&=\frac{1}{1+\gamma_{3}^{2}/C^{2}}\frac{1}{L_{0}-L},\notag\\
\gamma_{3}&=+C \sqrt{-1 + 8 C^2 + 4 \sqrt{1 - C^2 + 4 C^4}},\notag\\
\gamma_{4}&=\gamma_{5}=1-2C^2-\sqrt{1-C^2+4C^4}\label{fixed_point_case1},
\end{align}
which corresponds to Eq.~\eqref{FT_1} of the main text.

To obtain the second stable fixed trajectory, Eq.~\ref{FT_2} in the main text, we write the couplings as $u_{i}=\gamma_{i}\tilde{u}_{1}$, $\tilde{u}_{i}=\tilde{\gamma}_{i}\tilde{u}_{1}$. One finds the same structure of equations as above (Eq.~\eqref{case_1}) with  $\gamma_{3}$ replaced by  $\tilde{\gamma}_{3}$, $\gamma_{5}$ by $\gamma_{6}$, and $C$ by $\tilde C$. The stability analysis then is analogous to the one for Eq.~\eqref{case_1}, and we obtain as second stable fixed trajectory
\begin{align}
\tilde{u}_{1}(L)&=\frac{1}{1+\tilde{\gamma}_{3}^{2}/\tilde{C}^{2}}\frac{1}{L_{0}-L},\notag\\
\tilde{\gamma}_{3}&=+\tilde{C} \sqrt{-1 + 8 \tilde{C}^2 + 4 \sqrt{1 - \tilde{C}^2 + 4 \tilde{C}^4}},\notag\\
\gamma_{4}&=\gamma_{6}=1-2\tilde{C}^2-\sqrt{1-\tilde{C}^2+4\tilde{C}^4}.\label{fixed_point_case2}\\\notag
\end{align}

\subsubsection{Weakly unstable fixed trajectory}\label{sec:fixed_point_solution_detail_3}
Since in the case of the weakly unstable fixed trajectory the situation is more involved, we first consider the simpler case when $C=\tilde{C}$ (implying $A_0=1$). Then the parquet RG equations simplify to
\begin{align}
u_{1}&=\tilde{u}_{1}\notag\\
u_{2}&=\tilde{u}_{2}\notag\\
u_{3}&=\tilde{u}_{3}\notag\\
u_{5}&={u}_{6}\notag\\
\dot{u}_{1}&={u}_{1}^{2}+{u}_{3}^{2}/C^{2}\notag\\
\dot{u}_{2} & =2u_{1}u_{2}-2u_{2}^{2}\notag\\
\dot{u}_{3}&=-2u_{3}u_{5}(1+E)-2{u}_{3}{u}_{7}(1+E)\notag\\
&-2u_{3}u_{4}+4u_{1}u_{3}-2u_{2}u_{3}\notag\\
\dot{u}_{4}&=-4{u}_{3}^{2}(1+E)-2{u}_{4}^{2}\notag\\
\dot{u}_{5}&=-2u_{5}^{2}-2{u}_{7}^{2}-E(4u_{5}{u}_{7})-2u_{3}^{2}\notag\\
\dot{u}_{7}&=-4u_{5}{u}_{7}-E(2u_{5}^{2}+2{u}_{7}^{2})-2u_{3}^{2}.
\end{align}
We again reformulate these equations in terms of $u_{i}=\gamma_{i}u_{1}$ and determine the ratios $\gamma_i$. We solve the resulting algebraic set of equations numerically. We find that solutions with $\gamma_{3}=0$ are truly unstable and as above $\gamma_{3}<0$ cannot be reached with repulsive initial conditions.
For $\gamma_{3}>0$ and varying C and E, we find two solutions. One of them exhibits only one unstable directions, while the second one is more unstable. For example, when $m_h/m_e=1$ and $A_0=1$, we get $C=\sqrt{\frac{3}{2}}$ and $E=\frac{1}{3}$, and the two solutions are $\gamma_{2}=0$ and
\begin{align}
\gamma_{3}&=9.66, \gamma_{4}= -14.81,
\gamma_{5}=\gamma_{6}=-5.55, \gamma_{7}=-5.55;\notag\\
\gamma_{3}&=9.66, \gamma_{4}= -14.81,\gamma_{5}=\gamma_{6}=-29.27, \gamma_{7}=18.16
\end{align}
In this case the first fixed trajectory has one unstable direction and the second fixed trajectory has three such directions.

\begin{widetext}
Also in the general case, when $\tilde{C}\ne C$, there are two solutions for $\gamma_{3}>0$. Both are unstable with one and three unstable directions. We call the solution with only one unstable direction, the weakly unstable fixed trajectory. Explicitly the solutions for general $\tilde{C}\ne C$ are determined by
\begin{align}
\tilde{\gamma}_{1}&=1,\quad
\gamma_{2}=\tilde{\gamma}_{2}=0, \quad \tilde{\gamma}_{3}=\frac{\tilde{C}}{C}\gamma_{3} \notag\\
\gamma_{3}&=\sqrt{-C^2\left[\left(1 + E \frac{\tilde{C}}{C} + \alpha \left(\frac{\tilde{C}}{C}+E\right)\right) 2\gamma_{5}+ 2 \beta\left(\frac{\tilde{C}}{C} + E\right) \gamma_{6} + 2\gamma_{4}-3\right] }\notag\\
\gamma_{4}&=-\frac{1}{2}\left(c_{1}-3+\frac{c_{2}}{a_{1}\gamma_{5} + b_{1} \gamma_{6} + c_{3}}\right)\notag\\
\gamma_{7}&=\alpha\gamma_{5}+\beta\gamma_{6}\label{wFT_general},
\end{align}
where $\gamma_{5}$ and $\gamma_{6}$ are given by the solution of the following two equations of third order
\begin{align}
&(-a_{1}^2 + a_{1} x_{1})\gamma_{5}^3  + (-2 a_{1} b_{1} + b_{1} x_{1} + a_{1} z_{1})\gamma_{5}^2 \gamma_{6}  +
(-a_{1} - a_{1} c_{1} - a_{1} c_{3} - 2 a_{1}^2 C^2 + c_{3} x_{1})\gamma_{5}^2  \notag\\
&+(-b_{1}^2 + a_{1} y_{1} + b_{1} z_{1})\gamma_{5} \gamma_{6}^2  +
(-b_{1} - b_{1} c_{1} - b_{1} c_{3} - 4 a_{1} b_{1} C^2 + c_{3} z_{1}) \gamma_{5} \gamma_{6} \notag\\
& + (-c_{2} - c_{3} -
c_{1} c_{3} - 2 a_{1} c_{1} C^2 - 2 a_{1} c_{3} C^2) \gamma_{5} + b_{1}y_{1} \gamma_{6}^3+ (-2 b_{1}^2 C^2 + c_{3} y_{1})\gamma_{6}^2  \notag\\
&  + (-2 b_{1} c_{1} C^2 -
2 b_{1} c_{3} C^2) \gamma_{6} + (-2 c_{2} C^2 - 2 c_{1} c_{3} C^2) = 0\notag\\
& a_{1} x_{2}\gamma_{5}^3  +  (-a_{1}^2 + b_{1} x_{2} + a_{1} z_{2})\gamma_{5}^2 \gamma_{6} +
(-2 a_{1}^2 \tilde{C}^2 + c_{3} x_{2}) \gamma_{5}^2 +
(-2 a_{1} b_{1} + a_{1} y_{2} + b_{1} z_{2}) \gamma_{5} \gamma_{6}^2 \notag\\
& +  (-a_{1} - a_{1} c_{1} - a_{1} c_{3} - 4 a_{1} b_{1} \tilde{C}^2 +
c_{3} z_{2})\gamma_{5} \gamma_{6}  + (-2 a_{1} c_{1} \tilde{C}^2 - 2 a_{1} c_{3} \tilde{C}^2) \gamma_{5} + (-b_{1}^2 + b_{1} y_{2})
\gamma_{6}^3\notag\\
& +      (-b_{1} - b_{1} c_{1} - b_{1} c_{3} - 2 b_{1}^2 \tilde{C}^2 + c_{3} y_{2})\gamma_{6}^2 + (-c_{2} - c_{3} -
c_{1} c_{3}
- 2 b_{1} c_{1} \tilde{C}^2 - 2 b_{1} c_{3} \tilde{C}^2) \gamma_{6} \notag\\&+ (-2 c_{2} \tilde{C}^2 -
2 c_{1} c_{3} \tilde{C}^2) = 0\notag\\
\end{align}
\end{widetext}
In these expressions, we introduced the parameters
\begingroup
\allowdisplaybreaks
\begin{align}
\alpha&=-(1+E\frac{\tilde{C}}{C})/(\frac{\tilde{C}}{C}-\frac{C}{\tilde{C}})\notag\\
\beta&= (1 + E \frac{C}{\tilde{C}})/(\frac{\tilde{C}}{C} - \frac{C}{\tilde{C}})\notag\\
a_{1}&= -2 ((1 + E \frac{\tilde{C}}{C})+ \alpha(\frac{\tilde{C}}{C} + E))\notag\\
b_{1}&= -2 \beta (\frac{\tilde{C}}{C} + E)\notag\\
c_{1}&= 3 + 4 C^2 (1 + \frac{\tilde{C}^2}{C^2} + 2 E \frac{\tilde{C}}{C})\notag\\
c_{2}&=4C^2(1+\frac{\tilde{C}^2}{C^2}+2E\frac{\tilde{C}}{C})(4C^2(1+\frac{\tilde{C}^2}{C^2}+2E\frac{\tilde{C}}{C})-1)\notag\\
c_{3} &= 4 - 4 C^2 (1 + \frac{\tilde{C}^2}{C^2} + 2 E \frac{\tilde{C}}{C})\notag\\
x_{1} &= -2 (1 + \alpha^2 + 2 \alpha E)\notag\\
y_{1} &= -2 \beta^2;\notag\\
z_{1} &= -(4 \alpha \beta + 4 \beta E)\notag\\
x_{2} &= -2 \alpha^2\notag\\
y_{2} &= -2 (1 + \beta^2 + 2 \beta E)\notag\\
z_{2} &= -(4 \alpha \beta + 4 \alpha E)
\end{align}
\endgroup
As we said, there are two solutions for $\gamma_{5}$ and $\gamma_{6}$, which we obtain numerically with $A_0$ and $m_h/m_e$ as parameters. One of them leads to the weakly unstable trajectory with a single unstable direction and the other leads to the solution with three unstable directions.

\subsection{Stability analysis}\label{sec:stability_analysis}

As we have explained in App.~\ref{sec:fixed_point_solution_detail}, the calculation of fixed trajectories can conveniently be done by transforming the parquet RG equations for the couplings to equations for the ratios $\gamma_i,\tilde\gamma_i$. The ratios are determined by choosing one of the relevant couplings, e.g. $u_1$, and rewriting the other couplings as $u_i=\gamma_i u_1$ and $\tilde u_i=\tilde \gamma_i u_1$. We can hence analyze the stability of  different fixed trajectories in terms of the flow equations for the ratios.
Therefore we consider small deviations of $\gamma_i$ from their values on a fixed trajectory and determine whether these deviations increase or decrease during the RG flow.
 We label the flow equations for the ratios as $\beta_i=\frac{d}{dL}\gamma_i$, $\tilde\beta_i=\frac{d}{dL}\tilde\gamma_i$, and the deviations from a fixed trajectory as $\delta=\gamma-\gamma^{fix point}$. Linearizing the flow equations for small deviations from a fixed trajectory we obtain
\begin{align}\label{stability_analysis}
\begin{pmatrix}
\dot{\tilde{\delta}}_{1}        \\
\dot{\delta}_{2}        \\
\dot{\tilde{\delta}}_{2}        \\
\dot{\delta}_{3}        \\
\dot{\tilde{\delta}}_{3}        \\
\dot{\delta}_{4}     \\
\dot{\delta}_{5}     \\
\dot{\delta}_{6}     \\
\dot{\delta}_{7}
\end{pmatrix}
=
\begin{pmatrix}
\frac{\partial \tilde\beta_1}{\partial \tilde \gamma_1} & \frac{\partial \tilde\beta_1}{\partial \gamma_2} & \dots  & \frac{\partial \tilde\beta_1}{\partial \gamma_7} \\
\frac{\partial \beta_2}{\partial \tilde \gamma_1} & \frac{\partial \beta_2}{\partial \gamma_2} & \dots  & \frac{\partial \beta_2}{\partial \gamma_7} \\
\vdots & \vdots  & \ddots & \vdots \\
\frac{\partial \beta_7}{\partial \tilde \gamma_1} & \frac{\partial \beta_7}{\partial \gamma_2}  & \dots  & \frac{\partial \beta_7}{\partial \gamma_7}
\end{pmatrix}
\begin{pmatrix}
\tilde{\delta}_{1}        \\
\delta_{2}        \\
\tilde{\delta}_{2}        \\
\delta_{3}        \\
\tilde{\delta}_{3}        \\
\delta_{4}     \\
\delta_{5}     \\
\delta_{6}     \\
\delta_{7}
\end{pmatrix}
\end{align}
The eigenvalues of the stability matrix $\partial \beta_i/\partial \gamma_i$
at $\gamma_i$ taken on the fixed trajectory
 determine if deviations grow or flow back to a given fixed trajectory. A positive eigenvalue signals growing deviations, and therefore an unstable direction corresponding to the eigenvector of the positive eigenvalue. Attaining such an unstable fixed trajectory requires fine tuning of initial conditions, and with more positive eigenvalues, more initial couplings must be fined-tuned.
Only if all eigenvalues are negative, deviations in every direction will decrease during the flow. This is what happens for a stable fixed trajectory.

For example for the stable fixed trajectory of Eq.~(\ref{FT_1}) (i.e.~Eq.~(\ref{fixed_point_case1}) in the appendix), the eigenvalues for the stability matrix are
$-4u_1$, $ -4 ( 2 C^2 + \sqrt{1 - C^2 + 4 C^4} )u_1$, $ -4 ( 2 C^2 + \sqrt{1 - C^2 + 4 C^4} )u_1$, $ -4 ( 2 C^2 + \sqrt{1 - C^2 + 4 C^4} )u_1$, $ -4 (2 C^2-\frac{1}{2} + \sqrt{1 - C^2 + 4 C^4})u_1 $,
$ -2 (1 + 2 C^2 + \sqrt{1 - C^2 + 4 C^4} \pm C \sqrt{-1 + 8 C^2 + 4 \sqrt{1 - C^2 + 4 C^4}})u_1$ and $ -(1 + 8 C^2 +
4 \sqrt{1 - C^2 + 4 C^4} \pm \sqrt{-15 + 32 C^2 + 16 \sqrt{1 - C^2 + 4 C^4}})u_1 $. Since $u_{1}$ is positive, these eigenvalues are negative for any $C$, i.e. the fixed  trajectory is stable.  Analogously we determine that the fixed trajectory of Eq.~(\ref{FT_2}) (Eq.~(\ref{fixed_point_case2}) in the appendix) is stable and that the weakly unstable fixed trajectory Eq.~(\ref{FT_3}) (Eq.~(\ref{wFT_general}) in the appendix) has merely one unstable direction. Furthermore we find that this single unstable direction is weak in the sense that the corresponding eigenvalue is small.

\subsection{Subleading ordering tendencies}\label{app:subl}
In the main text, we have discussed instabilities towards superconducting order, $(\pi,0)$ and $(0,\pi)$ SDWs, and Pomeranchuk order in the $A_{1g}$ and $B_{1g}$ channel. We also considered further instabilities regarding CDW order, SDW order with momentum transfer $(\pi,\pi)$ and Pomeranchuk order in the $A_{2g}$ and $B_{2g}$ channel. However, we found them to be subleading. We discuss the details in the following.
 \subsubsection{Charge density wave (CDW) channel at $(0,\pi)$ and $(\pi,0)$}
 \label{sec:CDW0pi}
A CDW instability breaks the translational symmetry of the lattice and is characterized by particle-hole order parameters at finite momenta, $(0,\pi)$ and $(\pi,0)$.

In this case, we define the auxiliary complex test fields as follows,
 \begin{align}\label{CDW_test}
 H_{CDW}&=\notag\\
 \sum_{k\sigma}\big[\delta_{1}^{(0)}&\psi^{\dag}_{1,\sigma}(k)\psi_{5,\sigma}(k)+\delta_{1'}^{(0)}\psi^{\dag}_{1,\sigma}(k)\psi_{6,\sigma}(k)
 \notag \\
 +\delta_{2}^{(0)}&\psi^{\dag}_{2,\sigma}(k)\psi_{5,\sigma}(k)+\delta_{2'}^{(0)}\psi^{\dag}_{2,\sigma}(k)\psi_{6,\sigma}(k)\notag\\
 +\delta_{3}^{(0)}&\psi^{\dag}_{3,\sigma}(k)\psi_{6,\sigma}(k)+\delta_{3'}^{(0)}\psi^{\dag}_{3,\sigma}(k)\psi_{5,\sigma}(k)
 \notag \\
 +\delta_{4}^{(0)}&\psi^{\dag}_{4,\sigma}(k)\psi_{6,\sigma}(k)+\delta_{4'}^{(0)}\psi^{\dag}_{4,\sigma}(k)\psi_{5,\sigma}(k)
 +h.c.\big]
 \end{align}

 The parquet RG equations describing the renormalization of the real and imaginary parts of these vertices are
 \begin{align}
 \frac{d}{dL}\delta_{1(1')}^{re}&=\delta_{1(1')}^{re}(u_1-2u_2-\frac{u_3}{C})\notag\\
 \frac{d}{dL}\delta_{1(1')}^{im}&=\delta_{1(1')}^{im}(u_1-2u_2+\frac{u_3}{C})\notag\\
 \frac{d}{dL}\delta_{2(2')}^{re}&=\delta_{2(2')}^{re}(\tilde{u}_1-2\tilde{u}_2-\frac{\tilde{u}_3}{\tilde{C}})\notag\\
 \frac{d}{dL}\delta_{2(2')}^{im}&=\delta_{2(2')}^{im}(\tilde{u}_1-2\tilde{u}_2+\frac{\tilde{u}_3}{\tilde{C}})\label{pRG_CDW}
 \end{align}
 where we have already used the equivalences of Eq.~\eqref{anz} to simplify the expressions.
 The vertices $\delta_{3(3')}$ and $\delta_{4(4')}$ satisfy the same parquet RG equations as $\delta_{1(1')}$ and $\delta_{2(2')}$, respectively, due to the $C_4$ symmetry explained below Eq.~(\ref{Hubbard_relation}).

We perform the same procedure as in the main text, i.e.~we take the fixed-trajectory solutions for the couplings as input for the Eqs.~\eqref{pRG_CDW}, which allows to solve for the vertices. The vertices, in turn, determine the susceptibilities and signal the appearance or the absence of the corresponding order.
For the stable fixed trajectory in Eq.~\eqref{FT_1}, we then obtain 
 \begin{align}
 \delta_{1}^{re}(L)&=\delta_{1}^{re,(0)}(\frac{L_0}{L_0-L})^{\beta_{CDW,1}^{re}}\notag\\
 \beta_{CDW,1}^{re}&=\frac{1-\gamma_{3}/C}{1+\gamma_{3}^{2}/{C^2}}\, .\label{delta_CDW}
 \end{align}
 The parquet RG equation satisfied by the susceptibility reads
 \begin{align}
 \frac{d\chi_{CDW,1}^{re}}{dL}&=[\frac{\delta_{1}^{re}}{\delta_{1}^{re,(0)}}]^2\, .\label{chi_CDW}
 \end{align}
With Eq.~(\ref{delta_CDW}) on the stable fixed trajectory, this leads to 
 \begin{align}
 \chi_{CDW,1}^{re}(L)&=
 \frac{L_0^{2\beta_{CDW,1}^{re}}}{1-2\beta_{CDW,1}^{re}}(L_0-L)^{-\alpha_{CDW,1}^{re}}\, ,
 \label{charge_susceptibility}
 \end{align}
 where the scaling exponent of the susceptibility is
 \begin{align}
 \alpha_{CDW,1}^{re} = 2\beta_{CDW,1}^{re}-1\, .
 \end{align}
 Similarly, we determine $\alpha_{CDW,2}^{re}=\alpha_{CDW,2}^{im}=0 $ and $\alpha_{CDW,1}^{im}=2\beta_{CDW,1}^{im}-1$, where $\beta_{CDW,1}^{im}=\frac{1+\gamma_{3}/C}{1+\gamma_{3}^{2}/{C^2}}$. Furthermore,  as can be seen in \eqref{pRG_CDW}, we find $\alpha_{CDW,i}^{re,im}=\alpha_{CDW,i'}^{re,im}$.

We now perform the same analysis for the second stable fixed trajectory and the weakly unstable fixed trajectory.
 For the first stable fixed trajectory in Eq.~\eqref{FT_1}, we obtain the susceptibility exponents
 \begin{align}\label{CDW_1}
 \alpha_{CDW,1}^{re}&=2\frac{1-\gamma_{3}/C}{1+\gamma_{3}^{2}/{C^2}}-1\notag\\
 \alpha_{CDW,1}^{im}&=2\frac{1+\gamma_{3}/C}{1+\gamma_{3}^{2}/{C^2}}-1\notag\\
 \alpha_{CDW,2}^{re}&=0\notag\\
 \alpha_{CDW,2}^{im}&=0
 \end{align}

Furthermore we find that $\alpha_{CDW,1}^{re,im} =  \alpha_{CDW,1'}^{re,im} =  \alpha_{CDW,3}^{re,im} =  \alpha_{CDW,3'}^{re,im}$, and
 $\alpha_{CDW,2}^{re,im} =  \alpha_{CDW,2'}^{re,im} =  \alpha_{CDW,4}^{re,im} =  \alpha_{CDW,4'}^{re,im}$.
 Among the different exponents in Eq.~\eqref{CDW_1}, the largest one is $\alpha_{CDW,1}^{im}$.
 As a result the leading CDW instability is characterized by the order parameter $\langle \psi_1^{\dag} \psi_5 - \psi_5^{\dag} \psi_1\rangle$. Similarly the order parameters $\langle \psi_1^{\dag} \psi_6 - \psi_6^{\dag} \psi_1\rangle$, $ \langle \psi_3^{\dag} \psi_5 - \psi_5^{\dag} \psi_3\rangle$, and $ \langle \psi_3^{\dag} \psi_6 - \psi_6^{\dag} \psi_3\rangle$ lead to the same exponent on our level of approximation and are thus equivalent candidates for the instability.

 The susceptibility exponents of the second stable fixed trajectory \eqref{FT_2} are
 \begin{align}\label{CDW_2}
 \alpha_{CDW,1}^{re}&=0\notag\\
 \alpha_{CDW,1}^{im}&=0\notag\\
 \alpha_{CDW,2}^{re}&=2\frac{1-\tilde{\gamma}_{3}/\tilde{C}}{1+\tilde{\gamma}_{3}^{2}/{\tilde{C}^2}}-1\notag\\
 \alpha_{CDW,2}^{im}&=2\frac{1+\tilde{\gamma}_{3}/\tilde{C}}{1+\tilde{\gamma}_{3}^{2}/{\tilde{C}^2}}-1.
 \end{align}
 In this case the roles of the $d_{xz}/d_{yz}$ and $d_{xy}$ orbitals on electron pockets are interchanged, and
 the largest exponent is $ \alpha_{CDW,2}^{im}$.
Correspondingly, the leading CDW instability is characterized by the order parameter,  $\langle \psi_2^{\dag} \psi_5 - \psi_5^{\dag} \psi_2\rangle$, which is equivalent to the order parameters $ \langle \psi_2^{\dag} \psi_6 - \psi_6^{\dag} \psi_2\rangle$, $ \langle \psi_4^{\dag} \psi_5 - \psi_5^{\dag} \psi_4\rangle $, and $ \langle \psi_4^{\dag} \psi_6 - \psi_6^{\dag} \psi_4\rangle$.

 For the weakly unstable fixed trajectory, Eq. \eqref{FT_3}, we find
 \begin{align}
 \alpha_{CDW,1}^{re}&=2\frac{1-\gamma_{3}/C}{1+\gamma_{3}^{2}/{C^2}}-1\notag\\
 \alpha_{CDW,1}^{im}&=2\frac{1+\gamma_{3}/C}{1+\gamma_{3}^{2}/{C^2}}-1\notag\\
 \alpha_{CDW,2}^{re}&=2\frac{1-\tilde{\gamma}_{3}/\tilde{C}}{1+\tilde{\gamma}_{3}^{2}/{\tilde{C}^2}}-1=\alpha_{CDW,1}^{re}\notag\\
 \alpha_{CDW,2}^{im}&=2\frac{1+\tilde{\gamma}_{3}/\tilde{C}}{1+\tilde{\gamma}_{3}^{2}/{\tilde{C}^2}}-1=\alpha_{CDW,1}^{im},
 \end{align}
 where we used that $\gamma_3/C = \tilde{\gamma}_3/\tilde{C}$ for this fixed trajectory.
 The largest exponent in the CDW channel is again $\alpha_{CDW,1}^{im}$.

In summary we find that for all three fixed trajectories, the largest exponent occurs for charge current operators.
 The corresponding exponent
$ \alpha_{CDW}\equiv  2\frac{1+\gamma_{3}/C}{1+\gamma_{3}^{2}/{C^2}}-1$
is the same as in the SDW channel, i.e.~
$\alpha_{CDW}=\alpha_{SDW}$. The reason is that $\gamma_2=\tilde{\gamma}_2=0$ on the fixed trajectories. However, if $\gamma_2$ and $\tilde{\gamma}_2$ are non-zero and small, we can see in Eq.~\eqref{pRG_CDW} that SDW wins over CDW.

 \subsubsection{Spin density wave and charge density wave at $(\pi,\pi)$ }
 \label{sec:SDWCDWpipi}
Additionally, we consider  SDW and CDW channels with momentum transfer $(\pi,\pi)$. The corresponding coupling to fermion bilinears is given by
 \begin{align}
 H_{SDW,(\pi,\pi)} =&\sum_{k}[\bm{s}_{1,3}^{(0)}\cdot\psi^{\dag}_{1,\alpha}(k)\bm{\sigma}_{\alpha,\beta}\psi_{3,\beta}(k) \notag\\
 &+\bm{s}_{2,4}^{(0)}\cdot\psi^{\dag}_{2,\alpha}(k)\bm{\sigma}_{\alpha,\beta}\psi_{4,\beta}(k)\notag\\
 &+\bm{s}_{1,4}^{(0)}\cdot\psi^{\dag}_{1,\alpha}(k)\bm{\sigma}_{\alpha,\beta}\psi_{4,\beta}(k)+h.c.]
 \end{align}
and
 \begin{align}
 H_{CDW,(\pi,\pi)} =&\sum_{k\sigma}[\delta_{1,3}^{(0)}\psi^{\dag}_{1,\sigma}(k)\psi_{3,\sigma}(k)\notag\\
 &+\delta_{2,4}^{(0)}\psi^{\dag}_{2,\sigma}(k)\psi_{4,\sigma}(k)\notag\\
 &+\delta_{1,4}^{(0)}\psi^{\dag}_{1,\sigma}(k)\psi_{4,\sigma}(k)+h.c.]
 \end{align}
 The RG equations for the vertices then are
 \begin{align}
 \frac{ds_{1,3}^{re,im}}{dL}&=\pm\frac{A^{'}_{e}}{A_e}{u}_5\notag\\
 \frac{ds_{2,4}^{re,im}}{dL}&=\pm\frac{A^{''}_{e}}{\bar{A}_e}{u}_6\notag\\
 \frac{ds_{1,4}^{re,im}}{dL}&=\pm\frac{A^{'''}_{e}}{\tilde{A}_e}E{u}_7
 \end{align}
and
 \begin{align}
 \frac{d\delta_{1,3}^{re,im}}{dL}&=\frac{ds_{1,3}^{im,re}}{dL}\notag\\
 \frac{d\delta_{2,4}^{re,im}}{dL}&=\frac{ds_{2,4}^{im,re}}{dL}\notag\\
 \frac{d\delta_{1,4}^{re,im}}{dL}&=\frac{ds_{1,4}^{im,re}}{dL}
 \end{align}
We calculate the exponents of the susceptibilities for this SDW and CDW order similarly as before. We find that all exponents are smaller than zero.
The result for the largest exponent is shown in Fig.~\ref{fig:exponents_case3} in the main text as $\alpha_{SDW'(CDW')}$.

 \subsubsection{Pomeranchuk instability in $A_{2g}$ and $B_{2g}$ channel}\label{sec:Pom_A2g}
The order parameters in the $A_{2g}$ and $B_{2g}$ Pomeranchuk channel couple to the combination of vertices $\Gamma^{re(im)}_{1,2}= \Gamma_{1,2}\pm\Gamma_{1,2}^*$, $\Gamma^{re(im)}_{3,4}= \Gamma_{3,4}\pm\Gamma_{3,4}^*$ and
$\Gamma^{re(im)}_{5,6}= \Gamma_{5,6}\pm\Gamma_{5,6}^*$ as defined in Eq.~(\ref{test_Pom}). Let us recall that the indices $1-4$ label states on electron and 5-6 states on hole pockets. In the $A_{2g}$ channel the vertices are renormalized according to
\begin{align}\label{Pom_A2}
\Gamma_{1,2}^{re}&=-4Eu_7\Gamma_{1,2}^{re}+\Gamma_{1,2}^{re,(0)}\notag\\
\Gamma_{3,4}^{re}&=-4Eu_7\Gamma_{3,4}^{re}+\Gamma_{3,4}^{re,(0)}\notag\\
\Gamma_{5,6}^{re}&=-2u_4\Gamma_{5,6}^{re}+\Gamma_{5,6}^{re,(0)}\, .
\end{align}

On the weakly unstablefixed trajectory Eq.~\eqref{FT_3}, the susceptibilities in the $A_{2g}$ channel diverge as
\begin{align}
\chi_{A_{2g}}^{e}&\propto (L_{A_{2g}}^{e}-L)^{-1},\notag\\
\chi_{A_{2g}}^{h}&\propto (L_{A_{2g}}^{h}-L)^{-1},
\end{align}
and the divergence occurs on electron and hole pockets at scales $L_{A_{2g}}^{e}$ and $L_{A_{2g}}^{h}$, respectively,
\begin{align}
L_{A_{2g}}^{e}=L_0&+\frac{4E\gamma_7}{1+\gamma_3^{2}/C^2} \notag\\
L_{A_{2g}}^{h}=L_0&+\frac{2\gamma_4}{1+\gamma_3^{2}/C^2}\, . 
\end{align}
We see that $L_{A_{2g}}^{e}$ is larger than $L_{B_{1g}}$, i.e.~it is subleading to the $B_{1g}$ channel. However,
the critical scale in the $B_{1g}$ and $A_{2g}^{h}$ Pomeranchuk channel are formally the same.

But if we consider that the flow will only be close to and not exactly on the weakly unstable fixed trajectory, $\epsilon_1$ and $\epsilon_2$ will be small and non-zero. Then the $B_{1g}$ critical scale always appears before the $A_{2g}^{h}$ critical scale.

In the $B_{2g}$ channel, the vertices are renormalized by
\begin{align}\label{B2_Pom}
\Gamma_{1,2}^{im}&=\Gamma_{1,2}^{im,(0)}\notag\\
\Gamma_{3,4}^{im}&=\Gamma_{3,4}^{im,(0)}\notag\\
\Gamma_{5,6}^{im}&=2u_4\Gamma_{5,6}^{im}+\Gamma_{5,6}^{im,(0)}.
\end{align}
In this channel the vertices
 on electron pockets are not renormalized, while the vertex involving the hole pockets is reduced since
$u_4<0$. As a result, there is no instability in the $B_{2g}$ Pomeranchuk channel.

\end{document}